# Design and Analysis of Nonbinary LDPC Codes for Arbitrary Discrete-Memoryless Channels

Amir Bennatan, *Student Member, IEEE,* David Burshtein, *Senior Member, IEEE*

## Abstract

We present an analysis, under iterative decoding, of coset LDPC codes over GF($q$), designed for use over arbitrary discrete-memoryless channels (particularly nonbinary and asymmetric channels). We use a random-*coset* analysis to produce an effect that is similar to output-symmetry with binary channels. We show that the random selection of the nonzero elements of the GF($q$) parity-check matrix induces a *permutation-invariance* property on the densities of the decoder messages, which simplifies their analysis and approximation. We generalize several properties, including symmetry and stability from the analysis of binary LDPC codes. We show that under a Gaussian approximation, the entire $q - 1$ dimensional distribution of the vector messages is described by a single *scalar* parameter (like the distributions of binary LDPC messages). We apply this property to develop EXIT charts for our codes. We use appropriately designed signal constellations to obtain substantial shaping gains. Simulation results indicate that our codes outperform multilevel codes at short block lengths. We also present simulation results for the AWGN channel, including results within 0.56 dB of the unconstrained Shannon limit (i.e. not restricted to any signal constellation) at a spectral efficiency of 6 bits/s/Hz.

## I. Introduction

In their seminal work, Richardson *et al.* [29], [28] developed an extensive analysis of LDPC codes over memoryless binary-input output-symmetric (MBIOS) channels. Using this analysis, they designed edge-distributions for LDPC codes at rates remarkably close to the capacity of several such channels. However, their analysis is mostly restricted to MBIOS channels. This rules out many important channels, including bandwidth-efficient channels, which require nonbinary channel alphabets.

To design nonbinary codes, Hou *et al.* [18] suggested starting off with binary LDPC codes either as components of a multilevel code (MLC) or a bit-interleaved coded modulation (BICM) scheme. Nonbinary channels are typically

To appear, IEEE Trans. Inf. Theory, (submitted October 2004, revised and accepted for publication, November 2005). The authors are with the School of Electrical Engineering, Tel Aviv University, Ramat Aviv 69978, Tel Aviv, Israel (e-mail: abn@eng.tau.ac.il; burstyn@eng.tau.ac.il). This research was supported by the Israel Science Foundation, grant no. 22/01–1, by an equipment grant from the Israel Science Foundation to the school of Computer Science at Tel Aviv University and by a fellowship from The Yitzhak and Chaya Weinstein Research Institute for Signal Processing at Tel Aviv University. The material in this paper was presented in part at the 41st Annual Allerton Conference on Communications, Control and Computing, Monticello, Illinois, October 2003 and the 2005 IEEE International Symposium on Information Theory, Adelaide, Australia.

not output-symmetric, thus posing a problem to their analysis. To overcome this problem, Hou *et al*. used *coset* LDPC codes rather than plain LDPC codes. The use of coset-LDPC codes was first suggested by Kavčić *et al*. [19] in the context of LDPC codes for channels with intersymbol interference (ISI).

MLC and BICM codes are frequently decoded using multistage and parallel decoding, respectively. Both methods are suboptimal in comparison to methods that rely only on belief-propagation decoding[1]. Full belief-propagation decoding was considered by Varnica *et al*. [37] for MLC and by ourselves in [1] (using a variant of BICM LDPC called BQC-LDPC). However, both methods involve computations that are difficult to analyze.

An alternative approach to designing nonbinary codes starts off with nonbinary LDPC codes. Gallager [16] defined arbitrary-alphabet LDPC codes using modulo-$q$ arithmetic. Nonbinary LDPC codes were also considered by Davey and MacKay [10] in the context of codes for binary-input channels. Their definition uses Galois field (GF($q$)) arithmetic. In this paper we focus on GF($q$) LDPC codes similar to those suggested in [10].

In [1] we considered coset GF($q$) LDPC codes under maximum-likelihood (ML) decoding. We showed that appropriately designed coset GF($q$) LDPC codes are capable of achieving the capacity of any discrete-memoryless channel. In this paper, we examine coset GF($q$) LDPC codes under iterative decoding.

A straightforward implementation of the nonbinary belief-propagation decoder has a very large decoding complexity. However, we discuss an implementation method suggested by Richardson and Urbanke [28][Section V] that uses the multidimensional discrete Fourier transform (DFT). Coupled with an efficient algorithm for computing of the multidimensional DFT, this method reduces the complexity dramatically, to that of the above discussed binary-based MLC and BICM schemes (when full belief-propagation decoding is used).

With binary LDPC codes, the BPSK signals $\pm 1$ are typically used instead of the $\{0, 1\}$ symbols of the code alphabet, when transmitting over the AWGN channel. Similarly, with nonbinary LDPC codes, a straightforward choice would be to use a PAM or QAM signal constellation (which we indeed use in some of our simulations). However, with such constellations, the codes exhibit a *shaping* loss which, at high SNR, approaches 1.53 dB [13]. By carefully selecting the signal constellation, a substantial shaping gain can be achieved. Two approaches that we discuss are *quantization mapping*, which we have used in [1] (based on ideas by Gallager [17] and McEliece [25]) and *nonuniform spacing* (based on Sun and van Tilborg [33] and Fragouli *et al*. [14]).

An important aid in the analysis of binary LDPC codes is *density evolution*, proposed by Richardson and Urbanke [28]. Density evolution enables computing the exact threshold of binary LDPC codes asymptotically at large block lengths. Using density evolution, Chung *et al*. [8] were able to present irregular LDPC codes within 0.0045 dB of the Shannon limit of the binary-input AWGN channel. Efficient algorithms for computing density-evolution were proposed in [28] and [8].

Density evolution is heavily reliant on the output-symmetry of typical binary channels. In this paper, we show that focusing on coset-codes enables extension of the concepts of density-evolution to nonbinary LDPC codes. We examine our codes in a *random coset* setting, where the average performance is evaluated over all possible realizations of the coset vector. Our approach is similar to the one used by Kavčić *et al.* [19] for binary channels with ISI. Random-coset analysis enables us to generalize several properties from the analysis of binary LDPC, including the all-zero codeword assumption[2], and the symmetry property of densities.

---

[1]Multistage decoding involves transferring a hard decision on the decoded codeword (rather than a soft decision) from one component code to the next. It further does not benefit from feedback on this decision from subsequent decoders. Parallel decoding of BICM codes is bounded away from capacity as discussed in [7].

In [9] and [35], approximations of the density-evolution were proposed that use a Gaussian assumption. These approximations track one-dimensional surrogates rather that the true densities, and are easier to implement. A different approach was used in [6] to develop one-dimensional surrogates that can be used to compute lower-bounds on the decoding threshold.

Unlike binary LDPC codes, the problem of finding an efficient algorithm for computing density evolution for nonbinary LDPC codes remains open. This is a result of the fact that the messages transferred in nonbinary belief-propagation are multidimensional vectors rather than scalar values. Just storing the density of a non-scalar random variable requires an amount of memory that is exponential in the alphabet size. Nevertheless, we show that approximation using surrogates is very much possible.

With LDPC codes over GF($q$), the nonzero elements of the sparse parity-check matrix are selected at random from GF($q$)\{0}. In this paper, we show that this random selection induces an additional symmetry property on the distributions tracked by density-evolution, which we call *permutation-invariance*. We use permutation-invariance to generalize the stability property from binary LDPC codes.

Gaussian approximation of nonbinary LDPC was first considered by Li *et al.* [22] in the context of transmission over binary-input channels. Their approximation uses $q - 1$ dimensional vector parameters to characterize the densities of messages, under the assumption that the densities are approximately Gaussian. We show that assuming permutation-invariance, the densities may in fact be described by scalar, one-dimensional parameters, like the densities of binary LDPC.

Finally, binary LDPC codes are commonly designed using EXIT charts, as suggested by ten Brink *et al.* [35]. EXIT charts are based on the Gaussian approximation of density-evolution. In this paper, we therefore use the generalization of this approximation to extend EXIT charts to coset GF($q$) LDPC codes. Using EXIT charts, we design codes at several spectral efficiencies, including codes at a spectral efficiency of 6 bits/s/Hz within 0.56 dB of the unconstrained Shannon limit (i.e., when transmission is not restricted to any signal constellation). To the best of our knowledge, these are the best codes designed for this spectral efficiency. We also compare coset GF($q$) LDPC codes to codes constructed using multilevel coding and turbo-TCM, and provide simulation results that indicate that our codes outperform these schemes at short block-lengths.

Our work is organized as follows: We begin by introducing some notation in Section II [3]. In Section III we formally define coset LDPC codes over GF($q$) and ensembles of codes, and discuss mappings to the channel alphabet. In Section IV we present belief-propagation decoding of coset GF($q$) LDPC codes, and discuss its efficient implementation. In Section V we discuss the all-zero codeword assumption, symmetry and channel equivalence. In Section VI we present density evolution for nonbinary LDPC and permutation-invariance. We also develop the stability property and Gaussian approximation. In Section VII we discuss the design of LDPC codes using EXIT charts and present simulation results. In Section VIII, we compare our codes with multilevel coding and turbo-TCM. Section IX presents ideas for further research and concludes the paper.

---

[2]Note that in [38], an approach to generalizing density evolution to asymmetric binary channels was proposed that does not require the all-zero codeword assumption.

[3]We have placed this section first for easy reference, although none of the notations are required to understand Section III.

## II. NOTATION

### A. General Notation

Vectors are typically denoted by boldface e.g. $\mathbf{x}$. Random variables are denoted by upper-case letters, e.g. $X$ and their instantiations in lower-case, e.g. $x$. We allow an exception to this rule with random variables over GF($q$), to enable neater notation.

For simplicity, throughout this paper, we generally assume discrete random variables (with one exception involving Gaussian approximation). The generalization to continuous variables is immediate.

### B. Probability and LLR Vectors

An important difference between nonbinary and binary LDPC decoders is that the former use messages that are multidimensional vectors, rather than scalar values. Like the binary decoders, however, there are two possible representations for the messages: plain-likelihood probability-vectors or log-likelihood-ratio (LLR) vectors.

A $q$-dimensional probability-vector is a vector $\mathbf{x} = (x_0, ..., x_{q-1})$ of real numbers such that $x_i \geq 0$ for all $i$ and $\sum_{i=0}^{q-1} x_i = 1$. The indices $i = 0, ..., q-1$ of each message vector's components are also interpreted as elements of GF($q$). That is, each index $i$ is taken to mean the $i$th element of GF($q$), given some enumeration of the field elements (we assume that indices $0$ and $1$ correspond to the zero and one elements of the field, respectively).

Given a probability-vector $\mathbf{x}$, the LLR values associated with it are defined as $w_i \triangleq \log(x_0/x_i)$, $i = 0, ..., q-1$ (a definition borrowed from [22]).

Notice that for all $\mathbf{x}$, $w_0 = 0$. We define the LLR-*vector* representation of $\mathbf{x}$ as the $q - 1$ dimensional vector $\mathbf{w} = (w_1, ..., w_{q-1})$. For convenience, although $w_0$ is not defined as belonging to this vector, we will allow ourselves to refer to it with the implicit understanding that it is always equal to zero.

Given an LLR vector $\mathbf{w}$, the components of the corresponding probability-vector (the probability vector from which $\mathbf{w}$ was produced) can be obtained by

$$x_i = \mathrm{LLR}_i^{-1}(\mathbf{w}) = \frac{e^{-w_i}}{1 + \sum_{k=1}^{q-1} e^{-w_k}}, \quad i = 0, ..., q-1 \tag{1}$$

We use the shorthand notation $\mathbf{x}'$ to denote the LLR-vector representation of a probability-vector $\mathbf{x}$. Similarly, if $\mathbf{w}$ is an LLR-vector, then $\mathbf{w}'$ is its corresponding probability-vector representation.

A probability-vector random variable is defined to be a $q$-dimensional random variable $\mathbf{X} = (X_0, ..., X_{q-1})$, that takes only valid probability-vector values. An LLR-vector random variable is a $q - 1$-dimensional random variable $\mathbf{W} = (W_1, ..., W_{q-1})$.

### C. The Operations $\times g$ and $+g$

Given a probability vector $\mathbf{x}$ and an element $g \in \mathrm{GF}(q)$, we define the $+g$ operator in the following way (note that a different definition will shortly be given for LLR vectors)

$$\mathbf{x}^{+g} \triangleq (x_g, x_{1+g}, ..., x_{(q-1)+g}) \tag{2}$$

where addition is performed over $\mathrm{GF}(q)$. $\mathbf{x}^*$ is defined as the set

$$\mathbf{x}^* \triangleq \{\mathbf{x}, \mathbf{x}^{+1}, ..., \mathbf{x}^{+(q-1)}\} \tag{3}$$

We define $n(\mathbf{x})$ as the number of elements $g \in \mathrm{GF}(q)$ satisfying $\mathbf{x}^{+g} = \mathbf{x}$. For example, assuming GF(3) addition, $n([1, 0, 0]) = 1$, and $n([1/3, 1/3, 1/3]) = 3$. Note that $n(\mathbf{x}) \geq 1$ for all $\mathbf{x}$, because $\mathbf{x}^{+0} = \mathbf{x}$.

Similarly, we define

$$\mathbf{x}^{\times g} \quad \triangleq \quad (x_0, x_g, x_{2 \cdot g}, ..., x_{(q-1) \cdot g}) \tag{4}$$

Note that the operation $+g$ is reversible, and $(\mathbf{x}^{+g})^{-g} = \mathbf{x}$. Similarly, $\times g$ is reversible for all $g \neq 0$, and $(\mathbf{x}^{\times g})^{\times g^{-1}} = \mathbf{x}$. In Appendix I we summarize some additional properties of these operators that are used in this paper.

In the context of LLR vectors, we define the operation $+g$ differently. Given an LLR vector $\mathbf{w}$, we define $\mathbf{w}^{+g}$ using the corresponding probability vector. That is, $\mathbf{w}^{+g} \triangleq \mathrm{LLR}([\mathrm{LLR}^{-1}(\mathbf{w})]^{+g})$. Thus we obtain:

$$w_i^{+g} = w_{i+g} - w_g, \quad i = 1, ..., q-1 \tag{5}$$

The operation $\times g$ is similarly defined as $\mathbf{w}^{\times g} \triangleq \mathrm{LLR}([\mathrm{LLR}^{-1}(\mathbf{w})]^{\times g})$. However, unlike the $+g$ operation, the resulting definition coincides with the definition for probability vectors, and

$$w_i^{\times g} = w_{i \cdot g}, \quad i = 1, ..., q-1$$

## III. COSET GF($q$) LDPC CODES DEFINED

We begin in Section III-A by defining LDPC codes over GF($q$). We proceed in Section III-B to define coset GF($q$) LDPC codes. In Section III-C we define the concept of mappings, by which coset GF($q$) LDPC codes are tailored to specific channels. In Section III-D we discuss ensembles of coset GF($q$) LDPC codes.

### A. LDPC Codes over GF($q$)

A GF($q$) LDPC code is defined in a way similar to binary LDPC codes, using a bipartite Tanner graph [34]. The graph has $N$ *variable* (left) nodes, corresponding to codeword symbols, and $M$ *check* (right) nodes corresponding to parity-checks.

Two important differences distinguish GF($q$) LDPC codes from their binary counterparts. Firstly, the codeword elements are selected from the entire field GF($q$). Hence, each variable-node is assigned a symbol from GF($q$), rather than just a binary digit. Secondly, at each edge $(i, j)$ of the Tanner graph, a *label* $g_{i,j} \in \mathrm{GF}(q) \backslash \{0\}$ is defined. Figure 1 illustrates the labels at the edges adjacent to some check node of an LDPC code's bipartite graph (the digits 1, 2 and 5 represent nonzero elements of GF($q$)).

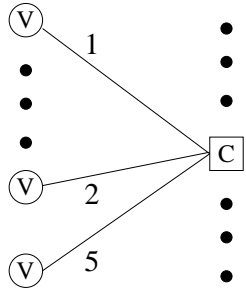

Fig. 1. Schematic diagram of a GF($q$) LDPC bipartite graph.

A word **c** with components from GF($q$) is a codeword if at each check-node $j$, the following equation holds:

$$\sum_{i \in \mathcal{N}(j)} g_{i,j} c_i = 0$$

where $\mathcal{N}(j)$ is the set of variable nodes adjacent to $j$. The GF($q$)-LDPC code's parity-check matrix can easily be obtained from its bipartite graph (see [1]).

As with binary LDPC codes, we say that a GF($q$) LDPC code is *regular* if all variable-nodes in its Tanner graph have the same degree, and all check-nodes have the same degree. Otherwise, we say it is *irregular*.

### B. Coset GF($q$) LDPC Codes

As mentioned in Section I, rather than use plain GF($q$) LDPC codes, it is useful instead to consider *coset* codes. In doing so, we follow the example of Elias [12] with binary codes.

*Definition 1:* Given a length $N$ linear code $C$ and a length $N$ vector **v** over GF($q$), the code $\{\mathbf{c} + \mathbf{v} : \mathbf{c} \in C\}$ (i.e. obtained by adding **v** to each of the codewords of $C$) is called a *coset code*. Note that the addition is performed componentwise over GF($q$). **v** is called the *coset vector*.

The use of coset codes, as we will later see, is a valuable asset to rigorous analysis and is easily accounted for in the decoding process.

### C. Mapping to the Channel Signal Set

With binary LDPC codes, the BPSK signals $\pm 1$ are typically used instead of the $\{0, 1\}$ symbols of the code alphabet. With nonbinary LDPC, we denote the signal constellation by $\mathcal{A}$ and the mapping from the code alphabet (GF($q$)) by $\delta(\cdot)$. When designing codes for transmission over an AWGN channel, a pulse amplitude modulation (PAM) or quadrature amplitude modulation (QAM) constellation is a straightforward choice for $\mathcal{A}$. In Section VIII we present codes where $\mathcal{A}$ is a PAM signal constellation. However, we now show that more careful attention to the design of the signal constellation can produce a substantial gain in performance.

In [1] we have shown that ensembles of GF($q$)-LDPC codes resemble uniform random-coding ensembles. That is, the empirical distribution of GF($q$) symbols in nearly all codewords is approximately uniform. Equivalently, for a given codeword **c**, $\Pr[c = g] \simeq \frac{1}{q}$ $\forall g \in \text{GF}(q)$, where $c$ is a randomly selected codeword symbol. Such codes are useful for transmission over symmetric channels, where the capacity-achieving distribution is uniform

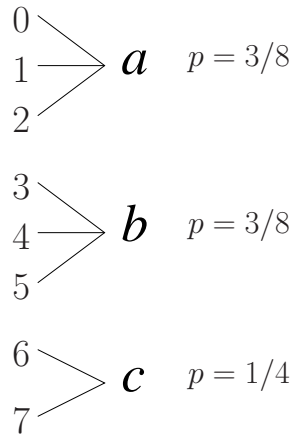

Fig. 2. An example of quantization mapping.

[17]. However, to approach capacity over asymmetric channels (and overcome the *shaping* gap [13]), we need the symbol distribution to be nonuniform. For example, to approach capacity over the AWGN channel, we need the distribution to resemble a Gaussian distribution.

One solution to this problem is a variant of an idea by Gallager [17]. The approach begins with a mapping of symbols from GF($q$) (the code alphabet) into the channel input alphabet. We typically use a code alphabet that is larger than the channel input alphabet. By mapping several GF($q$) symbols into each channel symbol (rather than using a one-to-one mapping), we can control the probability of each channel symbol. For example, in Fig. 2 we examine a channel alphabet $\mathcal{A} = \{a, b, c\}$, and a quantization mapping that is designed to achieve the distribution $Q(a) = Q(b) = 3/8, Q(c) = 1/4$ (The digits 0,...,7 represent elements of GF(8)). We call this a *quantizaion* mapping because the mapping is many-to-one.

Formally, we define quantization mapping as follows:

*Definition 2:* Let $Q(\cdot)$ be a rational probability assignment of the form $Q(a) = N_a/q$, for all $a \in \mathcal{A}$. A *quantization* $\delta(\cdot) = \delta_Q(\cdot)$ associated with $Q(a)$ is a mapping from a set of GF($q$) elements to $\mathcal{A}$ such that the number of elements mapped to each $a \in \mathcal{A}$ is $q \cdot Q(a)$.

Quantizations are designed for finite channel input alphabets and rational-valued probability assignments. However, other probability assignments can be approximated arbitrarily close. Independently of our work, a similar approach was developed by Ratzer and MacKay [26] (note that their approach does not involve coset codes).

A similar approach to designing mappings is based on Sun and van Tilborg [33] and Fragouli *et al.* [14] and is suitable for channels with continuous-input alphabets (like the AWGN channel). Instead of mapping many code symbols into each channel symbol, they used a one-to-one mapping to a set $\mathcal{A}$ of channel input signals that are *non-uniformly spaced*. To approximate a Gaussian input distribution, for example, the signals could be spaced more densely around zero.

Given a mapping $\delta(\cdot)$ over GF($q$), we define the mapping of a vector $\mathbf{v}$ with symbols in GF($q$), as the vector obtained by applying $\delta(\cdot)$ to each of its symbols. The mapping of a code is the code obtained by applying the mapping to each of the codewords.

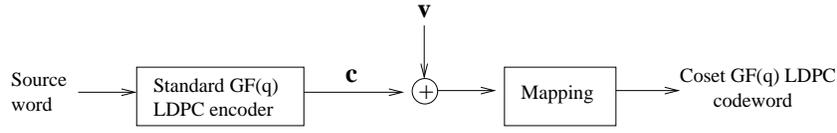

Fig. 3. Encoding of coset GF($q$) LDPC codes

It is useful to model coset GF($q$) LDPC encoding as a sequence of operations, as shown in Figure 3. An incoming message is encoded into a codeword of the *underlying* GF($q$) LDPC code $C$. The coset vector $\mathbf{v}$ is then added, and a mapping $\delta(\cdot)$ is applied. In the sequel, we will refer to the resulting codeword as a coset GF($q$) LDPC codeword, although strictly speaking, the mapping $\delta(\cdot)$ is not included in Definition 1. Finally, the resulting codeword is transmitted over the channel.

### D. $(\lambda, \rho, \delta)$ Ensembles of Coset GF($q$) LDPC Codes

As in the case of standard, binary LDPC codes, the analysis of coset GF($q$) LDPC focuses on the average behavior of codes selected at random from an *ensemble* of codes.

The following method, due to Luby *et al.* [24] is used to construct irregular bipartite Tanner graphs. The graphs are characterized by two probability vectors,

$$\boldsymbol{\lambda} = (\lambda_1, \ldots, \lambda_c) \qquad \boldsymbol{\rho} = (\rho_1, \ldots, \rho_d)$$

For convenience we also define the polynomials $\lambda(x) = \sum_{i=2}^{c} \lambda_i x^{i-1}$ and $\rho(x) = \sum_{j=2}^{d} \rho_j x^{j-1}$.

In a $(\lambda, \rho)$ Tanner graph, for each $i$ a fraction $\lambda_i$ of the edges has left degree $i$, and for each $j$ a fraction $\rho_j$ of the edges has right degree $j$. Letting $E$ denote the total number of edges, we obtain that there are $\lambda_i E/i$ left-nodes with degree $i$, and $\rho_j E/j$ right-nodes with degree $j$. Letting $N$ denote the number of left-nodes and $M$ denotes the number of right-nodes, we have

$$N = E \sum_{i=1}^{c} \frac{\lambda_i}{i} \qquad M = E \sum_{j=1}^{d} \frac{\rho_j}{j}$$

Luby *et al.* suggested the following method for constructing $(\lambda, \rho)$ bipartite graphs. The $E$ edges originating from left nodes are numbered from 1 to $E$. The same procedure is applied to the $E$ edges originating from right nodes. A permutation $\pi$ is then chosen with uniform probability from the space of all permutations of $\{1, 2, \ldots, E\}$. Finally, for each $i$, the edge numbered $i$ on the left side is associated with the edge numbered $\pi_i$ on the right side. Note that occasionally, multiple edges may link a pair of nodes.

A $(\lambda, \rho)$ GF($q$) LDPC code is constructed from a $(\lambda, \rho)$ Tanner graph by random i.i.d. selection of the labels with uniform probability from GF($q$)\{0}, at each edge. Given a mapping $\delta(\cdot)$, a $(\lambda, \rho, \delta)$ coset GF($q$) LDPC code is created by applying $\delta(\cdot)$ to a coset of a $(\lambda, \rho)$ GF($q$)-LDPC code. The coset vector $\mathbf{v}$ is generated by random uniform i.i.d selection of its components from GF($q$).

Summarizing, a random selection of a code from a $(\lambda, \rho, \delta)$ coset GF($q$) LDPC ensemble amounts to a random construction of its Tanner graph, a random selection of its labels and a random selection of a coset vector.

The rate of a $(\lambda, \rho, \delta)$ coset GF($q$) LDPC code is equal to the rate of its underlying GF($q$) LDPC code. The *design rate* $R$ of a $(\lambda, \rho)$ GF($q$) LDPC code is defined as

$$R \triangleq 1 - \frac{M}{N} = 1 - \frac{\sum_{j=1}^{d} \rho_j / j}{\sum_{i=1}^{c} \lambda_i / i} \tag{6}$$

This value is a lower bound on the true rate of the code, measured in $q$-ary symbols per channel use.

## IV. Belief-Propagation Decoding of Coset GF($q$) LDPC Codes

### A. Definition of the Decoder

The coset GF($q$) LDPC belief-propagation decoder is based on Gallager [16] and Kschischang *et al.* [21]. The decoder attempts to recover $\mathbf{c}$, the codeword of the underlying GF($q$) LDPC code. Decoding consists of alternating *rightbound* and *leftbound* iterations. In a rightbound iteration, messages are sent from variable-nodes to check-nodes. In a leftbound iteration, the opposite occurs. Note that with this terminology, a *rightbound* message is produced at a *left* node (a variable-node) and a *leftbound* message is produced at a *right* node (a check-node).

As mentioned in Section II, the decoder's messages are $q$ dimensional probability vectors, rather than scalar values as in standard binary LDPC.

*Algorithm 1:* Perform the following steps, alternately:

1) **Rightbound iteration.** For all edges $e = (i, j)$, do the following in parallel:

   If this is iteration zero, set the rightbound message $\mathbf{r} = \mathbf{r}(i, j)$ to the initial message $\mathbf{r^{(0)}} = \mathbf{r^{(0)}}(i)$, whose components are defined as follows:

   $$r_k^{(0)} = \frac{\Pr[y_i \mid \delta(k + v_i)]}{\sum_{k'=0}^{q-1} \Pr[y_i \mid \delta(k')]} \tag{7}$$

   $y_i$ and $v_i$ are the channel output and the element of the coset vector $\mathbf{v}$ corresponding to variable node $i$. The addition operation $k + v_i$ is performed over GF($q$).

   Otherwise (iteration number 1 and above),

   $$r_k = \frac{r_k^{(0)} \prod_{n=1}^{d_i - 1} l_k^{(n)}}{\sum_{k'=0}^{q-1} r_{k'}^{(0)} \prod_{n=1}^{d_i - 1} l_{k'}^{(n)}} \tag{8}$$

   where $d_i$ is the degree of the node $i$ and $\mathbf{l}^{(1)}, ..., \mathbf{l}^{(d_i - 1)}$ denote the incoming (leftbound) messages across the edges $\{(i, j') : j' \in \mathcal{N}(i) \setminus j\}$, $\mathcal{N}(i)$ denoting the set of nodes adjacent to $i$.

2) **Leftbound iteration.** For all edges $e = (i, j)$, do the following in parallel:

   Set the components of the leftbound message $\mathbf{l} = \mathbf{l}(j, i)$ as follows:

   $$l_k = \sum_{a_1, ..., a_{d_j - 1} \in GF(q), \; \sum_n g_n a_n = -g_{d_j} \cdot k} \; \prod_{n=1}^{d_j - 1} r_{a_n}^{(n)} \tag{9}$$

   where $d_j$ is the degree of node $j$, and $\mathbf{r}^{(1)}, ..., \mathbf{r}^{(d_j - 1)}$ denote the rightbound messages across the edges $\{(i', j) : i' \in \mathcal{N}(j) \setminus i\}$ and $g_1, ..., g_{d_j - 1}$ are the labels on those edges. $g_{d_j}$ denotes the label on the edge $(i, j)$. The summations and multiplications of the indices $a_n$ and the labels $g_n$ are performed over GF($q$). Note that an equivalent, simpler expression will be given shortly.

If $\mathbf{x}$ is a rightbound (leftbound) message from (to) a variable-node, then element $x_k$ represents an estimate of the *a-posteriori* probability (APP) that the corresponding code symbol is $k$, given the channel observations in a corresponding neighborhood graph (we will elaborate on this in Section IV-C). The *decision* associated with $\mathbf{x}$ is defined as follows: the decoder decides on the symbol $k$ that maximizes $x_k$. If the maximum was obtained at several indices, a uniform random selection is made among them.

In our analysis, we focus on the probability that a rightbound or leftbound message is erroneous (i.e., corresponds to an incorrect decision). However, in a practical setting, the decoder stops after a fixed number of decoding iterations and computes, at each variable-node $i$, a final vector $\hat{\mathbf{r}}(i)$ of APP values. The vector is computed using (8), replacing $\mathcal{N}(i)\backslash j$ with $\mathcal{N}(i)$. $\hat{\mathbf{r}}(i)$ is unique to each variable-node (unlike rightbound or leftbound messages), and can thus be used to compute a final decision on its value.

Consider expression (9) for computing the leftbound messages. A useful, equivalent expression is given by,

$$\mathbf{l} = \left[ \bigodot_{n=1}^{d_j-1} \left( \mathbf{r}^{(n)} \right)^{\times g_n^{-1}} \right]^{\times (-g_{d_j})} \tag{10}$$

where $\mathbf{l}$ is the entire leftbound vector (rather than a component as in (9)) and the $\times$ operator is defined as in (4). The GF($q$) convolution operator $\odot$ is defined as an operation between two vectors, which produces a vector whose components are given by,

$$\left[ \mathbf{x}^{(1)} \odot \mathbf{x}^{(2)} \right]_k = \sum_{a \in \mathrm{GF}(q)} x_a^{(1)} \cdot x_{k-a}^{(2)}, \qquad k \in \mathrm{GF}(q) \tag{11}$$

where the subtraction $k - a$ is evaluated over GF($q$). Throughout the paper, the following definitions are useful:

$$\bar{\mathbf{l}} \triangleq \mathbf{l}^{\times(-g_{d_j}^{-1})}, \quad \bar{\mathbf{r}}^{(n)} \triangleq \left( \mathbf{r}^{(n)} \right)^{\times g_n^{-1}}, \; n = 1, ..., d_j - 1 \tag{12}$$

Using these definitions, (10) may be further rewritten as,

$$\bar{\mathbf{l}} = \bigodot_{n=1}^{d_j-1} \bar{\mathbf{r}}^{(n)} \tag{13}$$

Like the standard binary LDPC belief-propagation decoder, the coset GF($q$) LDPC decoder also has an equivalent formulation using LLR messages.

*Algorithm 2:* Perform the following steps, alternately:

1) **Rightbound iteration.** For all edges $e = (i, j)$, do the following in parallel:

   If this is iteration zero, set the LLR rightbound message $\mathbf{r}' = \mathbf{r}'(i, j)$ to $\mathbf{r}'^{(0)} = \mathbf{r}'^{(0)}(i)$, whose components are defined as follows:

$$r_k'^{(0)} = \log \frac{\Pr[y_i \mid \delta(v_i)]}{\Pr[y_i \mid \delta(k + v_i)]} \tag{14}$$

   Otherwise (iteration number 1 and above),

$$\mathbf{r}' = \mathbf{r}'^{(0)} + \sum_{n=1}^{d_i-1} \mathbf{l}'^{(n)} \tag{15}$$

   where $d_i$ is the degree of the node $i$ and $\mathbf{l}'^{(1)}, ..., \mathbf{l}'^{(d_i-1)}$ denote the incoming (leftbound) LLR messages across the edges $\{(i, j') : j' \in \mathcal{N}(i) \setminus j\}$. Addition between vectors is performed componentwise.

2) **Leftbound iteration.** All rightbound messages are converted from LLR to plain-likelihood representation. Expression (9) is applied to obtain the plain-likelihood representation of the leftbound messages. Finally, the leftbound messages are converted back to their corresponding LLR representation.

Both versions of the decoder have similar execution times. However, the LLR representation is sometimes useful in the analysis of the decoders' performance. Note that Wymeersch *et al.* [39] have developed an alternative decoder that uses LLR representation, which does not require the conversion to plain-likelihood representation that is used in the leftbound iteration of the above algorithm.

### B. Efficient Implementation

To compute rightbound messages, we can save time by computing the numerators separately, and then normalizing the sum to 1. At a variable node of degree $d_i$, the computation of each rightbound message takes $O(q \cdot d_i)$ computations.

A straightforward computation of the leftbound messages at a check-node of degree $d_j$ has a complexity of $O(d_j q^{d_j-1})$ per leftbound-message, and a total of $O(d_j^2 q^{d_j-1})$ for all messages combined. We will now review a method due to Richardson and Urbanke [28] (developed for the decoding of standard GF($q$) LDPC codes) that significantly reduces this complexity. This method assumes plain-likelihood representation of messages. It is nonetheless relevant to the implementation of Algorithm 2, which uses LLR representation, because with this algorithm the leftbound messages are computed by converting them to plain-likelihood representation, applying (9) and converting back to LLR representation.

We first recount some properties of Galois fields (see e.g. [5] for a more extensive discussion). Galois fields GF($q$) exist for values of $q$ equal to $p^m$, where $p$ is a prime number and $m$ is a positive integer. Each element of GF($p^m$) can be represented as an $m$-dimensional vector over $0, ..., p-1$. The sum (difference) of two GF($p^m$) elements corresponds to the sum (difference) of the vectors, evaluated as the modulo-$p$ sums (differences) of the vectors' components.

Consider the GF($q$) convolution operator, defined by (11) and used in the process of computing the leftbound message in (10). We now replace the GF($q$) indices $a$ and $k$ in (11) with their vector representations, $\boldsymbol{\alpha}, \boldsymbol{\kappa} \in \{0, ..., p-1\}^m$. The expression can be rewritten as

$$\left[\mathbf{x}^{(1)} \odot \mathbf{x}^{(2)}\right]_{\boldsymbol{\kappa}} = \sum_{\boldsymbol{\alpha} \in \{0,...,p-1\}^m} x_{\boldsymbol{\alpha}}^{(1)} \cdot x_{\boldsymbol{\kappa}-\boldsymbol{\alpha} \bmod p}^{(2)}, \quad \boldsymbol{\kappa} \in \{0, ..., p-1\}^m \tag{16}$$

Consider, for example, the simple case of $m = 2$. (11) becomes

$$\left[\mathbf{x}^{(1)} \odot \mathbf{x}^{(2)}\right]_{\kappa_1, \kappa_2} = \sum_{\alpha_1=0}^{p-1} \sum_{\alpha_2=0}^{p-1} x_{\alpha_1, \alpha_2}^{(1)} \cdot x_{\kappa_1-\alpha_1 \bmod p, \kappa_2-\alpha_2 \bmod p}^{(2)}, \quad (\kappa_1, \kappa_2) \in \{0, ..., p-1\}^2 \tag{17}$$

The right hand side of (17) is the output of the two-dimensional cyclic convolution of $\mathbf{x}^{(1)}$ and $\mathbf{x}^{(2)}$, evaluated at $(\kappa_1, \kappa_2)$. In the general case we have the $m$-dimensional cyclic convolution. This convolution can equivalently be evaluated using the $m$-dimensional DFT ($m$-DFT) and IDFT ($m$-IDFT) [11][page 71]. Thus, (13) can be rewritten

as

$$\bar{\mathbf{l}} = \text{IDFT}\left(\prod_{n=1}^{d_j - 1} \text{DFT}(\bar{\mathbf{r}}^{(n)})\right)$$

Where the multiplication of the DFT vectors is performed componentwise ($\mathbf{l}$ can be evaluated from $\bar{\mathbf{l}}$ by $\mathbf{l} = \bar{\mathbf{l}}^{\times(-g_{d_j})}$).

Let $\mathbf{d}$ denote the DFT vector of a $q$-dimensional probability vector $\mathbf{x}$. The components of $\mathbf{d}$ and $\mathbf{x}$ are related by the equations [11][page 65]

$$d_{\beta_1,...,\beta_m} = \sum_{\alpha_1,...,\alpha_m \in \{0,...,p-1\}} x_{\alpha_1,...,\alpha_m} e^{j\frac{2\pi}{p}\sum_{i=1}^m \alpha_i \cdot \beta_i} \quad (m\text{-DFT})$$

$$x_{\alpha_1,...,\alpha_m} = \frac{1}{q}\sum_{\beta_1,...,\beta_m \in \{0,...,p-1\}} d_{\beta_1,...,\beta_m} e^{-j\frac{2\pi}{p}\sum_{i=1}^m \alpha_i \cdot \beta_i} \quad (m\text{-IDFT})$$

Efficient computation of the $m$-DFT is possible by successively applying the single-dimensional DFT on each of the dimensions in turn, as shown in the following algorithm [11][page 76]:

*Algorithm 3:*

for i = 1 to m

    for each vector $(\alpha_1, ..., \alpha_{i-1}, \alpha_{i+1}, ..., \alpha_m) \in \{0, ..., p-1\}^{m-1}$

        $\{d_{\alpha_1,...,\alpha_{i-1},\alpha_i,\alpha_{i+1},...,\alpha_m}\}_{\alpha_i=0}^{p-1} \leftarrow 1\text{-DFT}\{x_{\alpha_1,...,\alpha_{i-1},\alpha_i,\alpha_{i+1},...,\alpha_m}\}_{\alpha_i=0}^{p-1}$

    end

    if $i \neq m$ then $\mathbf{x} \leftarrow \mathbf{d}$

end

return $\mathbf{d}$

At each iteration of the above algorithm, $p^{m-1}$ 1-DFTs are computed. Each 1-DFT requires $p^2$ floating-point multiplications and $p \cdot (p-1)$ floating-point additions (to compute all components), and thus the entire algorithm requires $m \cdot p^{m+1} = m \cdot p \cdot q$ multiplications and $m \cdot (p-1)p^m = m \cdot (p-1) \cdot q$ additions. The $m$-IDFT can be computed in a similar manner. Note that a further reduction in complexity could be obtained by using number-theoretic transforms, such as the Winograd FFT.

We can use these results to reduce the complexity of leftbound computation at each check-node, by first computing the $m$-DFTs of all rightbound messages, then using the DFT vectors to compute convolutions. The resulting complexity at each check-node is now $O(d_j \cdot mpq + d_j(d_j-1) \cdot q)$. The first element of the sum is the computation of $m$-DFTs and $m$-IDFTs, the second is the multiplications of $m$-DFTs for all messages. This is a significant improvement in comparison to the straightforward approach.

Note that the $m$-DFT is particularly attractive when $p = 2$, i.e., when $q$ is $2^m$. The elements of the form $e^{j\frac{2\pi}{p}\sum_{i=1}^m \alpha_i \cdot \beta_i}$ become $(-1)^{\sum_{i=1}^m \alpha_i \cdot \beta_i}$. Thus, the floating-point multiplications are eliminated, and the DFT involves only additions and subtractions. The above complexity figure per check-node thus becomes $O(d_j \cdot mq + d_j(d_j-1) \cdot q)$. Furthermore, all quantities are real-valued and no complex-valued arithmetic is needed.

An additional improvement, to an order of $O(d_j \cdot mpq + 3 \cdot d_j \cdot q)$ (in the general case where $p$ is not necessarily 2) can be achieved using a method suggested by Davey and MacKay [10]. This method produces a negligible improvement except at very high values of $d_j$, and is therefore not elaborated here.

## C. Neighborhood Graphs and the Tree Assumption

Before we conclude this section, we briefly review the concepts neighborhood graphs and the tree assumption. These concepts were developed in the context of standard binary LDPC codes and carry over to coset GF($q$) LDPC codes as well.

*Definition 3:* (Richardson and Urbanke [28]) The *neighborhood graph of depth $d$*, spanned from an edge $e$, is the induced graph containing $e$ and all edges and nodes on directed paths of length $d$ that end with $e$.

At iteration $t$, a rightbound message produced from a variable-node $i$ to a check node $j$ is a vector of APP values for the code symbol at $i$, given information observed in the neighborhood of $e = (i, j)$ of depth $2t$. Similarly, a leftbound message from $j$ to $i$ is based on the information observed in the neighborhood of $e = (j, i)$, of depth $2t - 1$.

The APP values produced by belief-propagation decoders are computed under the *tree assumption*[4]. We say that the tree assumption is satisfied at a node $n$ in the context of computing a message $\mathbf{x}$, if the neighborhood graph on which the message is based is a tree. Asymptotically, at large block lengths $N$, the tree assumption is satisfied with high probability at any particular node [28].

At finite block lengths, the neighborhood graph frequently contains cycles and is therefore not a tree. Such cases are discussed in Appendix II. Nevertheless, simulation results indicate that the belief-propagation decoder produces remarkable performance even when the tree assumption is not strictly satisfied.

## V. Coset GF($q$) LDPC Analysis in a Random-Coset Setting

One important aid in the analysis of coset GF($q$) LDPC codes is the randomly selected coset vector that was used in their construction. Rather than examine the decoder of a single coset GF($q$) LDPC code, we focus on a set of codes. That is, given a fixed GF($q$)-LDPC code $C$ and a mapping $\delta(\cdot)$, we consider the behavior of a coset GF($q$) LDPC code constructed using a randomly selected coset vector $\mathbf{v}$. We refer to this as *random-coset analysis*.

With this approach, the random space consists of random channel transitions as well as random realizations of the coset vector $\mathbf{v}$. The random coset vector produces an effect that is similar to output-symmetry that is usually required in the analysis of standard LDPC codes [28], [29]. Note that although $\mathbf{v}$ is random, it is assumed to have been selected in advance and is thus known to the decoder.

Unlike the coset vector, in this section we keep the underlying GF($q$) LDPC code fixed. In Section VI, we will consider several of these concepts in the context of selecting the underlying LDPC code at random from an ensemble.

### A. The All-Zero Codeword Assumption

An important property of standard binary LDPC decoders [28] is that the probability of decoding error is equal for any transmitted codeword. This property is central to many analysis methods, and enables conditioning the analysis on the assumption that the all-zero[5] codeword was transmitted.

---

[4]In [28] it is called the *independence assumption*.

[5]In [28] a BPSK alphabet is used and thus the codeword is referred to as the "all-one" codeword.

With coset GF($q$) LDPC codes, we have the following lemma.

*Lemma 1:* Assume a discrete memoryless channel. Consider the analysis, in a random-coset setting, of a coset GF($q$) LDPC code constructed from a fixed GF($q$)-LDPC code $C$. For each $\mathbf{c} \in C$, let $\overline{P}_e^t(\mathbf{c})$ denote the conditional (bit or block) probability of decoding error after iteration $t$, assuming the codeword $\delta(\mathbf{c} + \mathbf{v})$ was sent, averaged over all possible values of the coset vector $\mathbf{v}$. Then $\overline{P}_e^t(\mathbf{c})$ is independent of $\mathbf{c}$.

The proof of the lemma is provided in Appendix III-B.

Lemma 1 enables us to condition our analysis results on the assumption that the transmitted codeword corresponds to $\mathbf{0}$ of the underlying LDPC code.

### B. Symmetry of Message Distributions

The *symmetry* property, introduced by Richardson and Urbanke [29] is a major tool in the analysis of standard binary LDPC codes. In this section we generalize its definition to $q$-ary random variables as used in the analysis of coset GF($q$) LDPC decoders. We provide two versions of the definition, the first using probability-vector random variables and the second using LLR-vector random variables.

*Definition 4:* A probability-vector random variable $\mathbf{X}$ is *symmetric* if for any probability-vector $\mathbf{x}$, the following expression holds:

$$\Pr[\mathbf{X} = \mathbf{x} \mid \mathbf{X} \in \mathbf{x}^*] = x_0 \cdot n(\mathbf{x}) \tag{18}$$

where $\mathbf{x}^*$ and $n(\mathbf{x})$ are as defined in Section I.

In the context of LLR-vector random variables, we have the following lemma.

*Lemma 2:* Let $\mathbf{W}$ be an LLR-vector random variable. The random variable $\mathbf{X} = \mathbf{W}' \triangleq \mathrm{LLR}^{-1}(\mathbf{W})$ is symmetric if and only if $\mathbf{W}$ satisfies

$$\Pr[\mathbf{W} = \mathbf{w}] = e^{w_i} \Pr[\mathbf{W} = \mathbf{w}^{+i}] \tag{19}$$

for all LLR-vectors $\mathbf{w}$ and all $i \in \mathrm{GF}(q)$.

The proof of this lemma is provided in Appendix III-C. In the sequel, we adopt the lemma as a definition of symmetry when discussing variables in LLR representation. Note that in the simple case of $q = 2$, the LLR vector degenerates to a scalar value and from (5) we have $w^{+1} = -w$. Thus, (19) becomes

$$\Pr[W = w] = e^w \Pr[W = -w] \tag{20}$$

This coincides with symmetry for binary codes as defined in [29].

We now examine the message produced at a node $n$.

*Theorem 1:* Assume a discrete memoryless channel and consider a coset GF($q$) LDPC code constructed in a random-coset setting from a fixed GF($q$)-LDPC code $C$. Let $\mathbf{X}$ denote the message produced at a node $n$ of the Tanner graph of $C$ (and of the coset GF($q$) LDPC code), at some iteration of belief-propagation decoding. Let the tree assumption be satisfied at $n$. Then under the all-zero codeword assumption, the random variable $\mathbf{X}$ is symmetric.

The proof of the theorem is provided in Appendix III-D.

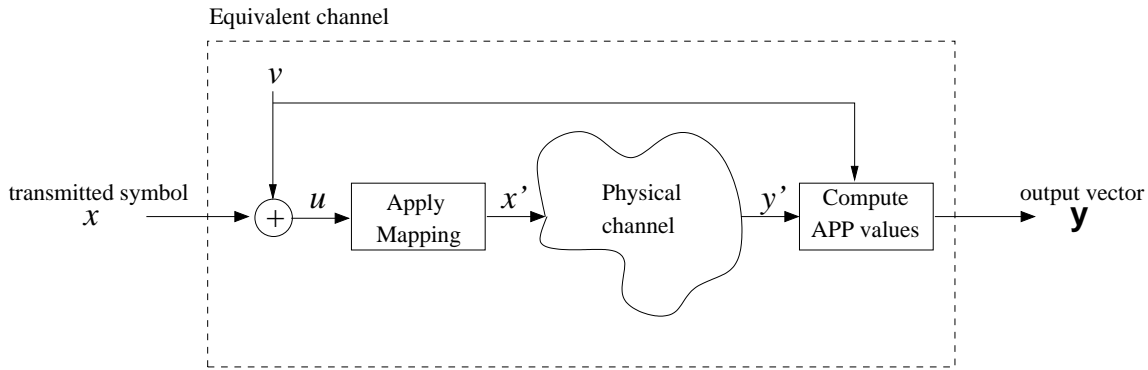

Fig. 4. Equivalent channel model for coset GF($q$) LDPC codes.

## C. Channel Equivalence

Simple GF($q$)-LDPC codes, although unsuitable for arbitrary channels, are simpler to analyze than coset GF($q$) LDPC codes and decoders. Fig. 4 presents the structure of coset GF($q$) LDPC encoding/decoding. $x$ is the transmitted symbol (of the underlying code) and $v$ is the coset symbol. $u = x + v$ (evaluated over GF($q$)) is the input to the mapper, $x' = \delta(u)$ is the mapper's output and $y'$ is the physical channel's output. $\mathbf{y}$ will be discussed shortly.

Comparing a coset GF($q$) LDPC decoder with the decoder of its underlying GF($q$) LDPC code we may observe that a difference exists only in the computation (7) of the initial messages $\mathbf{r^{(0)}}$. The messages $\mathbf{r^{(0)}}$ are APP values corresponding to a single channel observation. After they are computed, both decoders proceed in exactly the same way. It would thus be desirable to abstract the operations that are unique to coset GF($q$) LDPC codes into the channel, and examine an equivalent model, which employs simple GF($q$)-LDPC codes and decoders.

Consider the channel obtained by encapsulating the addition of a random coset symbol, the mapping and the computation of the APP values into the channel model. The input to the channel is a symbol $x$ from the code alphabet[6] and the output is a probability vector $\mathbf{y} = \mathbf{r^{(0)}}$ of APP values. The decoder of a GF($q$) LDPC code, if presented with $\mathbf{y}$ as raw channel output, would first compute a new vector of APP values. We will soon show that the computed vector would in fact be identical to $\mathbf{y}$.

We begin with the following definition:

*Definition 5:* Let $\Pr[\mathbf{y} \mid x]$ denote the transition probabilities of a channel whose input alphabet is GF($q$) and whose output alphabet consists of $q$-dimensional probability vectors. Then the channel is *cyclic-symmetric* if there exists a probability function $Q(\mathbf{y^*})$ (defined over sets of probability vectors (3)), such that

$$\Pr[\mathbf{Y} = \mathbf{y} \mid x = i] = y_i \cdot n(\mathbf{y}) \cdot Q(\mathbf{y^*}) \tag{21}$$

*Lemma 3:* Assume a cyclic-symmetric channel. Let $\mathrm{APP}(\mathbf{y})$ denote the APP values for the channel output $\mathbf{y}$. Then $\mathrm{APP}(\mathbf{y}) = \mathbf{y}$.

The proof of this lemma is provided in Appendix III-F. Returning to the context of our equivalent model, we have the following lemma,

---

[6] In most cases of interest, $x$ will be a symbol from a GF($q$) LDPC codeword. However, in this section we also consider the general, theoretical case, where the input to the channel is an arbitrary GF($q$) symbol.

*Lemma 4:* The equivalent channel of Fig. 4 is cyclic-symmetric.

The proof of this Lemma is provided in Appendix III-G.

Once the initial messages are computed, the performance of both the coset GF($q$) LDPC and GF($q$) LDPC decoding algorithms is a function of these messages alone. Therefore, we have obtained that the performance of a coset GF($q$) LDPC decoder in a random-coset setting over the original physical channel is identical to the performance of the underlying GF($q$) LDPC decoder over the equivalent channel. This result enables us to shift our discussion from coset GF($q$) LDPC codes over arbitrary channels to GF($q$) LDPC codes over cyclic-symmetric channels.

Note that a cyclic-symmetric channel is symmetric in the sense defined by Gallager [17][page 94]. Hence its capacity achieving distribution is uniform. This indicates that GF($q$) LDPC codes, which have an approximately uniformly distributed code spectrum (see [1]), are suitably designed for it.

We now relate the capacity of the equivalent channel to that of the physical channel. More precisely, we show that the equivalent channel's capacity is equal to the *equiprobable-signalling capacity* of the physical channel with the mapping $\delta(\cdot)$, denoted $C_\delta$ and defined below. Let $U$, $X'$ and $Y'$ be random variables corresponding to $u$, $x'$ and $y'$ in Fig. 4. $Y'$ is related to $X' = \delta(U)$ through the physical channel's transition probabilities. Assume that $U$ is uniformly distributed in $\{0, ..., q-1\}$, then we define $C_\delta$ by $C_\delta \triangleq I(U; Y')$. $C_\delta$ is equal to the capacity of transmission over the physical channel with an input alphabet $\{\delta(i)\}_{i=0}^{q-1}$ using a code whose codewords were generated by random uniform selection.

*Lemma 5:* The capacity of the equivalent channel of Fig. 4 is equal to $C_\delta$.

The proof of this lemma is provided in Appendix III-H.

Finally, the following lemma can be viewed as a generalization of the Channel Equivalence Lemma of [29].

*Lemma 6:* Let $P(\mathbf{y})$ be the probability function of a symmetric probability-vector random variable. Consider the cyclic-symmetric channel whose transition probabilities are given by $\Pr[\mathbf{y} \,|\, x = i] = P(\mathbf{y}^{+i})$. Then, assuming that the symbol zero is transmitted over this cyclic symmetric channel, then the initial messages of a GF($q$) LDPC decoder are distributed as $P(\mathbf{y})$.

The proof of this lemma is straightforward from Definitions 4 and 5 and from Lemma 3. We will refer to the cyclic-symmetric channel defined in Lemma 6 as the equivalent channel corresponding to $P(\mathbf{y})$.

*Remark 1:* Note that Lemma 6 remains valid if we switch to LLR representation. That is, we replace $\mathbf{y}$ with its LLR equivalent $\mathbf{w} = \text{LLR}(\mathbf{y})$ and define $\Pr[\mathbf{w} \,|\, x = i] = P(\mathbf{w}^{+i})$ (where $\mathbf{w}^{+i}$ is defined by (5)).

## VI. Analysis of Density Evolution

In this section we consider density-evolution for coset GF($q$) LDPC codes and its analysis. The precise computation of the coset GF($q$) LDPC version of the algorithm is generally not possible in practice. The algorithm is however valuable as a reference for analysis purposes. We begin by defining density evolution in Section VI-A and examine the application of the concentration theorem of [28] and of symmetry to it. We proceed in Section VI-B to consider permutation-invariance, which is an important property of the densities tracked by the algorithm. We then apply permutation-invariance in Section VI-C to generalize the stability property to coset GF($q$) LDPC codes and in Section VI-D to obtain an approximation of density-evolution under a Gaussian assumption.

### A. Density Evolution

The definition of coset GF($q$) LDPC density-evolution is based on that of binary LDPC codes. The description below is intended for completeness of this text, and focuses on the differences that are unique to coset GF($q$) LDPC codes. The reader is referred to [28] and [29] for a complete rigorous development.

Density evolution tracks the distributions of messages produced in belief-propagation, averaged over all possible neighborhood graphs on which they are based. The random space is comprised of random channel transitions, the random selection of the code from a $(\lambda, \rho, \delta)$ coset GF($q$) LDPC ensemble (see Section III-D) and the random selection of an edge from the graph. The random space does *not* include the transmitted codeword, which is assumed to be fixed at the all-zero codeword (following the discussion of Section V-A). We denote by $\mathbf{R}^{(\mathbf{0})}$ the initial message across the edge, by $\mathbf{R}_t$ the rightbound message at iteration $t$ and by $\mathbf{L}_t$ the leftbound message at iteration $t$. The neighborhood graph associated with $\mathbf{R}_t$ and $\mathbf{L}_t$ is always assumed to be tree-like, and the case that it is not so is neglected.

We will use the above notation when discussing plain-likelihood representation of density-evolution. When using LLR-vector representation, we let $\mathbf{R'}^{(\mathbf{0})}$, $\mathbf{R'}_t$ and $\mathbf{L'}_t$ denote the LLR-vector representations of $\mathbf{R}^{(\mathbf{0})}$, $\mathbf{R}_t$ and $\mathbf{L}_t$. To simplify our notation, we assume that all random variables are discrete-valued and thus track their probability-functions rather than their densities. The following discussion focuses on plain-likelihood representation. The translation to LLR representation is straightforward.

1) **The initial message.** The probability function of $\mathbf{R}^{(\mathbf{0})}$ is computed in the following manner.

$$\Pr[\mathbf{R}^{(\mathbf{0})} = \mathbf{x}] \;\; = \sum_{y \in \mathcal{Y}, v = 0, \ldots, q-1 \;:\; \mathbf{r}^{(0)}(y, v) = \mathbf{x}} \Pr[Y = y, V = v]$$

where $Y$ and $V$ are random variables denoting the channel output and coset-vector components, $\mathcal{Y}$ is the channel output alphabet and the components of $\mathbf{r}^{(\mathbf{0})}(y, v)$ are defined by (7), replacing $y_i$ and $v_i$ with $y$ and $v$. The expression is equal to,

$$\Pr[\mathbf{R}^{(\mathbf{0})} = \mathbf{x}] \;\; = \;\; \frac{1}{q} \sum_{y \in \mathcal{Y}, v = 0, \ldots, q-1 \;:\; \mathbf{r}^{(0)}(y, v) = \mathbf{x}} \Pr[y \text{ was received } \mid \delta(v) \text{ was transmitted}]$$

2) **Leftbound messages.** $\mathbf{L}_t$ is obtained from (9). The rightbound messages in (9) are replaced by independent random variables, distributed as $\mathbf{R}_{t-1}$ and assumed to be independent. Similarly, the labels in (9) are also replaced by independent random variables uniformly distributed in GF($q$)\{0}.

Formally, Let $d$ be the maximal right-degree. Then for each $d_j = 2, \ldots, d$ we first define,

$$\Pr[\mathbf{L}_t^{(d_j)} = \mathbf{x}] \;\; = \sum_{\substack{\mathbf{r}^{(1)}, \ldots, \mathbf{r}^{(d_j-1)} \in \mathcal{P}, g_1, \ldots, g_{d_j} \in \mathrm{GF}(q) \;:\\ \mathbf{l}(\mathbf{r}^{(1)}, \ldots, \mathbf{r}^{(d_j-1)}, g_1, \ldots, g_{d_j}) = \mathbf{x}}} \prod_{n=1}^{d_j} \Pr[G_n = g_n] \cdot \prod_{n=1}^{d_j-1} \Pr[\mathbf{R}_{t-1} = \mathbf{r}^{(n)}]$$

where $\mathcal{P}$ is the set of all probability vectors, and the components of $\mathbf{l}(\mathbf{r}^{(1)}, ..., \mathbf{r}^{(d_j-1)}, g_1, ..., g_{d_j})$ are defined as in (9). $G_n$ is a random variable corresponding to the $n$th label, and thus $\Pr[G_n = g] = 1/(q-1)$ for all $g$. $\Pr[\mathbf{R}_{t-1} = \mathbf{r}^{(n)}]$ is obtained recursively from the previous iteration of belief propagation.

The probability function of $\mathbf{L}_t$ is now obtained by,

$$\Pr[\mathbf{L}_t = \mathbf{x}] = \sum_{d_j=2}^{d} \rho_{d_j} \cdot \Pr[\mathbf{L}_t^{(d_j)} = \mathbf{x}]$$

3) **Rightbound messages.** The probability function of $\mathbf{R}_0$ is equal to that of $\mathbf{R}^{(0)}$. For $t > 0$, $\mathbf{R}_t$ is obtained from (8). The leftbound messages and the initial message in (8) are replaced by independent random variables, distributed as $\mathbf{L}_t$ and $\mathbf{R}^{(0)}$, respectively.

Formally, let $c$ be the maximal left-degree. Then for each $d_i = 2, ..., c$ we first define,

$$\Pr[\mathbf{R}_t^{(d_i)} = \mathbf{x}] = \sum_{\substack{\mathbf{r}^{(0)}, \mathbf{l}^{(1)}, ..., \mathbf{l}^{(d_i-1)} \in \mathcal{P} \ : \\ \mathbf{r}(\mathbf{r}^{(0)}, \mathbf{l}^{(1)}, ..., \mathbf{l}^{(d_i-1)}) = \mathbf{x}}} \Pr[\mathbf{R}^{(0)} = \mathbf{r}^{(0)}] \cdot \prod_{n=1}^{d_i-1} \Pr[\mathbf{L}_t = \mathbf{l}^{(n)}]$$

where the components of $\mathbf{r}(\mathbf{r}^{(0)}, \mathbf{l}^{(1)}, ..., \mathbf{l}^{(d_i-1)})$ are defined as in (8). $\Pr[\mathbf{R}^{(0)} = \mathbf{r}^{(0)}]$ and $\Pr[\mathbf{L}_t = \mathbf{l}^{(n)}]$ are obtained recursively from the previous iterations of belief propagation.

The probability function of $\mathbf{R}_t$ is now obtained by,

$$\Pr[\mathbf{R}_t = \mathbf{x}] = \sum_{d_i=2}^{c} \lambda_{d_i} \cdot \Pr[\mathbf{R}_t^{(d_i)} = \mathbf{x}]$$

Theoretically, the above algorithm is sufficient to compute the desired densities. In practice, a major problem is the fact that the quantities of memory required to store the probability density of a $q$-dimensional message grows exponentially with $q$. For instance, with 100 quantization[7] levels per dimension, the amount of memory required for a 7-ary code is of the order of $100^7$. Hence, unless an alternative method for describing the densities is found, the algorithm is not realizable. It is noteworthy, however, that the algorithm can be approximated using Monte Carlo simulations.

We now discuss the probability that a message examined in density-evolution is erroneous. That is, the message corresponds to an incorrect decision regarding the variable-node to whom it is directed or from which it was sent. Under the all-zero codeword assumption, the true transmitted code symbol (of the underlying LDPC code), at the relevant variable-node, is assumed to be zero.

We first assume that the message is a fixed probability-vector $\mathbf{x}$. Suppose $x_0$ is greater than all other elements $x_i$, $i = 1, ..., q-1$. Given the decision criterion used by the belief propagation decoder, described in Section IV-A, the decoder will correctly decide zero. Similarly, if there exists an index $i \neq 0$ such that $x_i > x_0$, then the decoder will incorrectly decide $i$. However, if the maximum is achieved at 0 as well as $k-1$ other indices, the decoder will correctly decide zero with probability $1/k$.

*Definition 6:* Given a probability vector $\mathbf{x}$, $P_e(\mathbf{x})$ is the probability of error in a decision according to the vector $\mathbf{x}$.

Thus, for example $P_e([1/2, 1/4, 1/4, 0]) = 0$, $P_e([1/4, 1/2, 1/4, 0]) = 1$ and $P_e([3/10, 3/10, 3/10, 1/10]) = 2/3$.

---

[7]"Quantization" here means the operation performed by a discrete quantizer, not in the context of Definition 2.

Given a random variable $\mathbf{X}$, we define

$$P_e(\mathbf{X}) \triangleq \sum_{\mathbf{x}} P_e(\mathbf{x}) \Pr[\mathbf{X} = \mathbf{x}] \tag{22}$$

where the sum is over all probability vectors.

Consider $P_e(\mathbf{R}_t)$. This corresponds to the probability of error at a randomly selected edge at iteration $t$. Richardson and Urbanke [28] proved a *concentration theorem* that states that as the block length $N$ approaches infinity, the bit error rate at iteration $t$ converges to a similarly defined probability of error. The convergence is in probability, exponentially in $N$. Replacing bit- with symbol- error rate, this theorem carries over to coset GF$(q)$ LDPC density-evolution unchanged.

Let $P_e^t \triangleq P_e(\mathbf{R}_t)$ be a sequence of error probabilities produced by density evolution. A desirable property of this sequence is given by the following theorem.

*Theorem 2:* $P_e^t$ is nonincreasing with $t$.

The proof of this theorem is similar to that of Theorem 7 of [29] and is omitted.

Finally, in Section V-B we considered symmetry in the context of the message corresponding to a fixed underlying GF$(q)$ LDPC code and across a fixed edge of its Tanner graph. We now consider its relevance in the context of density-evolution, which assumes a random underlying LDPC code and a random edge.

*Theorem 3:* The random variables $\mathbf{R}^{(0)}$, $\mathbf{R}_t$ and $\mathbf{L}_t$ (for all $t$) are symmetric.

The proof of this theorem is provided in Appendix IV-A.

### B. Permutation-Invariance Induced by Labels

Permutation-invariance is a key property of coset GF$(q)$ LDPC codes that allows the approximation of their densities using one-dimensional functionals, thus greatly simplifying their analysis. The definition is based on the permutation, inferred by the operation $\times g$, on the elements of a probability vector.

Before we provide the definition, let us consider (10), by which a leftbound message $\mathbf{l}$ is computed in the process of belief propagation decoding. Let $h \in \text{GF}(q) \backslash \{0\}$, and consider $\mathbf{l}^{\times h}$.

$$\mathbf{l}^{\times h} = \left\{ \left[ \bigodot_{n=1}^{d_j-1} \left( \mathbf{r}^{(n)} \right)^{\times g_n^{-1}} \right]^{\times (-g_{d_j})} \right\}^{\times h} = \left[ \bigodot_{n=1}^{d_j-1} \left( \mathbf{r}^{(n)} \right)^{\times g_n^{-1}} \right]^{\times (-g_{d_j} \cdot h)} \tag{23}$$

With density evolution, the label $g_{d_j}$ is a random variable, independent of the other labels, of the rightbound messages $\{\mathbf{R}^{(n)}\}$ and consequently of $\bigodot_{n=1}^{d_j-1} \left( \mathbf{R}^{(n)} \right)^{\times g_n^{-1}}$. Similarly, $g_{d_j} \cdot h$ (where $h$ is fixed) is distributed identically with $g_{d_j}$, and is independent of $\bigodot_{n=1}^{d_j-1} \left( \mathbf{R}^{(n)} \right)^{\times g_n^{-1}}$. Thus, the random variable $\mathbf{L}^{\times h}$ is distributed identically with $\mathbf{L}$. This leads us to the following definition:

*Definition 7:* A probability-vector random variable $\mathbf{X}$ is *permutation-invariant* if for any fixed $h \in \text{GF}(q) \backslash \{0\}$, the random variable $\mathbf{\Xi} \triangleq \mathbf{X}^{\times h}$ is distributed identically with $\mathbf{X}$.

Although this definition assumes plain-likelihood representation, it carries over straightforwardly to LLR representation, and the following lemma is easy to verify:

*Lemma 7:* Let $\mathbf{W}$ be an LLR-vector random-variable and $\mathbf{X} = \mathbf{W}' = \mathrm{LLR}^{-1}(\mathbf{W})$. Then $\mathbf{X}$ is permutation-invariant if and only if, for any fixed $h \in \mathrm{GF}(q)\backslash\{0\}$, the random variable $\mathbf{\Omega} \triangleq \mathbf{W}^{\times h}$ is distributed identically with $\mathbf{W}$.

To give an idea of why permutation-invariance is so useful, we now present two important lemmas involving permutation-invariant random variables. Both lemmas examine marginal random variables. The first lemma is valid for both probability-vector and LLR-vector representation.

*Lemma 8:* Let $\mathbf{X}$ ($\mathbf{W}$) be a probability-vector (LLR-vector) random variable. If $\mathbf{X}$ ($\mathbf{W}$) is permutation-invariant then for any $i, j = 1, ..., q - 1$, the random variables $X_i$ and $X_j$ ($W_i$ and $W_j$) are identically distributed.

The proof of this lemma is provided in Appendix IV-B.

*Lemma 9:* Let $\mathbf{W}$ be a symmetric LLR-vector random variable. Assume that $\mathbf{W}$ is also permutation-invariant. Then for all $k = 1, ..., q - 1$, $W_k$ is symmetric in the binary sense, as defined by (20).

Note that this lemma does *not* apply to plain-likelihood representation. The proof of the lemma is provided in Appendix IV-C. Consider the following definition,

*Definition 8:* Given a probability-vector random variable $\mathbf{X}$, we define the *random-permutation* of $\mathbf{X}$, denoted $\tilde{\mathbf{X}}$, as the random variable equal to $\mathbf{X}^{\times g}$ where $g$ is randomly selected from $\mathrm{GF}(q)\backslash\{0\}$ with uniform probability, and is independent of $\mathbf{X}$.

The definition with LLR-vector representation is identical. The following lemma links permutation-invariance with random-permutation.

*Lemma 10:* A probability-vector (LLR-vector) random-variable $\mathbf{X}$ ($\mathbf{W}$) is permutation-invariant if and only if there exists a probability-vector (LLR-vector) random-variable $\mathbf{T}$ ($\mathbf{S}$) such that $\mathbf{X} = \tilde{\mathbf{T}}$ ($\mathbf{W} = \tilde{\mathbf{S}}$).

In Appendix IV-E we present some additional useful lemmas that involve permutation-invariance.

Finally, the following theorem discusses permutation-invariance's relevance to the distributions tracked by density evolution.

*Theorem 4:* Let $\mathbf{R^{(0)}}$, $\mathbf{R}_t$ and $\mathbf{L}_t$ be defined as in Section VI-A. Then,

1) $\mathbf{L}_t$ is permutation-invariant.

2) Let $\bar{\mathbf{R}}_t \triangleq (\mathbf{R}_t)^{\times g^{-1}}$, where $g$ is the label on the edge associated with the message. Then $\bar{\mathbf{R}}_t$ is symmetric, permutation-invariant and satisfies $P_e(\bar{\mathbf{R}}_t) = P_e(\mathbf{R}_t)$.

3) Let $\bar{\mathbf{R}}^{(0)}$ be a random-permutation of $\mathbf{R^{(0)}}$. Then replacing $\mathbf{R^{(0)}}$ by $\bar{\mathbf{R}}^{(0)}$ in the computation of density-evolution does not affect the densities of $\mathbf{L}_t$ and $\bar{\mathbf{R}}_t$. The random variable $\bar{\mathbf{R}}^{(0)}$ is symmetric, permutation-invariant and satisfies $P_e(\bar{\mathbf{R}}^{(0)}) = P_e(\mathbf{R^{(0)}})$.

The proof of this theorem is provided in Appendix IV-F. Although not all distributions involved in density-evolution are permutation-invariant, Theorem 4 enables us to focus our attention on permutation-invariant random variables alone. Our interest in the distribution of the rightbound message $\mathbf{R}_t$ is confined to the error probability implied by it. Thus we may instead examine $\bar{\mathbf{R}}_t$. Similarly, our interest in the initial message $\mathbf{R^{(0)}}$ is confined to its effect on the distribution of $\bar{\mathbf{R}}_t$ and $\mathbf{L}_t$. Thus we may instead examine $\bar{\mathbf{R}}^{(0)}$.

*C. Stability*

The *stability condition*, introduced by Richardson *et al.* [29], is a necessary and sufficient condition for the probability of error to approach arbitrarily close to zero, assuming it has already dropped below some value at some iteration. Thus, this condition is an important aid in the design of LDPC codes with low error floors. In this section we generalize the stability condition to coset GF($q$) LDPC codes.

Given a discrete memoryless channel with transition probabilities $\Pr[y \mid x]$ and a mapping $\delta(\cdot)$, we define the following channel parameter.

$$\Delta \triangleq \frac{1}{q(q-1)} \sum_{i,j \in \mathrm{GF}(q),\, i \neq j} \sum_y \sqrt{\Pr[y|\delta(i)] \Pr[y|\delta(j)]} \tag{24}$$

For example, consider an AWGN channel with a noise variance of $\sigma$. $\Delta$ for this case is obtained in a similar manner to that of [29][Example 12].

$$\Delta = \frac{1}{q(q-1)} \sum_{i \neq j} \exp\left( -\frac{1}{2\sigma^2} \left( \frac{\delta(i) - \delta(j)}{2} \right)^2 \right)$$

In Appendix IV-G, we present the concept of non-degeneracy for mappings $\delta(\cdot)$ and channels (taken from [1]). Under these assumptions, $\Delta$ is strictly smaller than 1. We assume these non-degeneracy definitions in the following theorem.

Finally, we are now ready to state the stability condition for coset GF($q$) LDPC codes:

*Theorem 5:* Assume we are given the triplet $(\lambda, \rho, \delta)$ for a coset GF($q$) LDPC ensemble designed for the above discrete memoryless channel. Let $P_0$ denote the probability distribution function of $\mathbf{R^{(0)}}$, the initial message of density evolution. Let $P_e^t \triangleq P_e(\mathbf{R}_t)$ denote the average probability of error at iteration $t$ under density evolution.

Assume $E \exp(s \cdot \bar{R}_1'^{(0)}) < \infty$ in some neighborhood of zero (where $\bar{R}_1'^{(0)}$ denotes element 1 of the LLR representation of $\bar{\mathbf{R}}^{(0)}$). Then

1) If $\lambda'(0)\rho'(1) > 1/\Delta$ then there exists a positive constant $\xi = \xi(\rho, \lambda, P_0)$ such that $P_e^t > \xi$ for all iterations $t$.
2) If $\lambda'(0)\rho'(1) < 1/\Delta$ then there exists a positive constant $\xi = \xi(\rho, \lambda, P_0)$ such that if $P_e^t < \xi$ at some iteration $t$, then $P_e^t$ approaches zero as $t$ approaches infinity.

Note that the requirement $E \exp(s \cdot \bar{R}_1'^{(0)}) < \infty$ is typically satisfied in channel of interest. The proof of Part 1 of the theorem is provided in Appendix V and the proof of Part 2 is provided in Appendix VI. Outlines of both proofs are provided below.

The proof of Part 1 is a generalization of a proof provided by Richardson *et al.* [29]. The proof [29] begins by observing that since the distributions at some iteration $t$ are symmetric, they may equivalently be modelled as APP values corresponding to the outputs of a MBIOS channel. By an erasure decomposition lemma, the output of an MBIOS channel can be modelled as the output of a degraded erasure channel. The proof proceeds by replacing the distributions at iteration $t$ by erasure-channel equivalents, and shows that the probability of error with the new distributions is lower bounded by some nonzero constant. Since the true MBIOS channel is a degraded version of the erasure channel, the true probability of error must be lower-bounded by the same nonzero constant as well.

Returning to the context of coset GF($q$) LDPC codes, we first observe that by Theorem 1 the random variable $\mathbf{R}_t$ at iteration $t$ is symmetric and hence by Lemma 6 it can be modelled as APP values of the outputs of a cyclic-symmetric channel. We then show that any cyclic-symmetric channel can be modelled as a degraded *erasurized* channel, appropriately defined. The continuation of the proof follows in the lines of [29].

The proof of Part 2 is a generalization of a proof by Khandekar [20]. As in [20] (and also [6]), our proof tracks a one-dimensional functional of the distribution of a message $\mathbf{X}$, denoted $D(\mathbf{X})$. We show that the rightbound messages at two consecutive iterations, satisfy $D(\mathbf{R}_{t+1}) \leq \Delta \cdot \lambda \left(1 - \rho(1 - D(\mathbf{R}_t)) + O(D(\mathbf{R}_t)^2)\right)$. Using first-order Taylor expansions of $\lambda(\cdot)$ and $\rho(\cdot)$, we proceed to show $D(\mathbf{R}_{t+1}) \leq \Delta \cdot \lambda'(0)\rho'(1) \cdot D(\mathbf{R}_t) + O(D(\mathbf{R}_t)^2)$. Since $\Delta \cdot \lambda'(0)\rho'(1) < 1$ by the theorem's conditions, for small enough $D(\mathbf{R}_t)$ we have $D(\mathbf{R}_{t+1}) \leq K \cdot D(\mathbf{R}_t)$ where $K < 1$, and thus $D(\mathbf{R}_t)$ descends to zero. Further details, including the relation between $D(\mathbf{R}_t)$ and $P_e^t$, are provided in Appendix VI.

### D. Gaussian Approximation

With binary LDPC, Chung *et al.* [9] observed that the rightbound messages of density-evolution are well approximated by Gaussian random variables. Furthermore, the symmetry of the messages in binary LDPC decoding implies that the mean $m$ and variance $\sigma^2$ of the random variable are related by $\sigma^2 = 2m$. Thus, the distribution of a symmetric Gaussian random variable may be described by a single parameter: $\sigma$. This property was also observed by ten Brink *et al.* [35] and is essential to their development of EXIT charts. In the context of nonbinary LDPC codes, Li *et al.* [22] obtained a description of the $q - 1$-dimensional messages, under a Gaussian assumption, by $q - 1$ parameters.

In the following theorem, we use symmetry and permutation-invariance as defined in Sections V-B and VI-B to reduce the number of parameters from $q - 1$ to one. This is a key property that enables the generalization of EXIT charts to coset GF($q$) LDPC codes.

Note that the theorem assumes a continuous Gaussian distribution. The definition of symmetry for LLR-vector random variables (Lemma 2) is extended to continuous distributions by replacing the probability function in (19) with a probability density function.

*Theorem 6:* Let $\mathbf{W}$ be an LLR-vector random-variable, Gaussian distributed with a mean $\mathbf{m}$ and covariance matrix $\Sigma$. Assume that the probability density function $f(\mathbf{w})$ of $\mathbf{W}$ exists and that $\Sigma$ is nonsingular. Then $\mathbf{W}$ is both symmetric and permutation-invariant if and only if there exists $\sigma > 0$ such that,

$$\mathbf{m} = \begin{bmatrix} \sigma^2/2 \\ \sigma^2/2 \\ ... \\ \sigma^2/2 \end{bmatrix} \quad \Sigma = \begin{bmatrix} \sigma^2 & & & \sigma^2/2 \\ & \sigma^2 & & \\ & & ... & \\ \sigma^2/2 & & & \sigma^2 \end{bmatrix} \tag{25}$$

That is, $m_i = \sigma^2/2$, $i = 1, ..., q - 1$, and $\Sigma_{i,j} = \sigma^2$ if $i = j$ and $\sigma^2/2$ otherwise.

The proof of this theorem is provided in Appendix VII. A Gaussian symmetric and permutation-invariant random variable, is thus completely described by a single parameter $\sigma$. In Sections VII-B and VII-D we discuss the validity of the Gaussian assumption with coset GF($q$) LDPC codes.

## VII. Design of Coset GF($q$) LDPC Codes

With binary LDPC codes, design of edge distributions is frequently done using extrinsic information transfer (EXIT) charts [35]. EXIT charts are particularly suited for designing LDPC codes for AWGN channels. In this section we develop EXIT charts for coset GF($q$) codes. We assume throughout the section transmission over AWGN channels.

### A. EXIT Charts

Formally, EXIT charts track the mutual information $I(C; \mathbf{W})$ between the transmitted code symbol $C$ at an average variable node[8] and the rightbound (leftbound) message $\mathbf{W}$ transmitted across an edge emanating from it. If this information is zero, then the message is independent of the transmitted code symbol and thus the probability of error is $(q-1)/q$. As the information approaches 1, the probability of error approaches zero. Note that we assume that the base of the $\log$ function in the mutual information is $q$, and thus $0 \le I(C; \mathbf{W}) \le 1$.

$I(C; \mathbf{W})$ is taken to represent the distribution of the message $\mathbf{W}$. That is, unlike density evolution, where the entire distribution of the message $\mathbf{W}$ at each iteration is recorded, with EXIT charts, $I(C; \mathbf{W})$ is assumed to be a faithful surrogate (we will shortly elaborate how this is done).

With EXIT charts, two curves (functions) are computed: The VND (variable node decoder) curve and the CND (check node decoder) curve, corresponding to the rightbound and leftbound steps of density-evolution, respectively. The argument to each curve is denoted $I_A$ and the value of the curve is denoted $I_E$. With the VND curve, $I_A$ is interpreted as equal to the functional $I(C; \mathbf{L}_t)$ when applied to the distribution of the leftbound messages $\mathbf{L}_t$ at a given iteration $t$. The output $I_E$ is interpreted as equal to $I(C; \mathbf{R}_t)$ where $\mathbf{R}_t$ is the rightbound message produced at the following rightbound iteration. With the CND curve, the opposite occurs.

Note that unlike density-evolution, where the densities are tracked from one iteration to another, the VND and CND curves are evaluated for every possible value of their argument $I_A$. However, a decoding *trajectory* that produces an approximation of the functionals $I(C; \mathbf{L}_t)$ and $I(C; \mathbf{R}_t)$ at each iteration, may be computed (see [36] for a discussion of the trajectory).

The decoding process is predicted to converge if after each decoding iteration (comprised of a leftbound and rightbound iteration), the resulting $I_E = I(C; \mathbf{R}_{t+1})$ is increased in comparison to $I_A = I(C; \mathbf{R}_t)$ of the previous iteration. We therefore require $I_{E,VND}(I_{E,CND}(I_A)) > I_A$ for all $I_A \in [0, 1]$. Equivalently, $I_{E,VND}(I_A) > I_{E,CND}^{-1}(I_A)$. In an EXIT chart, the CND curve is plotted with its $I_A$ and $I_E$ axes reversed (see, for example, Fig. 7). The decoding process is thus predicted to converge if and only if the VND curve is strictly greater than the reversed-axes CND curve.

### B. Using $I(C; \mathbf{W})$ as a Surrogate

Let $\mathbf{W}$ be a leftbound or rightbound message at some iteration of belief-propagation. Strictly speaking, an approximation of $I(C; \mathbf{W})$ requires not only the knowledge of the distribution of $\mathbf{W}$ but primarily the knowledge of the conditional distribution $\Pr[\mathbf{W} \mid C = i]$ for all $i = 0, ..., q-1$ (we assume that $C$ is uniformly distributed). However, as shown in Lemma 17 (Appendix III-A), the messages of the coset GF($q$) LDPC decoder satisfy

$$\Pr[\mathbf{W} = \mathbf{w} \mid C = i] = \Pr[\mathbf{W} = \mathbf{w}^{+i} \mid C = 0]$$

---

[8] In Definition 1, the notation $C$ was used to denote a code rather than a codeword symbol. The distinction between the two meanings is to be made based on the context of the discussion.

Thus, we may restrict ourselves to an analysis of the conditional distribution $\Pr[\mathbf{W} \mid C = 0]$.

*Lemma 11:* Under the tree-assumption, the above defined $\mathbf{W}$ satisfies:

$$I(C; \mathbf{W}) = 1 - E\left[\log_q\left(1 + \sum_{i=1}^{q-1} e^{-W_i}\right) \mid C = 0\right] \tag{26}$$

The proof of this lemma is provided in Appendix VIII-A. Note that by Lemma 16 (Appendix III-A), we may replace the conditioning on $C = 0$ in (26) by a conditioning on the transmission of the all-zero codeword. In the remainder of this section, we will assume that all distributions are conditioned on the all-zero codeword assumption.

In their development of EXIT charts for binary LDPC codes, ten Brink *et al.* [35] confine their attention to LLR message distributions that are Gaussian and symmetric. Under these assumptions, a message distribution is uniquely described by its variance $\sigma^2$. For every value of $\sigma$, they evaluate (26) (with $q = 2$) when applied to the corresponding Gaussian distribution. The result, denoted $J(\sigma)$, is shown to be monotonically increasing in $\sigma$. Thus $J^{-1}(\cdot)$ is well-defined. Given $I = I(C; W)$, $J^{-1}(I)$ can be applied to obtain the $\sigma$ that describes the corresponding distribution of $W$. Thus, $I(C; W)$ uniquely defines the entire distribution of $W$.

The Gaussian assumption is not strictly true. With binary LDPC codes, assuming transmission over an AWGN channel, the distributions of rightbound messages are approximately Gaussian mixtures (with irregular codes). The distributions of the leftbound messages, resemble "spikes". The EXIT method in [35] nonetheless continues to model the distributions as Gaussian. Simulation results are provided, which indicate that this approach still produces a very close prediction of the performance of binary LDPC codes.

With coset GF($q$) LDPC codes, we discuss two methods for designing EXIT charts. The first method models the LLR-vector messages distributions as Gaussian random variables, following the example of [35]. This modelling also enables us to evaluate the VND and CND curves using approximations that were developed in [35], thus greatly simplifying their computation.

However, the modelling of the rightbound message distributions of coset GF($q$) LDPC as Gaussian is less accurate than it is with binary LDPC codes. As we will explain in Section VII-D, this results from the distribution of the initial messages, which is not Gaussian even on an AWGN channel. In Section VII-D we will therefore develop an alternative approach, which models the rightbound distributions more accurately. We will then apply this approach in Section VII-E, to produce an alternative method for computing EXIT charts. With this method, the VND and CND curves are more difficult to compute. However, the method produces codes with approximately 1dB better performance.

*C. Computation of EXIT Charts, Method 1*

With this method, we confine our attention to distributions that are permutation-invariant[9], symmetric and Gaussian. By Theorem 6, under these assumptions, a $q-1$ dimensional LLR-vector message distribution is uniquely defined by a parameter $\sigma$. We proceed to define $J(\sigma)$ in a manner similar to that of [35]. In Appendix VIII-D we show that $J(\sigma)$ is monotonically increasing and thus $J^{-1}(\sigma)$ is well defined. Given $I = I(C; \mathbf{W})$, the distribution of $\mathbf{W}$ may be obtained in the same way as [35].

We use the following method to compute the VND and CND curves, based on a development of ten Brink *et al.* [35] for binary LDPC codes.

1) **The VND curve.** By (15), a rightbound message is a sum of incoming leftbound messages and an initial message. Let $I_A$ and $I^{(0)}$ denote the mutual-information functionals of the incoming leftbound messages and initial messages, respectively. By Lemma 5, $I^{(0)}$ equals the equiprobable-signalling capacity of the channel with the mapping $\delta(\cdot)$. It may be obtained by numerically evaluating $I(U, Y')$ as defined in Section V-C.

For each left-degree $i$, we let $I_{E,VND}(I_A; i, I^{(0)})$ denote the value of the VND curve when confined to the distribution of rightbound messages across edges whose left-degree is $i$. We now employ the following approximation, which holds under the tree assumption, when both the initial and the incoming leftbound messages are Gaussian.

$$I_{E,VND}(I_A; i, I^{(0)}) \approx J\left(\sqrt{(i-1)[J^{-1}(I_A)]^2 + [J^{-1}(I^{(0)})]^2}\right)$$

The validity of this approximation relies on the observation that a rightbound message (15) is equal to a sum of $i-1$ i.i.d. leftbound messages and an independently distributed initial message (under the tree assumption). $[J^{-1}(I_A)]^2$ is the variance of each of the leftbound messages and $[J^{-1}(I^{(0)})]^2$ is the variance of the initial message, and hence the variance of the rightbound message is $(i-1)[J^{-1}(I_A)]^2 + [J^{-1}(I^{(0)})]^2$.

2) **The CND curve.** Let $I_{E,CND}(I_A; j)$ denote the value of the CND curve when confined to the distribution of leftbound messages across edges whose right-degree is $j$.

$$I_{E,CND}(I_A; j) \approx 1 - J\left(\sqrt{j-1} \cdot J^{-1}(1 - I_A)\right)$$

This approximation is based on a similar approximation that was used in [35] and relies on Sharon *et al.* [31]. In the context of coset GF($q$) LDPC codes, we have verified its effectiveness empirically.

Given an edge distribution pair $(\lambda, \rho)$, we have

$$
\begin{aligned}
I_{E,VND}(I_A; I^{(0)}) &= \sum_{i=2}^{c} \lambda_i I_{E,VND}(I_A; i, I^{(0)}) \\
I_{E,CND}(I_A) &= \sum_{j=2}^{d} \rho_j I_{E,CND}(I_A; j)
\end{aligned}
\tag{27}
$$

Code design may be performed by fixing the right-distribution $\rho$ and computing $\lambda$. Like [35], the following constraints are used in the design.

---

[9] Strictly speaking, rightbound messages are not permutation-invariant. However, in Appendix VIII-B, we show that this does not pose a problem to the derivation of EXIT charts.

1) $\boldsymbol{\lambda}$ is required be a valid probability vector. That is $\lambda_i \geq 0 \; \forall i$, and $\sum_i \lambda_i = 1$.

2) To ensure decoding convergence, we require $I_{E,VND}(I, I^{(0)}) > I_{E,CND}^{-1}(I)$ (as explained in Section VII-A) for all $I$ belonging to a discrete, fine grid over $(0, 1)$.

The design process seeks to maximize $\sum \lambda_i / i$, which by (6) is equivalent to maximizing the design rate of the code. Typically, this can be done using a linear program. A similar process can be used to design $\boldsymbol{\rho}$ with $\boldsymbol{\lambda}$ fixed.

### D. More Accurate Modelling of Message Distributions

We now provide a more accurate model for the rightbound messages, as mentioned in Section VII-B. We focus, for simplicity, on regular LDPC codes. Observe that the computation of the rightbound message using (15) involves the summation of i.i.d leftbound messages, $\mathbf{l}'^{(n)}$. This sum is typically well-approximated by a Gaussian random variable[10]. To this sum, the initial message $\mathbf{r}'^{(0)}$ is added. With binary LDPC codes, transmission over an AWGN channel results in an initial message $\mathbf{r}'^{(0)}$ which is also Gaussian distributed (assuming the all-zero codeword was transmitted). Thus, the rightbound messages are very closely approximated by a Gaussian random variable.

With coset GF($q$) LDPC codes, the initial message is not well approximated by a Gaussian random variable, as illustrated in the following lemma:

*Lemma 12:* Consider the initial message produced at some variable node, under the all-zero codeword assumption, using LLR representation. Assume the transmission is over an AWGN channel with noise variance $\sigma_z^2$ and with a mapping $\delta(\cdot)$. Let the coset symbol at the variable node be $v$. Then the initial message $\mathbf{r}'^{(0)}$ is given by $\mathbf{r}'^{(0)} = \boldsymbol{\alpha}(v) + \boldsymbol{\beta}(v) \cdot z$, where $z$ is the noise produced by the channel and $\boldsymbol{\alpha}(v)$ and $\boldsymbol{\beta}(v)$ are $q - 1$ dimensional vectors, dependent on $v$, whose components are given by,

$$\alpha(v)_i = \frac{1}{2\sigma_z^2}(\delta(v) - \delta(v + i))^2, \quad \beta(v)_i = \frac{1}{\sigma_z^2}(\delta(v) - \delta(v + i))$$

The proof of this lemma is straightforward from the observation that the received channel output is $y = \delta(v) + z$.

In our analysis, we assume a random coset symbol $V$ that is uniformly distributed in GF($q$). Thus, $\boldsymbol{\alpha}(V)$ and $\boldsymbol{\beta}(V)$ are random variables, whose values are determined by the mapping $\delta(\cdot)$ and by the noise variance $\sigma_z^2$. The distribution of the channel noise $Z$ is determined by $\sigma_z^2$. The distribution of the initial messages is therefore determined by $\delta(\cdot)$ and $\sigma_z^2$.

Fig. 5 presents the empirical distribution of LLR messages at several stages of the decoding process, as observed by simulations. The code was a $(3, 6)$ coset GF(3) LDPC. Since $q = 3$, the LLR messages in this case are two-dimensional. The distribution of the initial messages (Fig. 5(a)) is seen to be a mixture of one-dimensional Gaussian curves, as predicted by Lemma 12. The leftbound messages at the first iteration are shown in Fig. 5(b). We model their distribution as Gaussian, although it resembles a "spike" and not the distribution of a Gaussian random variable (this situation is similar to the one with binary LDPC [9]). Fig. 5(c) presents the sum of leftbound messages computed in the process of evaluating (15). As predicted, this sum is well approximated by a Gaussian random variable. Finally, the rightbound messages at the first iteration are given in Fig. 5(d).

---

[10]Quantification of the quality of the approximation is beyond the scope of this discussion. "Well approximated" is to be understood in a heuristic sense, in the context of suitability to design using EXIT charts.

Following the above discussion, we model the distribution of the rightbound messages as the sum of two random vectors. The first is distributed as the initial messages above, and the second (the intermediate sum of leftbound messages) is modelled as Gaussian[11].

The intermediate value (the second random variable) is symmetric and permutation-invariant. This may be seen from the fact that the leftbound messages are symmetric and permutation-invariant (by Theorems 3 and 4) and from Lemmas 18 (Appendix III-E) and 22 (Appendix IV-E). Thus, by Theorem 6, it is characterized by a single parameter $\sigma$.

Summarizing, the approximate distribution of rightbound messages is determined by three parameters: $\sigma_z^2$ and $\delta(\cdot)$, which determine the distribution of the initial message, and $\sigma$, which determines the distribution of the intermediate value.

### E. Computation of EXIT Charts, Method 2

The second method for designing EXIT charts differs from the first (Section VII-C) in its modelling of the initial and rightbound message distributions, following the discussion in Section VII-D. We continue, however, to model the leftbound messages as Gaussian.

For every value of $\sigma$, we define $J_R(\sigma; \sigma_z, \delta)$ ($\sigma_z$ and $\delta$ are fixed parameters) in a manner analogous to $J(\sigma)$ as discussed in Section VII-C. That is, $J_R(\sigma; \sigma_z, \delta)$ equals (26) when applied to the rightbound distribution corresponding to $\sigma$, $\sigma_z^2$ and $\delta$. In an EXIT chart, $\sigma_z$ and $\delta(\cdot)$ are fixed. The remaining parameter that determines the rightbound distribution is thus $\sigma$, and $\sigma = J_R^{-1}(I; \sigma_z, \delta)$ is well-defined[12]. The computation of $J_R$ and $J_R^{-1}$ is discussed in Appendix VIII-E.

The following method is used to compute the VND and CND curves.

1) **The VND curve.** For each left-degree $i$, we evaluate $I_{E,VND}(I_A; i, \sigma_z, \delta)$ (defined in a manner analogous to $I_{E,VND}(I_A; i, I^{(0)})$ of Section VII-C) using the following approximation:

$$I_{E,VND}(I_A; i, \sigma_z, \delta) \approx J_R\left(\sqrt{(i-1)[J^{-1}(I_A)]^2}; \ \sigma_z, \delta\right)$$

2) **The CND curve.** Let $I_{E,CND}(I_A; j, \sigma_z, \delta)$ be defined in a manner analogous to $I_{E,CND}(I_A; j)$ of Section VII-C. The parameters $\sigma_z$ and $\delta$ are used in conjunction with $\sigma = J_R^{-1}(I_A; \sigma_z, \delta)$ to characterize the distribution of the rightbound messages at the input of the check-nodes. The computation of $I_{E,CND}(I_A; j, \sigma_z, \delta)$ is done empirically and is elaborated in Appendix VIII-F.

Given an edge distribution pair $(\lambda, \rho)$ we evaluate $I_{E,VND}(I_A; \sigma_z, \delta)$ and $I_{E,CND}(I_A; \sigma_z, \delta)$ from the above computed $\{I_{E,VND}(I_A; i, \sigma_z, \delta)\}_{i=1}^c$ and $\{I_{E,CND}(I_A; j, \sigma_z, \delta)\}_{j=1}^d$ using expressions similar to (27).

Note that $J_R(\sigma; \sigma_z, \delta)$ needs to be computed once for each choice of $\sigma_z$ and $\delta(\cdot)$. $I_{E,CND}(\sigma; j, \sigma_z, \delta)$ needs to be computed also for each value of $j$. $J(\sigma)$ needs to be computed once for each choice of $q$.

---

[11]Note that with irregular codes, the number of i.i.d leftbound variables that is summed is a random variable itself (distributed as $\{\lambda_i\}_{i=1}^c$), and thus the distribution of this random variable resembles a Gaussian mixture rather than a Gaussian random variable. However, we continue to model it as Gaussian, following the example that was set with binary codes [35].

[12]See Appendix VIII-E for a more accurate discussion of this matter.

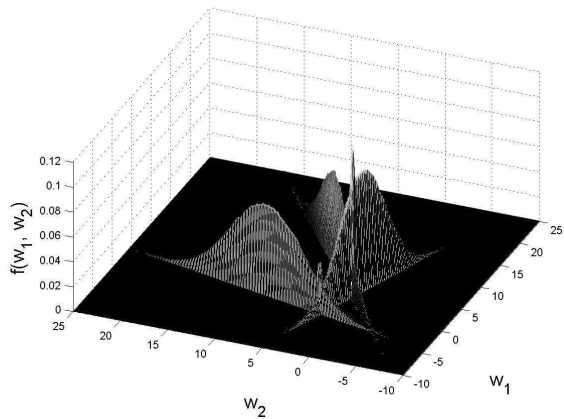

(a) Initial messages

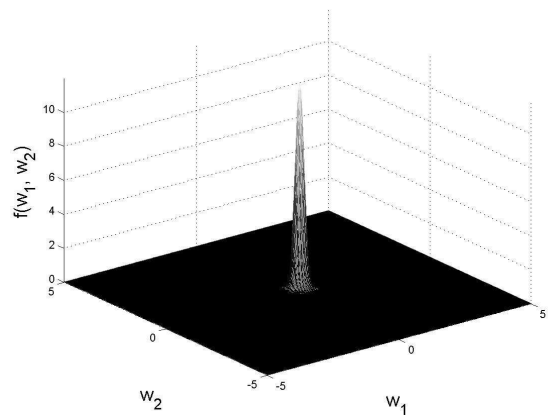

(b) Leftbound messages

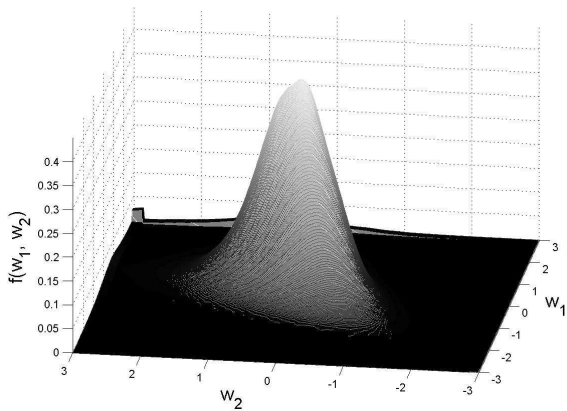

(c) Sum of leftbound messages, prior to the addition of the initial message.

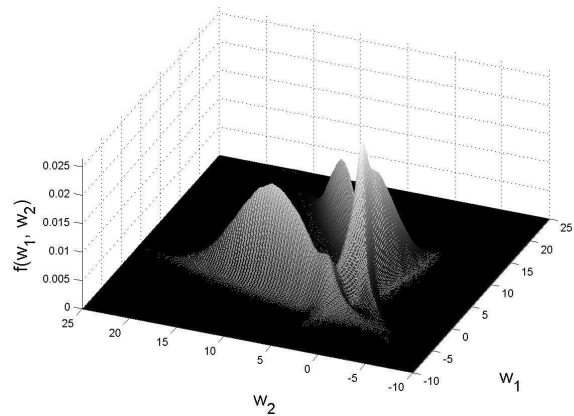

(d) Rightbound messages.

Fig. 5.   Empirical distributions of the messages of a (3,6) ternary coset LDPC code

Design of edge-distributions $\lambda$ and $\rho$ may be performed by linear programming in the same manner as in Section VII-C. Further details are provided in Section VII-F below.

*F. Design Examples*

We designed codes for spectral efficiencies of 6 bits/s/Hz (3 bits per dimension) and 8 bits/s/Hz (4 bits per dimension) over the AWGN channel. In all our constructions, we used the above method 2 (Section VII-E) to compute the EXIT charts. Our Matlab source code is provided at [4].

For the code at 6 bits/s/Hz, we set the alphabet size at $q = 32$. We used a nonuniformly-spaced signal constellation $\mathcal{A}$ (following the discussion of Section III-C). The constellation was obtained by applying the following method, which is a variation of a method suggested by Sun and van Tilborg [33]. First, the unique points $x_0 < x_1 < ... <$ $x_{q-1}$ were computed such that for $X \sim \mathcal{N}(0, 1)$, $\Pr[x_i < X < x_{i+1}] = 1/(q+1)$ $i = 0, ..., q-2$ and $\Pr[X < x_0] = \Pr[X > x_{q-1}] = 1/(q+1)$. The signal constellation was obtained by scaling the result so that the average energy was 1. The mapping $\delta$ from the code alphabet is given below, with its elements listed in ascending order using the representation of GF(32) elements as binary numbers (e.g. $\delta(00000) = -2.0701$, $\delta(00001) = -1.7096$). Note, however, that our simulations indicate that for a given $\mathcal{A}$, different mappings $\delta$ typically render the same performance.

$$\delta = [-2.0701, -1.7096, -1.473, -1.2896, -1.1362, -1.0022, -0.88161, -0.77061, -0.66697, -0.569,$$
$$-0.47523, -0.38474, -0.29689, -0.21075, -0.12592, -0.041887, 0.041887, 0.12592, 0.21075, 0.29689,$$
$$0.38474, 0.47523, 0.569, 0.66697, 0.77061, 0.88161, 1.0022, 1.1362, 1.2896, 1.473, 1.7096, 2.0701]$$

We fixed $\rho(7) = 1$ and iteratively applied linear programming, first to obtain $\boldsymbol{\lambda}$, and then, fixing $\boldsymbol{\lambda}$, to obtain a better $\boldsymbol{\rho}$.

Rather than require $I_{E,VND}(I; \sigma_z, \delta) > I_{E,CND}^{-1}(I; \sigma_z, \delta)$ as in Sections VII-A and VII-C, we enforced a more stringent condition when designing $\boldsymbol{\lambda}$. We required $I_{E,VND}(I; \sigma_z, \delta) > I_{E,CND}^{-1}(I; \sigma_z, \delta) + \epsilon(I)$ where $\epsilon(I)$ equals $5 \cdot 10^{-3}$ when $I \in (0, 0.5)$, equals $4 \cdot 10^{-3}$ when $I \in [0.5, 0.6)$ and is zero elsewhere. Similarly, when designing $\boldsymbol{\rho}$, we required $I_{E,CND}(I; \sigma_z, \delta) > I_{E,VND}^{-1}(I; \sigma_z, \delta) + 5 \cdot 10^{-3}$.

After a few linear programming iterations, we obtained the edge-distributions $\lambda(2, 5, 6, 16, 30) = (0.5768, 0.1498, 0.07144, 0.1045, 0.09752)$, $\rho(5, 6, 7, 8, 20) = (0.09973, 0.02331, 0.5885, 0.1833, 0.1051)$. The code rate is $3/5$ GF(32) symbols per channel use, equal to 3 bits per channel use, and a spectral efficiency of 6 bits/s/Hz. Interestingly, this code is right-irregular, unlike typical binary LDPC codes. Fig. 6 presents the EXIT chart for the code (computed by method 2). Note that the CND curve in Fig. 6 does not begin at $I_A = 0$. This is discussed in Appendix VIII-F.

Simulation results indicate successful decoding at an SNR of 18.55 dB. The block length was $1.8 \cdot 10^5$ symbols, and decoding typically converged after approximately 150–200 iterations. The symbol error rate, after 50 simulations, was approximately $10^{-6}$. The unconstrained Shannon limit (i.e. not restricted to any signal constellation) at this rate is 17.99 dB, and thus our gap from this limit is 0.56 dB. This result is well beyond the shaping gap, which at 6 bits/s/Hz is approximately 1.1 dB.

We can obtain some interesting insight on these figures by considering the equiprobable-signalling Shannon-limit for our constellation (defined based on the equiprobable-signalling capacity, which was introduced in Section V-C). At 6 bits/s/Hz, this limit equals 18.25 dB. The equiprobable-signalling Shannon limit is the best we can hope for with any design method for the edge-distributions of our code. The gap between our code's threshold and this limit is just 0.3 dB, indicating the effectiveness of our EXIT chart design method.

The equiprobable-signalling Shannon limit for a 32-PAM constellation, at 6 bits/s/Hz is 19.11 dB. The gap between this limit and the above-discussed limit for our constellation, is 0.86 dB. This is the shaping gain obtained from the use of a nonuniform signal constellation.

For the code at 8 bits/s/Hz, we set the alphabet size at $q = 64$. We used the same method to construct a nonuniformly-spaced signal constellation. The mapping to the signal constellation is given below.

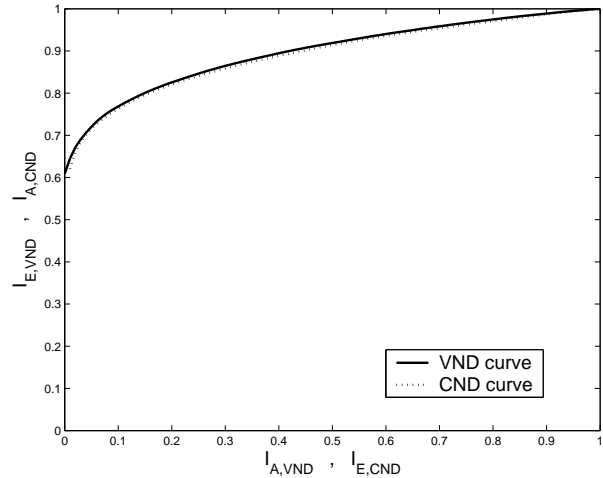

Fig. 6. An EXIT chart, computed using method 2, for a code at a spectral efficiency of 6 bits/s/Hz and an SNR of 18.5 dB.

$$\delta = [-2.29, -1.98, -1.78, -1.63, -1.51, -1.4, -1.31, -1.23, -1.15, -1.08, -1.01, -0.951, -0.891, -0.834,$$
$$-0.78, -0.727, -0.676, -0.627, -0.579, -0.532, -0.486, -0.441, -0.397, -0.354, -0.311, -0.268, -0.226,$$
$$-0.185, -0.143, -0.102, -0.0613, -0.0204, 0.0204, 0.0613, 0.102, 0.143, 0.185, 0.226, 0.268, 0.311, 0.354, 0.397,$$
$$0.441, 0.486, 0.532, 0.579, 0.627, 0.676, 0.727, 0.78, 0.834, 0.891, 0.951, 1.01, 1.08, 1.15, 1.23, 1.31, 1.4, 1.51,$$
$$1.63, 1.78, 1.98, 2.29]$$

We fixed $\rho(8) = 1$ and applied one iteration of linear programming to obtain $\lambda(2, 9, 29) = (0.7087, 0.1397, 0.1516)$. The code rate is $2/3$ GF(64) symbols per channel use, equal to 4 bits per channel use, and a spectral efficiency of 8 bits/s/Hz. Fig. 7 presents the EXIT charts for the code using the two methods.

Simulation results indicate successful decoding at an SNR of 25.06 dB over the AWGN channel. The block length was $10^5$ symbols, and decoding typically converged after approximately 70 iterations. The symbol error rate, after 100 simulations, was exactly zero. We also applied an approximation of density-evolution by Monte-Carlo simulations, as mentioned in Section VI-A, and obtained similar results. The gap between our code's threshold and the unconstrained Shannon limit, which at 8 bits/s/Hz is 24.06 dB, is 1 dB. This result is beyond the shaping gap, which at 8 bits/s/Hz is 1.3 dB. The equiprobable-signalling Shannon limit for our signal constellation at 8 bits/s/Hz is 24.34 dB. The gap between our code's threshold and this limit is thus only 0.72 dB.

## VIII. Comparison with other Bandwidth-Efficient Coding Schemes

The simulation results presented in Section VII-F indicate that coset GF($q$) LDPC codes have remarkable performance over bandwidth-efficient channels. In this section, we compare their performance with multilevel coding using binary LDPC component codes and with with turbo-TCM.

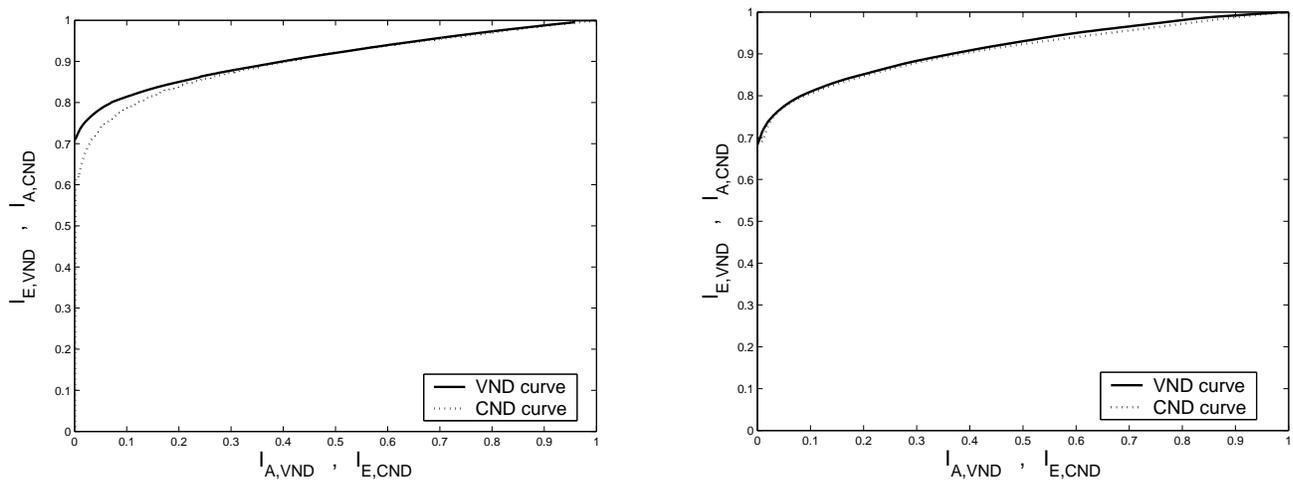

(a) An EXIT chart computed using method 1 at an SNR of 26.06 dB.

(b) An EXIT chart computed using method 2 at an SNR of 25.06 dB.

Fig. 7. EXIT charts for a code at a spectral efficiency of 8 bits/s/Hz.

### A. Comparison with Multilevel Coding (MLC)

Hou *et al.* [18] presented simulations for MLC over the AWGN channel at a spectral efficiency of 2 bits/s/Hz (equal to 1 bit per dimension), using a 4-PAM constellation. The equiprobable-signalling Shannon limit[13] for 4-PAM and at this rate is 5.12 dB (SNR). Their best results were obtained using multistage decoding (MSD). At a block length of $10^4$ symbols, their best code is capable of transmission at 1 dB of the Shannon limit with an average BER of about $10^{-5}$. It is composed of binary LDPC component codes with maximum left-degrees of 15.

We designed edge-distributions for two coset GF(4) LDPC codes at the same spectral efficiency, signal constellation and BER as [18]. Our first code's edge-distributions are given by $\lambda(2, 3, 4, 5, 6, 7, 15, 16, 20, 21) = (0.341895, 0.172092, 0.081613, 0.064992, 0.043213, 0.000037, 0.029562, 0.140071, 0.000002, 0.126522)$ and $\rho(7) = 1$. Our simulations at a block-length of $10^4$ indicate that this code is capable of transmission within 0.55 dB of the Shannon limit (100 simulations), and thus has a substantial advantage over the above MLC LDPC code, which is capable of transmission only within 1 dB of the Shannon limit.

Our above code has obtained its superior performance at the price of increased decoding complexity, in comparison with the MLC code of [18]. We also designed a second code, with a lower decoding complexity, in order to compare the two schemes when the complexity is restricted. This code's edge distributions are given by $\lambda(2, 3, 6) = (0.3978, 0.2853, 0.3169)$ and $\rho(5, 6) = (0.203, 0.797)$. Our simulation results indicate that the code is capable of reliable transmission within 0.8 dB of the Shannon limit. The code's maximum left-degree is 6, and is thus lower than the MLC code of [18]. Consequently, it has a lower level of connectivity in its Tanner graph, implying that its slightly better performance was achieved at a comparable decoding complexity. A precise comparison between the decoding complexities of the two codes must account for the entire edge-distributions (rather than just the maximum left-degrees), and for the number of decoding iterations. Such a comparison is beyond the scope of this work.

---

[13] Throughout this section, we assume equiprobable-signalling whenever we refer to the Shannon limit.

Hou *et al.* [18] also experimented at a large block length of $10^6$ symbols. Their best code is capable of transmission within 0.14 dB of the Shannon limit. At a slightly smaller block length ($5 \cdot 10^5$ symbols), our above-discussed first code is capable of transmission within 0.2 dB of the Shannon limit (14 simulations), and thus has a slightly inferior performance. This may be attributed either to the smaller block-length that we used, or to the availability of density-evolution for the design of binary MLC component LDPC codes at large block lengths.

Hou *et al.* [18] obtained their remarkable performance at large block lengths also at the price of increased decoding complexity (the maximum left-degrees of their component codes are 50). It could be argued that increasing the decoding complexity could produce improved performance also at the above mentioned block length of $10^4$. We believe this not to be true, because increasing the maximum left-degree would also result in an increase in the Tanner graph connectivity. This, at short block lengths, would dramatically increase the number of cycles in the graph, thus reducing performance.

Summarizing, our simulations indicate that coset GF($q$) LDPC have an advantage over MLC LDPC codes at short block lengths in terms of the gap from the Shannon limit. This result assumes no restriction on decoding complexity. The simulations also indicate that when decoding complexity is restricted, both schemes admit comparable performance. In this case, however, further research is required in order to provide a more accurate comparison of the two schemes.

### B. Comparison with Turbo Trellis-Coded Modulation (Turbo-TCM)

Robertson and Wörz [30] experimented with turbo-TCM at several spectral efficiencies and block lengths. The highest spectral-efficiency they experimented at was 5 bits/s/Hz. They used a 64-QAM constellation, and their best results were achieved at a block length of 3000 QAM symbols. They obtained a BER of $10^{-4}$ at an SNR of about 16.85 dB. The equiprobable-signalling Shannon-limit at 5 bits/s/Hz is 16.14 dB, and thus their result is within approximately 0.7 dB of the Shannon limit.

We experimented with an 8-PAM constellation and a block length of 6000 PAM symbols, which are the one-dimensional equivalents of two-dimensional 64-QAM and of 3000 QAM symbols. Our code's edge distributions are $\lambda(2, 3, 4, 18) = (0.375115, 0.049623, 0.255708, 0.319554)$ and $\rho(21) = 1$. Simulation results indicate a symbol error rate of less than $10^{-4}$ at an SNR of 16.6 dB (100 simulations). This result is within 0.46 dB of the Shannon limit, and thus exceeds the above result of 0.7 dB.

## IX. CONCLUSION

### A. Suggestions for Further Research

1) **Nonuniform labels.** The labels of GF($q$) LDPC codes, as defined in Section III-A, are randomly selected from GF($q$)\{0} with uniform probability. Davey and MacKay [10], in their work on GF($q$) LDPC codes for binary channels, suggested selecting them differently. It would be interesting to investigate their approach (and possibly other approaches to the selection of the labels) when applied to coset GF($q$) LDPC codes for nonbinary channels.

2) **Density evolution.** In Section VI-A, we discussed the difficulty in efficiently computing density evolution for nonbinary codes. An assumption in that discussion is that the densities would be represented on a grid of the form $\{-M/2 \cdot \Delta, ..., M/2 \cdot \Delta\}^{q-1}$ (assuming LLR-vector representation), requiring an amount of memory of the order of $(M + 1)^{q-1}$ . However, a more efficient approach would be to experiment with other forms of quantization, perhaps tailored to each density. We have tried applying the Lloyd-Max algorithm to design such quantizers for each density. However, the computation of the algorithm, coupled with the actual application of the quantizer, are too computationally complex. An alternative approach would perhaps make use of a Gaussian approximation as described in Section VI-D to design effective quantizers.

3) **Other surrogates for distributions.** In [6], the functional $EX$ ($X$ denoting a message of a binary LDPC decoder) was used to lower-bound (rather than approximate) the asymptotic performance of binary LDPC codes. It would be interesting to find a similar, scalar, functional that can be used to bound the performance of coset GF($q$) LDPC codes. Another possibility is to experiment with the function $D(\mathbf{X})$, which is defined in Appendix VI.

4) **Comparison with the $q$-ary erasure channel (QEC).** In a $\mathrm{QEC}(\epsilon)$ channel, the output symbol is equal to the input with a probability of $1 - \epsilon$ and to an erasure with a probability of $\epsilon$. Much of the analysis of Luby *et al.* [23] for LDPC codes over binary erasure channels is immediately applicable to GF($q$) LDPC codes over QEC channels. It may be possible to gain insight on coset GF($q$) LDPC codes from an analysis of their use over the QEC.

5) **Better mappings.** The mapping function $\delta(\cdot)$ that was presented in Section VII-F was designed according to a concept that was developed heuristically. Further research may perhaps uncover better mapping methods.

6) **Additional channels.** The development in Section VII focuses on AWGN channels. It would be interesting to extend this development to additional types of channel.

7) **Additional applications.** In [3], coset GF($q$) LDPC codes were used for transmission over the binary dirty-paper channel. Applying an appropriately designed quantization mapping (as discussed in Section III-C), a binary code was produced whose codewords' empirical distribution was approximately Bernoulli(1/4). There are many other applications, beside bandwidth-efficient transmission, that could similarly profit from codewords with a nonuniform empirical distribution.

### B. Other Coset LDPC Codes

In [1], other nonbinary LDPC ensembles, called BQC-LDPC and MQC-LDPC, are considered (beside coset GF($q$) LDPC). Random-coset analysis, as defined in Section V, applies to these codes as well. Similarly, the all-zero codeword assumption (Lemma 1) and the symmetry of message distributions (Definition 4 and Theorem 1) apply to these codes. With MQC-LDPC, $+i$ in (2) is evaluated using modulo-$q$ arithmetic instead of over GF($q$). With BQC-LDPC decoders, which use scalar messages, symmetry coincides with the standard binary definition of [29]. Channel equivalence as defined in Section V-C applies to MQC-LDPC codes, but not to BQC-LDPC.

### C. Concluding Remarks

Coset GF($q$) LDPC codes are a natural extension of binary LDPC codes to nonbinary channels. Our main contribution in this paper is the generalization of much of the analysis that was developed by Richardson *et al.* [28], [29], Chung *et al.* [9], ten Brink *et al.* [35] and Khandekar [20] from binary LDPC codes to coset GF($q$) LDPC codes.

Random-coset analysis helps overcome the absence of output-symmetry. With it, we have generalized the all-zero codeword assumption, the symmetry property and channel equivalence. The random selection of the nonzero elements of the parity-check matrix (the labels) induces permutation-invariance on the messages. Although density-evolution is not realizable, permutation-invariance enables its analysis (e.g. the stability property) and approximation (e.g. EXIT charts).

Analysis of GF($q$) LDPC codes would not be interesting if their decoding complexity was prohibitive. Richardson and Urbanke [28] have suggested using the multidimensional DFT. This, coupled with an efficient recursive algorithm for the computation of the DFT, dramatically reduces the decoding complexity and makes coset GF($q$) LDPC an attractive option.

Although our focus in this work has been on the decoding problem, it is noteworthy that the work done by Richardson and Urbanke [27] on efficient *encoding* of binary LDPC codes is immediately applicable to coset GF($q$) LDPC codes. For simulation purposes, however, a pleasing side-effect of our generalization of the all-zero codeword assumption is that no encoder needs to be implemented. In a random coset setting, simulations may be performed on the all-zero codeword alone (of the underlying LDPC code).

Using quantization or non-uniform spaced mapping produces a substantial shaping gain. This, coupled with our generalization of EXIT charts has enabled us to obtain codes at 0.56 dB of the Shannon limit, at a spectral efficiency of 6 bits/s/Hz. To the best of our knowledge, these are the best codes found for this spectral efficiency. However, further research (perhaps in the lines of Section IX-A) may possibly narrow this gap to the Shannon limit even further.

### APPENDIX I

### PROPERTIES OF THE $+g$ AND $\times g$ OPERATORS

*Lemma 13:* For $g \in \mathrm{GF}(q) \backslash \{0\}$ and $i \in \mathrm{GF}(q)$,

1) $\left(\mathbf{x}^{+i}\right)^{\times g} = \left(\mathbf{x}^{\times g}\right)^{+(i \cdot g^{-1})}$
2) $\left(\mathbf{x}^{\times g}\right)^{+i} = \left(\mathbf{x}^{+(g \cdot i)}\right)^{\times g}$
3) $n(\mathbf{x}^{\times g}) = n(\mathbf{x})$
4) $n(\mathbf{x}^{+i}) = n(\mathbf{x})$

**Proof:** The first two identities are proved by examining the $k$th index of both side of the equation. The third identity is obtained from the second by observing that $(\mathbf{x}^{\times g})^{+j} = \mathbf{x}^{\times g}$ if and only if $\mathbf{x}^{+j \cdot g} = \mathbf{x}$. The fourth identity is straightforward. $\square$

*Lemma 14:* For $g \in \mathrm{GF}(q) \backslash \{0\}$ and $i \in \mathrm{GF}(q)$,

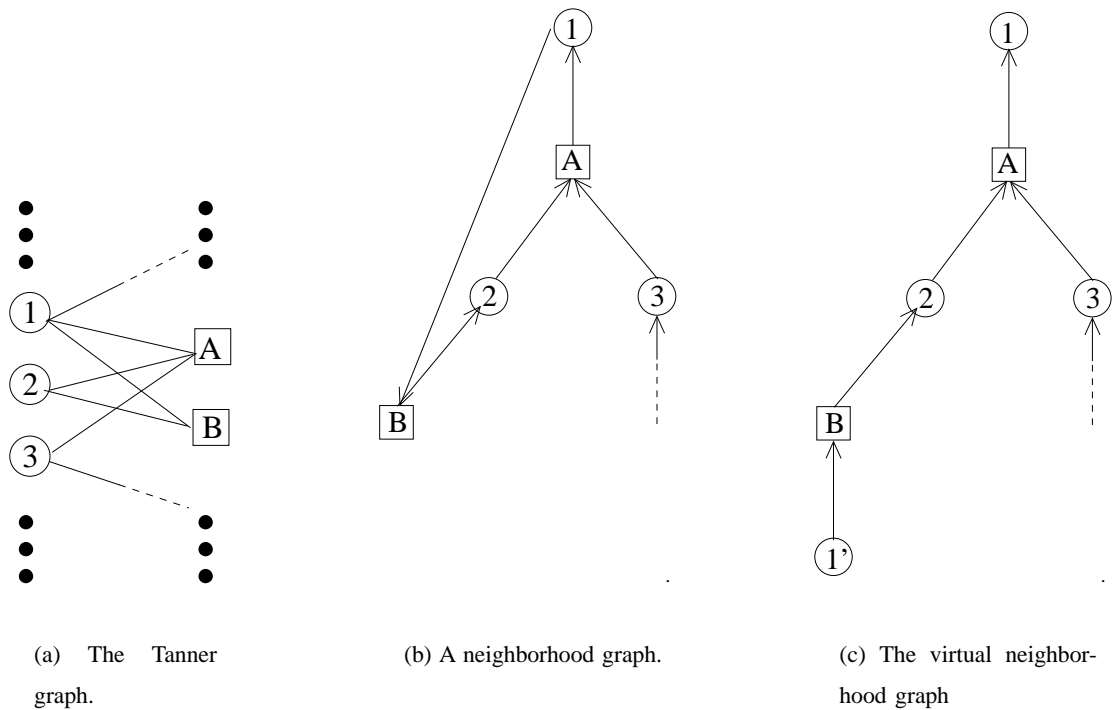

(a) The Tanner graph.

(b) A neighborhood graph.

(c) The virtual neighborhood graph

Fig. 8.   A neighborhood graph with cycles

1) $(\mathbf{x}^{\times g})^* = (\mathbf{x}^*)^{\times g}$ where $(\mathbf{x}^*)^{\times g}$ denotes the result of applying the operation $\times g$ on all elements of $\mathbf{x}^*$.

2) $(\mathbf{x}^{+i})^* = \mathbf{x}^*$

The proof of the first identity is obtained from Lemma 13, identity 2. The second identity is straightforward. □

## Appendix II

### Neighborhood Graphs with Cycles

Fig. 8(b) gives an example of a case where a neighborhood graph contains cycles. The neighborhood graph corresponds to the Tanner graph of Fig. 8(a).

When the neighborhood graph contains cycles, the APP values computed by a belief-propagation decoder correspond to a *virtual neighborhood graph*. In this graph, nodes that are contained in cycles are duplicated to artificially create a tree structure. For example, in Fig. 8(c) a variable-node $1'$ was produced by duplicating 1. The APP values are computed according to the virtual code[14] $\tilde{C}$ implied by this graph. $\tilde{C}$ is virtual in the sense that it is based on false assumptions regarding the channel model and the transmitted code. In Fig. 8(c), the channel model falsely assumes that the nodes $1'$ and 1 correspond to different channel observations.

## Appendix III

### Proofs for Section V

#### A. Preliminary Lemmas

The proofs in this section focus on the properties of a message produced at some iteration $t$ of coset GF$(q)$ LDPC belief propagation at a node $n$. Assuming the underlying code $C$ is fixed, this message is a function of the channel output $\mathbf{y}$ and the coset vector $\mathbf{v}$. We therefore denote it by $\mathbf{m}(\mathbf{y}, \mathbf{v})$.

$\mathbf{m}(\mathbf{y}, \mathbf{v})$ may be either a rightbound message from a variable-node or a leftbound message to a variable-node. In both cases, we denote the variable-node involved by $i$. We begin with the following lemma.

*Lemma 15:* Let $\mathbf{c}$ be a codeword of $C$, $\mathbf{y}$ some given channel output, and $\mathbf{v}$ an arbitrary coset vector. Then

$$\mathbf{m}(\mathbf{y}, \mathbf{v} - \mathbf{c}) = \mathbf{m}(\mathbf{y}, \mathbf{v})^{-c_i} \tag{28}$$

where $c_i$ is the value of $\mathbf{c}$ at the codeword position $i$.

In the left hand side of (28), $\mathbf{v} - \mathbf{c}$ is evaluated componentwise over $\mathrm{GF}(q)$. In the right hand side, we are using the notation of (2).

**Proof:** $m_k(\mathbf{y}, \mathbf{v})$ satisfies,

$$m_k(\mathbf{y}, \mathbf{v}) = \Pr[\sigma_i = k \mid \boldsymbol{\sigma} \in \tilde{C}, \, \delta(\boldsymbol{\sigma} + \tilde{\mathbf{v}}) \text{ was transmitted and } \tilde{\mathbf{y}} \text{ was received}] \tag{29}$$

The above expression is only an estimate of the true APP value. The code used by the decoder is not the LDPC code $C$, but rather the code $\tilde{C}$ defined by the parity-checks of the neighborhood graph spanned from $n$, as defined in Section IV-C and Appendix II. $\boldsymbol{\sigma}$ is a random variable representing the transmitted codeword of $\tilde{C}$ (prior to the addition of the coset vector) and $\sigma_i$ is its value at position $i$. The vectors $\tilde{\mathbf{v}}$ and $\tilde{\mathbf{y}}$ are constructed from $\mathbf{v}$ and $\mathbf{y}$ by including only values at nodes contained in the neighborhood graph of node $n$. We define $\tilde{\mathbf{c}}$ similarly. If the neighborhood graph contains cycles, we use the virtual neighborhood graph defined in Appendix II. For each variable-node that has duplicate copies in this graph, elements of the true $\mathbf{y}$, $\mathbf{v}$ and $\mathbf{c}$ will have duplicate entries in $\tilde{\mathbf{y}}$, $\tilde{\mathbf{v}}$ and $\tilde{\mathbf{c}}$.

The decoder assumes that all codewords are equally likely, hence (29) becomes

$$m_k(\mathbf{y}, \mathbf{v}) \quad = \quad \frac{\sum_{\sigma_i = k, \boldsymbol{\sigma} \in \tilde{C}} \Pr[\tilde{\mathbf{y}} \text{ was received} \mid \delta(\boldsymbol{\sigma} + \tilde{\mathbf{v}}) \text{ was transmitted}]}{\sum_{\boldsymbol{\sigma} \in \tilde{C}} \Pr[\tilde{\mathbf{y}} \text{ was received} \mid \delta(\boldsymbol{\sigma} + \tilde{\mathbf{v}}) \text{ was transmitted}]}$$

Equivalently, we obtain

$$m_k(\mathbf{y}, \mathbf{v} - \mathbf{c}) \quad = \quad \frac{\sum_{\sigma_i = k, \boldsymbol{\sigma} \in \tilde{C}} \Pr[\tilde{\mathbf{y}} \mid \delta(\boldsymbol{\sigma} + \tilde{\mathbf{v}} - \tilde{\mathbf{c}})]}{\sum_{\boldsymbol{\sigma} \in \tilde{C}} \Pr[\tilde{\mathbf{y}} \mid \delta(\boldsymbol{\sigma} + \tilde{\mathbf{v}} - \tilde{\mathbf{c}})]}$$

The word $\tilde{\mathbf{c}}$, having being constructed from a true codeword $\mathbf{c} \in C$, satisfies all parity-checks in the neighborhood graph and is therefore a codeword of $\tilde{C}$. Changing variables, we set $\boldsymbol{\sigma}' = \boldsymbol{\sigma} - \tilde{\mathbf{c}}$. Thus, for any $\boldsymbol{\sigma} \in \tilde{C}$, we have $\boldsymbol{\sigma}' \in \tilde{C}$. The condition $\sigma_i = k$ now becomes $\sigma'_i = k - c_i$ and we have

$$
\begin{aligned}
m_k(\mathbf{y}, \mathbf{v} - \mathbf{c}) \quad &= \quad \frac{\sum_{\sigma'_i = k - c_i, \boldsymbol{\sigma}' \in \tilde{C}} \Pr[\tilde{\mathbf{y}} \mid \delta(\boldsymbol{\sigma}' + \tilde{\mathbf{v}})]}{\sum_{\boldsymbol{\sigma}' \in \tilde{C}} \Pr[\tilde{\mathbf{y}} \mid \delta(\boldsymbol{\sigma}' + \tilde{\mathbf{v}})]} \\
&= \quad m_{k - c_i}(\mathbf{y}, \mathbf{v}) = m_k(\mathbf{y}, \mathbf{v})^{-c_i}
\end{aligned}
$$

$\square$

We now examine $\mathbf{X} \triangleq \mathbf{m}(\mathbf{Y}, \mathbf{V})$, which denotes the rightbound (leftbound) message from (to) a variable-node $i$, at some iteration of belief-propagation. $\mathbf{V}$ and $\mathbf{Y}$ are random variables representing the coset vector and channel-output vectors, respectively.

*Lemma 16:* For any $k \in \mathrm{GF}(q)$, the value $\Pr[\mathbf{X} = \mathbf{x} \mid c_i = k]$ is well-defined in the sense that for any two codewords $\mathbf{c}^{(1)}, \mathbf{c}^{(2)} \in C$ that satisfy $c_i^{(1)} = c_i^{(2)} = k$,

$$\Pr[\mathbf{X} = \mathbf{x} \mid \mathbf{c}^{(1)} \text{ was transmitted}] = \Pr[\mathbf{X} = \mathbf{x} \mid \mathbf{c}^{(2)} \text{ was transmitted}]$$

for all probability vectors $\mathbf{x}$.

*Proof:* Let $\tilde{\mathbf{c}} \triangleq \mathbf{c}^{(2)} - \mathbf{c}^{(1)}$. Consider transmission of $\mathbf{c}^{(1)}$ with an arbitrary coset vector of $\mathbf{v}$, compared to transmission of $\mathbf{c}^{(2)}$ with a coset vector of $\mathbf{v} - \tilde{\mathbf{c}}$. In both cases, the transmitted signal over the channel is $\delta(\mathbf{v} + \mathbf{c}^{(1)})$, and hence the probability of obtaining any particular $\mathbf{y}$ is identical. The word $\tilde{\mathbf{c}}$ satisfies $\tilde{c}_i = 0$. Since $C$ is linear, we have $\tilde{\mathbf{c}} \in C$. Therefore, Lemma 15 (Appendix III-A) implies

$$\mathbf{m}(\mathbf{y}, \mathbf{v} - \tilde{\mathbf{c}}) = \mathbf{m}(\mathbf{y}, \mathbf{v})^{-\tilde{c}_i} = \mathbf{m}(\mathbf{y}, \mathbf{v}) \tag{30}$$

We therefore obtain that

$$\Pr[\mathbf{X} = \mathbf{x} \mid \mathbf{V} = \mathbf{v}, \mathbf{c}^{(1)} \text{ was transmitted}] = \Pr[\mathbf{X} = \mathbf{x} \mid \mathbf{V} = \mathbf{v} - \tilde{\mathbf{c}}, \mathbf{c}^{(2)} \text{ was transmitted}]$$

Since $\mathbf{V}$ is uniformly distributed, averaging over all possible values of $\mathbf{V}$ completes the proof. □

The following lemma will be useful in Section VII-A.

*Lemma 17:* For any $k \in \mathrm{GF}(q)$,

$$\Pr[\mathbf{X} = \mathbf{x} \mid c_i = k] = \Pr[\mathbf{X} = \mathbf{x}^{+k} \mid c_i = 0]$$

*Proof:* The proof follows almost in direct lines as Lemma 16. Let $\mathbf{c}^{(2)}$ be the all-zero codeword, and $\mathbf{c}^{(1)}$ a codeword that satisfies $\mathbf{c}_i^{(1)} = k$. Thus

$$\Pr[\mathbf{X} = \mathbf{x} \mid c_i = k] = \Pr[\mathbf{X} = \mathbf{x} \mid \mathbf{c}^{(1)} \text{ was transmitted}]$$
$$\Pr[\mathbf{X} = \mathbf{x} \mid c_i = 0] = \Pr[\mathbf{X} = \mathbf{x} \mid \mathbf{c}^{(2)} \text{ was transmitted}]$$

$\tilde{c}_i = -k$, and thus (30) now becomes

$$\mathbf{m}(\mathbf{y}, \mathbf{v} - \tilde{\mathbf{c}}) = \mathbf{m}(\mathbf{y}, \mathbf{v})^{-\tilde{c}_i} = \mathbf{m}(\mathbf{y}, \mathbf{v})^{+k}$$

Thus,

$$\Pr[\mathbf{X} = \mathbf{x} \mid \mathbf{V} = \mathbf{v}, \mathbf{c}^{(1)} \text{ was transmitted}] = \Pr[\mathbf{X} = \mathbf{x}^{+k} \mid \mathbf{V} = \mathbf{v} - \tilde{\mathbf{c}}, \mathbf{c}^{(2)} \text{ was transmitted}]$$

Averaging over all possible values of $\mathbf{V}$ completes the proof. □

### B. Proof of Lemma 1

Let $\mathbf{c}$ be some codeword. Let $E_{\mathbf{y}, \mathbf{v}}^t(\mathbf{c})$ denote the event of error at a message produced at a variable-node $i$ after iteration $t$, assuming the channel output was $\mathbf{y}$, the coset vector was $\mathbf{v}$ and the true codeword was $\mathbf{c}$. Recalling the decision rule of Section IV-A, the decoder decides $\mathrm{argmax}_k\{m_k(\mathbf{y}, \mathbf{v})\}$ (where $m_k(\mathbf{y}, \mathbf{v})$ is defined as in Appendix III-A). Using Lemma 15 (Appendix III-A), we obtain that the maximum of $\{m_k(\mathbf{y}, \mathbf{v})\}$ is obtained at 0 if and only if the maximum of $\{m_k(\mathbf{y}, \mathbf{v} - \mathbf{c})\}$ is obtained at $c_i$. Therefore

$$E_{\mathbf{y}, \mathbf{v} - \mathbf{c}}^t(\mathbf{c}) = E_{\mathbf{y}, \mathbf{v}}^t(\mathbf{0})$$

In both cases, the word transmitted over the channel is $\delta(\mathbf{v})$ and hence the probability of obtaining any channel output $\mathbf{y}$ is the same. Therefore we obtain

$$\overline{P}^t_{e|\mathbf{v}-\mathbf{c}}(\mathbf{c}) = \overline{P}^t_{e|\mathbf{v}}(\mathbf{0})$$

Finally, averaging over all instances of $\mathbf{v}$, we obtain

$$\overline{P}^t_e(\mathbf{c}) = \overline{P}^t_e(\mathbf{0})$$

□

### C. Proof of Lemma 2

We first assume that $\mathbf{X}$ is symmetric and prove (19). Let $\mathbf{w}$ be an arbitrary LLR-vector, $\mathbf{x} \triangleq \mathrm{LLR}^{-1}(\mathbf{w})$ and $\mathbf{x}^{+i}, \mathbf{w}^{+i}$ be defined using (2) and (5), respectively.

$$
\begin{aligned}
e^{w_i}\Pr[\mathbf{W} = \mathbf{w}^{+i}] &= \frac{x_0}{x_i}\Pr[\mathbf{X} = \mathbf{x}^{+i}] = \frac{x_0}{x_i}\cdot x_0^{+i}\cdot n(\mathbf{x}^{+i})\Pr[\mathbf{X} \in (\mathbf{x}^{+i})^*] = x_0\cdot n(\mathbf{x})\Pr[\mathbf{X} \in \mathbf{x}^*]\\
&= \Pr[\mathbf{X} = \mathbf{x}] = \Pr[\mathbf{W} = \mathbf{w}]
\end{aligned}
$$

where we have relied on Lemmas 13 and 14 (Appendix I). This proves (19).

We now assume (19) and prove that $\mathbf{X}$ is symmetric. Let $\mathbf{x}$ and $\mathbf{w}$ be defined as above.

$$\Pr[\mathbf{X} \in \mathbf{x}^*] = \sum_{\mathbf{z} \in \mathbf{x}^*}\Pr[\mathbf{X} = \mathbf{z}] = \frac{1}{n(\mathbf{x})}\sum_{i=0}^{q-1}\Pr[\mathbf{X} = \mathbf{x}^{+i}] \tag{31}$$

The last equality is obtained from the fact that $n(\mathbf{z}) = n(\mathbf{x})$ (Lemma 13, Appendix I), and hence each $\mathbf{z} \in \mathbf{x}^*$ is added in $\sum_{i=0}^{q-1}\Pr[\mathbf{X} = \mathbf{x}^{+i}]$ exactly $n(\mathbf{x})$ times. We continue,

$$
\begin{aligned}
\Pr[\mathbf{X} \in \mathbf{x}^*] &= \frac{1}{n(\mathbf{x})}\sum_{i=0}^{q-1}\Pr[\mathbf{W} = \mathbf{w}^{+i}] = \frac{1}{n(\mathbf{x})}\sum_{i=0}^{q-1}e^{-w_i}\Pr[\mathbf{W} = \mathbf{w}]\\
&= \frac{1}{n(\mathbf{x})}\Pr[\mathbf{W} = \mathbf{w}]\left(\sum_{i=0}^{q-1}e^{-w_i}\right) = \frac{1}{n(\mathbf{x})\cdot x_0}\Pr[\mathbf{W} = \mathbf{w}] = \frac{\Pr[\mathbf{X} = \mathbf{x}]}{n(\mathbf{x})\cdot x_0}
\end{aligned}
$$

The equality before last results from (1), recalling that $w_0 = 0$ in all LLR vectors. We thus obtain that $\mathbf{X}$ is symmetric as desired. □

### D. Proof of Theorem 1

Let $i$ be a variable-node associated with the message produced at $n$, defined as in Lemma 15 (Appendix III-A). Let $\tilde{C}$, $\tilde{\mathbf{v}}$ and $\tilde{\mathbf{y}}$ be defined as in the proof of the lemma. Using this notation, we may equivalently denote the message produced at $n$ by $\mathbf{m}(\tilde{\mathbf{y}}, \tilde{\mathbf{v}})$. This is because the message is in fact a function only of the channel observations and coset vector elements contained in the neighborhood graph spanning from $n$. The following corollary follows immediately from the proof of Lemma 15.

*Corollary 1:* Let $\boldsymbol{\sigma}$ be a codeword of $\tilde{C}$. Then for any $\tilde{\mathbf{y}}$ and $\tilde{\mathbf{v}}$ as defined above,

$$\mathbf{m}(\tilde{\mathbf{y}}, \tilde{\mathbf{v}} - \boldsymbol{\sigma}) = \mathbf{m}(\tilde{\mathbf{y}}, \tilde{\mathbf{v}})^{-\sigma_i} \tag{32}$$

where $\sigma_i$ is the value of $\boldsymbol{\sigma}$ at the codeword position corresponding to the variable-node $i$.

We now return to $\mathbf{X}$, a random variable corresponding to the message produced at $n$ and equal to $\mathbf{m}(\tilde{\mathbf{Y}}, \tilde{\mathbf{V}})$. We assume plain-likelihood representation of messages. Let $\mathbf{x}$ be an arbitrary probability vector. Since we assume the all-zero codeword was transmitted, the random space consists of random selection of $\tilde{\mathbf{v}}$ and the random channel transitions. Therefore,

$$\Pr[\mathbf{X} \in \mathbf{x}^*] = \sum_{\tilde{\mathbf{y}}, \tilde{\mathbf{v}}: \mathbf{m}(\tilde{\mathbf{y}}, \tilde{\mathbf{v}}) \in \mathbf{x}^*} \Pr[\tilde{\mathbf{V}} = \tilde{\mathbf{v}}, \tilde{\mathbf{Y}} = \tilde{\mathbf{y}}] \tag{33}$$

Let $\tilde{N}$ be the block length of code $\tilde{C}$ (note that like $\tilde{C}$, $\tilde{N}$ is a function of the neighborhood graph spanning from $n$, which is also a function of the iteration number). The set of all vectors $\tilde{\mathbf{v}} \in \{\mathrm{GF}(q)\}^{\tilde{N}}$ can be presented as a union of nonintersecting cosets of $\tilde{C}$. That is

$$\{\mathrm{GF}(q)\}^{\tilde{N}} = \bigcup_{\mathbf{r} \in \mathcal{R}} \{\mathbf{r} + \tilde{C}\}$$

where $\mathcal{R}$ is a set of coset representatives with respect to $\tilde{C}$. For each vector $\tilde{\mathbf{v}} \in \{\mathrm{GF}(q)\}^{\tilde{N}}$, we let $\mathbf{r} \in \mathcal{R}$ and $\boldsymbol{\sigma} \in \tilde{C}$ denote the unique vectors that satisfy $\tilde{\mathbf{v}} = \mathbf{r} + \boldsymbol{\sigma}$.

Let $\tilde{\mathbf{y}}$ be a channel output portion and $\tilde{\mathbf{v}}$ a coset vector. From Corollary 1, we have that $\mathbf{m}(\tilde{\mathbf{y}}, \tilde{\mathbf{v}}) = \mathbf{m}(\tilde{\mathbf{y}}, \mathbf{r} + \boldsymbol{\sigma}) = \mathbf{m}(\tilde{\mathbf{y}}, \mathbf{r})^{+\sigma_i}$. Therefore, $\mathbf{m}(\tilde{\mathbf{y}}, \tilde{\mathbf{v}}) \in \mathbf{x}^*$ if and only if $\mathbf{m}(\tilde{\mathbf{y}}, \mathbf{r}) \in \mathbf{x}^*$. We can thus rewrite (33) as

$$\Pr[\mathbf{X} \in \mathbf{x}^*] = \sum_{\tilde{\mathbf{y}}, \mathbf{r} \in \mathcal{R}: \mathbf{m}(\tilde{\mathbf{y}}, \mathbf{r}) \in \mathbf{x}^*} \sum_{\tilde{\mathbf{v}} \in \{\mathbf{r} + \tilde{C}\}} \Pr[\tilde{\mathbf{V}} = \tilde{\mathbf{v}}, \tilde{\mathbf{Y}} = \tilde{\mathbf{y}}] = \sum_{\tilde{\mathbf{y}}, \mathbf{r} \in \mathcal{R}: \mathbf{m}(\tilde{\mathbf{y}}, \mathbf{r}) \in \mathbf{x}^*} \Pr[\tilde{\mathbf{V}} \in \{\mathbf{r} + \tilde{C}\}, \tilde{\mathbf{Y}} = \tilde{\mathbf{y}}] \tag{34}$$

Examining $\Pr[\mathbf{X} = \mathbf{x}]$, we have

$$\begin{aligned}
\Pr[\mathbf{X} = \mathbf{x}] &= \Pr[\mathbf{X} = \mathbf{x}, \mathbf{X} \in \mathbf{x}^*] = \sum_{\tilde{\mathbf{y}}, \mathbf{r} \in \mathcal{R}: \mathbf{m}(\tilde{\mathbf{y}}, \mathbf{r}) \in \mathbf{x}^*} \Pr[\mathbf{X} = \mathbf{x}, \tilde{\mathbf{V}} \in \{\mathbf{r} + \tilde{C}\}, \tilde{\mathbf{Y}} = \tilde{\mathbf{y}}] \\
&= \sum_{\tilde{\mathbf{y}}, \mathbf{r} \in \mathcal{R}: \mathbf{m}(\tilde{\mathbf{y}}, \mathbf{r}) \in \mathbf{x}^*} \Pr[\mathbf{X} = \mathbf{x} \mid \tilde{\mathbf{V}} \in \{\mathbf{r} + \tilde{C}\}, \tilde{\mathbf{Y}} = \tilde{\mathbf{y}}] \cdot \Pr[\tilde{\mathbf{V}} \in \{\mathbf{r} + \tilde{C}\}, \tilde{\mathbf{Y}} = \tilde{\mathbf{y}}]
\end{aligned} \tag{35}$$

We now examine $\Pr[\mathbf{X} = \mathbf{x} \mid \tilde{\mathbf{V}} \in \{\mathbf{r} + \tilde{C}\}, \tilde{\mathbf{Y}} = \tilde{\mathbf{y}}]$ for $\tilde{\mathbf{y}}$ and $\mathbf{r}$ such that $\mathbf{m}(\tilde{\mathbf{y}}, \mathbf{r}) \in \mathbf{x}^*$. The random space is confined to the random selection of the coset vector $\tilde{\mathbf{V}}$ from $\{\mathbf{r} + \tilde{C}\}$ or, equivalently, a random selection of $\boldsymbol{\Sigma} \in \tilde{C}$ such that $\tilde{\mathbf{V}} = \mathbf{r} + \boldsymbol{\Sigma}$.

Applying Corollary 1 again, we have for $\tilde{\mathbf{V}} \in \{\mathbf{r} + \tilde{C}\}$ and assuming $\tilde{\mathbf{Y}} = \tilde{\mathbf{y}}$,

$$\mathbf{X} = \mathbf{m}(\tilde{\mathbf{y}}, \tilde{\mathbf{V}}) = \mathbf{m}(\tilde{\mathbf{y}}, \mathbf{r} + \boldsymbol{\Sigma}) = \mathbf{m}(\tilde{\mathbf{y}}, \mathbf{r})^{+\Sigma_i} = \mathbf{z}^{+\Sigma_i} \tag{36}$$

where $\mathbf{z} \triangleq \mathbf{m}(\tilde{\mathbf{y}}, \mathbf{r})$. We assumed $\mathbf{m}(\tilde{\mathbf{y}}, \mathbf{r}) \in \mathbf{x}^*$ and therefore there exists some index $l$ such that $\mathbf{z} = \mathbf{x}^{-l}$ (or equivalently $\mathbf{x} = \mathbf{z}^{+l}$). We first assume, for simplicity, that $n(\mathbf{x}) = 1$. Therefore, $l$ is unique, and no other index $l'$ satisfies $\mathbf{z} = \mathbf{x}^{-l'}$. From (36) we have that $\mathbf{X} = \mathbf{x}$ if and only if $\Sigma_i = l$. Therefore,

$$\begin{aligned}
\Pr[\mathbf{X} = \mathbf{x} \mid \tilde{\mathbf{V}} \in \{\mathbf{r} + \tilde{C}\}, \tilde{\mathbf{Y}} = \tilde{\mathbf{y}}] &= \Pr[\Sigma_i = l \mid \tilde{\mathbf{V}} \in \{\mathbf{r} + \tilde{C}\}, \tilde{\mathbf{Y}} = \tilde{\mathbf{y}}] \\
&= \Pr[\Sigma_i = l \mid \boldsymbol{\Sigma} \in \tilde{C}, \ \delta(\mathbf{r} + \boldsymbol{\Sigma}) \text{ was transmitted and } \tilde{\mathbf{Y}} = \tilde{\mathbf{y}} \text{ was received}]
\end{aligned}$$

Now the key observation in this proof is that under the tree assumption, the above corresponds to $m_l(\tilde{\mathbf{y}}, \mathbf{r}) = z_l$. Therefore

$$\Pr[\mathbf{X} = \mathbf{x} \mid \tilde{\mathbf{V}} \in \{\mathbf{r} + \tilde{C}\}, \tilde{\mathbf{Y}} = \tilde{\mathbf{y}}] = z_l = x_l^{-l} = x_0$$

We now consider the general case of $n(\mathbf{x}) = K$, for arbitrary $K$. In this case there are exactly $K$ indices $l_1, ..., l_K$ satisfying $\mathbf{z} = \mathbf{x}^{-l_k}, k = 1, ..., K$. Using the same arguments as before, we have

$$\Pr[\mathbf{X} = \mathbf{x} \mid \tilde{\mathbf{V}} \in \{\mathbf{r} + \tilde{C}\}, \tilde{\mathbf{Y}} = \tilde{\mathbf{y}}] =$$
$$= \Pr[\Sigma_i \in \{l_1, ..., l_K\} \mid \boldsymbol{\Sigma} \in \tilde{C}, \delta(\mathbf{r} + \boldsymbol{\Sigma}) \text{ was transmitted and } \tilde{\mathbf{Y}} = \tilde{\mathbf{y}} \text{ was received}]$$
$$= \sum_{k=1}^{K} \Pr[\Sigma_i = l_k \mid \boldsymbol{\Sigma} \in \tilde{C}, \delta(\mathbf{r} + \boldsymbol{\Sigma}) \text{ was transmitted and } \tilde{\mathbf{Y}} = \tilde{\mathbf{y}} \text{ was received}]$$
$$= \sum_{k=1}^{K} z_{l_k} = \sum_{k=1}^{K} x_{l_k}^{-l_k} = \sum_{k=1}^{K} x_0 = n(\mathbf{x}) \cdot x_0$$

Recalling (34) and (35), we now have

$$\begin{aligned}
\Pr[\mathbf{X} = \mathbf{x}] &= x_0 \cdot n(\mathbf{x}) \sum_{\tilde{\mathbf{y}}, \mathbf{r} \in \mathcal{R}: \mathbf{m}(\tilde{\mathbf{y}}, \mathbf{r}) \in \mathbf{x}^*} \Pr[\tilde{\mathbf{V}} \in \{\mathbf{r} + \tilde{C}\}, \tilde{\mathbf{Y}} = \tilde{\mathbf{y}}] \\
&= x_0 \cdot n(\mathbf{x}) \Pr[\mathbf{X} \in \mathbf{x}^*]
\end{aligned}$$

This proves (18). □

### E. The Sum of Two Symmetric Variables

The following lemma is used in Section VI-D.

*Lemma 18:* Let $\mathbf{A}$ and $\mathbf{B}$ be two independent LLR-vector random-variables. If $\mathbf{A}$ and $\mathbf{B}$ are symmetric, then $\mathbf{A} + \mathbf{B}$ is symmetric too.

*Proof:* The proof relies on the observation that for all $i \in \mathrm{GF}(q)$ and LLR vectors $\mathbf{a}$ and $\mathbf{b}$, $(\mathbf{a} + \mathbf{b})^{+i} = \mathbf{a}^{+i} + \mathbf{b}^{+i}$. Let $\mathbf{w}$ be an LLR-vector and $i \in \mathrm{GF}(q)$ an arbitrary element.

$$\begin{aligned}
\Pr[\mathbf{A} + \mathbf{B} = \mathbf{w}] &= \sum_{\mathbf{a} + \mathbf{b} = \mathbf{w}} \Pr[\mathbf{A} = \mathbf{a}] \cdot \Pr[\mathbf{B} = \mathbf{b}] \\
&= \sum_{(\mathbf{a} + \mathbf{b})^{+i} = \mathbf{w}^{+i}} e^{a_i} \Pr[\mathbf{A} = \mathbf{a}^{+i}] \cdot e^{b_i} \Pr[\mathbf{B} = \mathbf{b}^{+i}] \\
&= e^{w_i} \sum_{\mathbf{a}^{+i} + \mathbf{b}^{+i} = \mathbf{w}^{+i}} \Pr[\mathbf{A} = \mathbf{a}^{+i}] \cdot \Pr[\mathbf{B} = \mathbf{b}^{+i}] \\
&= e^{w_i} \Pr[\mathbf{A} + \mathbf{B} = \mathbf{w}^{+i}]
\end{aligned}$$

□

### F. Proof of Lemma 3

By definition, component $i$ of $\mathrm{APP}(\mathbf{y})$ satisfies

$$\mathrm{APP}(\mathbf{y})_i = \alpha \Pr[\mathbf{Y} = \mathbf{y} \mid x = i]$$

Where $\alpha$ is some constant, independent of $i$ (but dependent on $\mathbf{y}$), selected such that the sum of the vector components is 1. Using (21), we have

$$\begin{aligned}
\mathrm{APP}(\mathbf{y})_i &= \alpha \cdot y_i \cdot n(\mathbf{y}) \cdot Q(\mathbf{y}^*) \\
&= (\alpha \cdot n(\mathbf{y}) \cdot Q(\mathbf{y}^*)) \cdot y_i
\end{aligned}$$

$\mathbf{y}$, being the output of the equivalent channel, is a probability vector. Thus the sum of all $\mathbf{y}$ components is 1. Hence $\alpha \cdot n(\mathbf{y}) \cdot Q(\mathbf{y}^*) = 1$. We therefore obtain our desired result

$$\text{APP}(\mathbf{y})_i = y_i$$

$\square$

### G. Proof of Lemma 4

Let $\mathbf{Y}$ be a random variable denoting the equivalent channel output, and assume the equivalent channel's input (denoted $x$ in Fig. 4) was zero. $\mathbf{Y}$ thus corresponds to a vector of APP probabilities, computed using the physical channel output $y'$ and the coset vector component $v$. We can therefore invoke Theorem 1 and obtain that for any probability vector $\mathbf{y}$,

$$\Pr[\mathbf{Y} = \mathbf{y} \mid x = 0] = y_0 \cdot n(\mathbf{y}) \cdot \Pr[\mathbf{Y} \in \mathbf{y}^* \mid x = 0]$$

Note that Theorem 1 requires that the entire transmitted codeword be zero and not only the symbol at a particular discrete channel time. However, since the initial message is a function of a single channel output, we can relax this requirement by considering a code that contains a single symbol.

Let $i$ be an arbitrary symbol from the code alphabet. Applying Lemma 17 (Appendix III-A) to the single-symbol code we obtain,

$$
\begin{aligned}
\Pr[\mathbf{Y} = \mathbf{y} \mid x = i] &= \Pr[\mathbf{Y} = \mathbf{y}^{+i} \mid x = 0] \\
&= y_0^{+i} \cdot n(\mathbf{y}^{+i}) \cdot \Pr[\mathbf{Y} \in \mathbf{y}^*] \\
&= y_i \cdot n(\mathbf{y}) \cdot \Pr[\mathbf{Y} \in \mathbf{y}^*]
\end{aligned}
$$

Therefore the equivalent channel is cyclic-symmetric. $\square$

### H. Proof of Lemma 5

Consider the following set of random variables, defined as in Fig. 4. $X$ is the input to the equivalent channel. $V$ is the coset symbol, and $U = X + V$, evaluated over GF($q$). $X' = \delta(U)$ is the physical channel input and $Y'$ is the physical channel output, related to $X'$ through the channel transition probabilities. $\mathbf{Y} \triangleq \text{APP}(Y', V)$ equals the output of the equivalent channel, which is a deterministic function of $Y'$ and $V$.

Since the equivalent channel is symmetric, a choice of $X$ that is uniformly distributed renders $I(X; \mathbf{Y})$ that is equal to the equivalent channel's capacity. This choice of $X$ renders $U$ uniformly distributed as well, and thus $C_\delta = I(U; Y')$. We will now show that $I(U; Y') = I(X; \mathbf{Y})$.

$$
\begin{aligned}
I(U; Y') &= E \log \frac{\Pr[Y' \mid U]}{\Pr[Y']} \\
&= \sum_{u=0}^{q-1} \sum_{y' \in \mathcal{Y}} \Pr[Y' = y', U = u] \log \frac{\Pr[Y' = y' \mid X' = \delta(u)]}{\frac{1}{q} \sum_{u'=0}^{q-1} \Pr[Y' = y' \mid X' = \delta(u')]} \\
&= \sum_{i=0}^{q-1} \sum_{v=0}^{q-1} \sum_{y' \in \mathcal{Y}} \Pr[X = i, V = v, Y' = y'] \log \frac{\Pr[Y' = y' \mid X' = \delta(v+i)]}{\frac{1}{q} \sum_{u'=0}^{q-1} \Pr[Y' = y' \mid X' = \delta(u')]} \\
&= \sum_{i=0}^{q-1} \sum_{v=0}^{q-1} \sum_{y' \in \mathcal{Y}} \Pr[X = i, V = v, Y' = y'] \log(q \cdot Y_i) \\
&= E \log(q \cdot Y_X)
\end{aligned}
\tag{37}
$$

where $\mathcal{Y}$ denotes the physical channel's output alphabet, and $Y_X$ denotes the element of $\mathbf{Y}$ at index number $X$.

$$
\begin{aligned}
I(X; \mathbf{Y}) &= E \log \frac{\Pr[\mathbf{Y} \mid X]}{\Pr[\mathbf{Y}]} \\
&= \sum_{i=0}^{q-1} \sum_{\mathbf{y} \in \mathcal{P}} \Pr[\mathbf{Y} = \mathbf{y}, X = i] \log \frac{\Pr[\mathbf{Y} = \mathbf{y} \mid X = i]}{\frac{1}{q} \sum_{i'=0}^{q-1} \Pr[\mathbf{Y} = \mathbf{y} \mid X = i']}
\end{aligned}
$$

where $\mathcal{P}$ is the set of all probability vectors. Using Lemma 4 and Definition 5 we have, for some probability function $Q(\mathbf{y}^*)$,

$$
I(X; \mathbf{Y}) = \sum_{i=0}^{q-1} \sum_{\mathbf{y} \in \mathcal{P}} \Pr[\mathbf{Y} = \mathbf{y}, X = i] \log \frac{y_i n(\mathbf{y}) Q(\mathbf{y}^*)}{\frac{1}{q} \sum_{i'=0}^{q-1} y_{i'} n(\mathbf{y}) Q(\mathbf{y}^*)}
$$

By definition of $\mathbf{y}$ as a probability vector, we have $\sum_{i'=0}^{q-1} y_{i'} = 1$ and thus,

$$
\begin{aligned}
I(X; \mathbf{Y}) &= \sum_{i=0}^{q-1} \sum_{\mathbf{y} \in \mathcal{P}} \Pr[\mathbf{Y} = \mathbf{y}, X = i] \log \frac{y_i n(\mathbf{y}) Q(\mathbf{y}^*)}{\frac{1}{q} \cdot 1 \cdot n(\mathbf{y}) Q(\mathbf{y}^*)} \\
&= \sum_{i=0}^{q-1} \sum_{\mathbf{y} \in \mathcal{P}} \Pr[\mathbf{Y} = \mathbf{y}, X = i] \log(q \cdot y_i) \\
&= E \log(q \cdot Y_X)
\end{aligned}
\tag{38}
$$

Combining (37) with (38) completes the proof. $\square$

## Appendix IV

## Proofs for Section VI

### A. *Proof of Theorem 3*

We prove the theorem for $\mathbf{R}_t$. $\mathbf{R}_t$ is the message at iteration $t$ averaged over all possibilities of the neighborhood tree $\mathcal{T}_t$.

$$
\begin{aligned}
\Pr[\mathbf{R}_t = \mathbf{x}] &= \sum_{\mathcal{T}_t} \Pr[\mathbf{R}_t = \mathbf{x} \mid \mathcal{T}_t] \cdot \Pr[\mathcal{T}_t] \\
&= \sum_{\mathcal{T}_t} x_0 \cdot n(\mathbf{x}) \Pr[\mathbf{R}_t \in \mathbf{x}^* \mid \mathcal{T}_t] \cdot \Pr[\mathcal{T}_t]
\end{aligned}
$$

The last equation was obtained from Theorem 1.

$$\begin{aligned} \Pr[\mathbf{R}_t = \mathbf{x}] &= x_0 \cdot n(\mathbf{x}) \sum_{\mathcal{T}_t} \Pr[\mathbf{R}_t \in \mathbf{x}^* \mid \mathcal{T}_t] \cdot \Pr[\mathcal{T}_t] \\ &= x_0 \cdot n(\mathbf{x}) \Pr[\mathbf{R}_t \in \mathbf{x}^*] \end{aligned}$$

Hence $\mathbf{R}_t$ is symmetric as desired ($\mathbf{R}^{(0)} = \mathbf{R}_0$ is obtained as a special case). The proof for $\mathbf{L}_t$ is similar. $\square$

### B. Proof of Lemma 8

Let $g' \triangleq j/i$ (evaluated over GF($q$)),

$$\Pr[X_i = x] = \Pr[X_i^{\times g'} = x] = \Pr[X_{i \cdot g'} = x] = \Pr[X_j = x]$$

The proof for $\mathbf{W}$ is identical. $\square$

### C. Proof of Lemma 9

First, we observe that $w_{-k}^{+k} = w_0 - w_k = -w_k$. We now have

$$\begin{aligned} \Pr[W_k = w] &= \sum_{\mathbf{w}:w_k=w} \Pr[\mathbf{W} = \mathbf{w}] = \sum_{\mathbf{w}:w_k=w} e^{w_k} \Pr[\mathbf{W} = \mathbf{w}^{+k}] = e^w \sum_{\mathbf{w}:w_{-k}^{+k}=-w} \Pr[\mathbf{W} = \mathbf{w}^{+k}] \\ &= e^w \sum_{\mathbf{w}:w_{-k}=-w} \Pr[\mathbf{W} = \mathbf{w}] = e^w \Pr[W_{-k} = -w] = e^w \Pr[W_k = -w] \end{aligned}$$

The last result having been obtained from Lemma 8. $\square$

### D. Proof of Lemma 10

We prove the lemma for the probability-vector representation. The proof for LLR-vector representation is identical. We first assume $\mathbf{X} = \tilde{\mathbf{T}}$ and show that $\mathbf{X}$ is permutation-invariant. Let $g \in \text{GF}(q) \backslash \{0\}$ be randomly selected as in Definition 8, such that $\mathbf{X} = \mathbf{T}^{\times g}$. Let $g' \in \text{GF}(q) \backslash \{0\}$ be arbitrary such that $\mathbf{\Xi} \triangleq \mathbf{X}^{\times g'}$.

$$\mathbf{\Xi} = (\mathbf{T}^{\times g})^{\times g'} = \mathbf{T}^{\times (g \cdot g')} \tag{39}$$

$g \cdot g'$ is a random variable, independent of $\mathbf{T}$ that is distributed identically with $g$. Thus, $\mathbf{\Xi}$ is identically distributed with $\mathbf{T}^{\times g} = \tilde{\mathbf{T}} = \mathbf{X}$. Since $g'$ was arbitrary, we obtain that $\mathbf{X}$ is permutation-invariant.

We now assume that $\mathbf{X}$ is permutation-invariant. Consider $\mathbf{T} \triangleq \mathbf{X}^{\times g^{-1}}$, where $g$ is uniformly random in $\text{GF}(q) \backslash \{0\}$ and independent of $\mathbf{X}$. Equivalently, $\mathbf{X} = \mathbf{T}^{\times g}$. We now show that $\mathbf{T}$ is independent of $g$,

$$\Pr[\mathbf{T} = \mathbf{t} \mid g] = \Pr[\mathbf{X}^{\times g^{-1}} = \mathbf{t} \mid g] = \Pr[\mathbf{X} = \mathbf{t}]$$

the last result having been obtained by the definition of $\mathbf{X}$ as permutation-invariant. Since the above is true for all $g$, $\mathbf{T}$ is independent of $g$. Thus, $\mathbf{X} = \tilde{\mathbf{T}}$ as desired. $\square$

*E. Some Lemmas Involving Permutation-Invariance*

We now present some lemmas that are used in Appendices IV-F, VI and V and in Section VI-D. The first three lemmas apply to both the probability-vector and LLR representations of vectors.

*Lemma 19:* If $\tilde{\mathbf{X}}$ is a random-permutation of $\mathbf{X}$, then $P_e(\tilde{\mathbf{X}}) = P_e(\mathbf{X})$.

The proof of this lemma is obtained from the fact that the operation $\times g$, for all $g$, leaves element $X_0$ unchanged.

*Lemma 20:* If $\mathbf{X}$ is a symmetric random variable, and $\tilde{\mathbf{X}}$ is a random-permutation of $\mathbf{X}$, then $\tilde{\mathbf{X}}$ is also symmetric.

*Proof:*

$$\Pr[\tilde{\mathbf{X}} = \mathbf{x} \mid \tilde{\mathbf{X}} \in \mathbf{x}^*] \; = \sum_{g \in \mathrm{GF}(q) \backslash \{0\}} \Pr[\tilde{\mathbf{X}} = \mathbf{x} \mid \tilde{\mathbf{X}} \in \mathbf{x}^*, g] \Pr[g \mid \tilde{\mathbf{X}} \in \mathbf{x}^*] \tag{40}$$

In the following derivation, we make use of the fact that $n(\mathbf{x}^{\times g}) = n(\mathbf{x})$ (see Lemma 13, Appendix I) and $(\mathbf{x}^*)^{\times g} = (\mathbf{x}^{\times g})^*$ (see Lemma 14, Appendix I).

$$\begin{aligned}
\Pr[\tilde{\mathbf{X}} = \mathbf{x} \mid \tilde{\mathbf{X}} \in \mathbf{x}^*, g] &= \Pr[\mathbf{X}^{\times g} = \mathbf{x} \mid \mathbf{X}^{\times g} \in \mathbf{x}^*] = \Pr[\mathbf{X} = \mathbf{x}^{\times g^{-1}} \mid \mathbf{X} \in (\mathbf{x}^*)^{\times g^{-1}}] \\
&= \Pr[\mathbf{X} = \mathbf{x}^{\times g^{-1}} \mid \mathbf{X} \in (\mathbf{x}^{\times g^{-1}})^*] = x_0^{\times g^{-1}} \cdot n(\mathbf{x}^{\times g^{-1}}) = x_0 \cdot n(\mathbf{x})
\end{aligned} \tag{41}$$

Combining (40) and (41) we obtain

$$\Pr[\tilde{\mathbf{X}} = \mathbf{x} \mid \tilde{\mathbf{X}} \in \mathbf{x}^*] = x_0 \cdot n(\mathbf{x})$$

and thus conclude the proof. $\qquad\blacksquare$

*Lemma 21:* If $\mathbf{X}$ is permutation-invariant and $\tilde{\mathbf{X}}$ is a random-permutation of $\mathbf{X}$, then $\tilde{\mathbf{X}}$ and $\mathbf{X}$ are identically distributed.

The proof of this lemma is straightforward from Definitions 7 and 8.

The following lemmas discuss permutation-invariance in the context of the LLR representation of random-variables.

*Lemma 22:* Let $\mathbf{A}$ and $\mathbf{B}$ be two independent, permutation-invariant LLR-vector random-variables. Then $\mathbf{W} = \mathbf{A} + \mathbf{B}$ is also permutation-invariant.

*Proof:* Let $g \in \mathrm{GF}(q) \backslash \{0\}$ and $\mathbf{\Omega} = \mathbf{W}^{\times g}$. Let $\mathbf{w}$ be an arbitrary LLR-vector.

$$\begin{aligned}
\Pr[\mathbf{\Omega} = \mathbf{w}] &= \Pr[(\mathbf{A} + \mathbf{B})^{\times g} = \mathbf{w}] = \Pr[\mathbf{A}^{\times g} + \mathbf{B}^{\times g} = \mathbf{w}] = \sum_{\mathbf{a} + \mathbf{b} = \mathbf{w}} \Pr[\mathbf{A}^{\times g} = \mathbf{a}] \cdot \Pr[\mathbf{B}^{\times g} = \mathbf{b}] \\
&= \sum_{\mathbf{a} + \mathbf{b} = \mathbf{w}} \Pr[\mathbf{A} = \mathbf{a}] \cdot \Pr[\mathbf{B} = \mathbf{b}] = \Pr[\mathbf{A} + \mathbf{B} = \mathbf{w}] = \Pr[\mathbf{W} = \mathbf{w}]
\end{aligned}$$

Since $g$ and $\mathbf{w}$ are arbitrary, this implies that $\mathbf{W}$ is permutation-invariant, as desired. $\qquad\blacksquare$

*Lemma 23:* Let $\mathbf{A}$ and $\mathbf{B}$ be two LLR-vector random variables. Let $g$, $h$ and $k$ be independent random variables, uniformly distributed in $\mathrm{GF}(q) \backslash \{0\}$ and independent of $\mathbf{A}$ and $\mathbf{B}$. Let $\tilde{\mathbf{A}} = \mathbf{A}^{\times g}$, $\mathbf{W} = \tilde{\mathbf{A}} + \mathbf{B}$ and $\tilde{\mathbf{W}} = \mathbf{W}^{\times h}$, $\tilde{\mathbf{B}} = \mathbf{B}^{\times k}$, $\mathbf{\Omega} = \tilde{\mathbf{A}} + \tilde{\mathbf{B}}$, $\tilde{\mathbf{\Omega}} = \mathbf{\Omega}^{\times h}$. Then $\tilde{\mathbf{W}}$, $\mathbf{\Omega}$ and $\tilde{\mathbf{\Omega}}$ are identically distributed.

*Proof:* We begin with the following equalities,

$$\tilde{\mathbf{W}} = (\mathbf{A}^{\times g} + \mathbf{B})^{\times h} = \mathbf{A}^{\times g \cdot h} + \mathbf{B}^{\times h}, \quad \mathbf{\Omega} = \mathbf{A}^{\times g} + \mathbf{B}^{\times k}, \quad \tilde{\mathbf{\Omega}} = (\mathbf{A}^{\times g} + \mathbf{B}^{\times k})^{\times h} = \mathbf{A}^{\times g \cdot h} + \mathbf{B}^{\times k \cdot h}$$

Consider the expressions for $\tilde{\mathbf{W}}$ and $\boldsymbol{\Omega}$. $g \cdot h$ is identically distributed with $g$, and $h$ is identically distributed with $k$. $g \cdot h$ is independent of $h$, and both are independent of $\mathbf{A}$ and $\mathbf{B}$. The same holds if we replace $g \cdot h$ and $h$ with $g$ and $k$. Thus $\tilde{\mathbf{W}}$ and $\boldsymbol{\Omega}$ are identically distributed. The proof for $\tilde{\boldsymbol{\Omega}}$ is similar. $\qquad\square$

### F. Proof of Theorem 4

$\mathbf{L}_t$ is permutation-invariant following the discussion at the beginning of Section VI-B, and thus Part 1 of the theorem is proved.

$\bar{\mathbf{R}}_t \triangleq (\mathbf{R}_t)^{\times g^{-1}}$ where the label $g$ is randomly selected, uniformly from $\mathrm{GF}(q)\backslash\{0\}$. Thus $\bar{\mathbf{R}}_t$ is a random-permutation of $\mathbf{R}_t$, and by Lemma 10 it is permutation-invariant. $\bar{\mathbf{R}}_t$ is symmetric by Lemma 20 (Appendix IV-E), and $P_e(\bar{\mathbf{R}}_t) = P_e(\mathbf{R}_t)$ by Lemma 19 (Appendix IV-E). This proves part 2 of the theorem.

$\bar{\mathbf{R}}^{(0)}$ is permutation-invariant by its construction. $\bar{\mathbf{R}}_t$ is a random-permutation of $\mathbf{R}_t$. Switching to LLR representation, $\mathbf{R}'_t$ is obtained by applying expression (15). The leftbound messages are permutation-invariant, hence, by Lemma 22 (Appendix IV-E) the sum $\sum_{k=1}^{d_i-1} \mathbf{L}'^{(k)}_t$ is also permutation-invariant. Using Lemma 23 (Appendix IV-E), the distribution of $\bar{\mathbf{R}}'_t$ may equivalently be computed by replacing the instantiation $\mathbf{r}'^{(0)}$ of $\mathbf{R}'^{(0)}$ in (15) with an instantiation of $\bar{\mathbf{R}}'^{(0)}$.

The distribution of $\mathbf{L}_t$ is computed in density evolution recursively from $\bar{\mathbf{R}}_t$, using (10). Thus, the above discussion implies that replacing $\mathbf{R}^{(0)}$ with $\bar{\mathbf{R}}^{(0)}$ would not affect this density either. The remainder of Part 3 of the theorem is obtained from Lemmas 20 and 19. $\square$

### G. Non-Degeneracy of Channels and Mappings

A mapping is non-degenerate if there exists no integer $n > 1$ such that for all $a \in \mathcal{A}$, the number of elements satisfying $\delta(x) = a$ is a multiple of $n$. With quantization-mapping, such a mapping could be replaced by a simpler quantization over an alphabet of size $q/n$ that would equally attain the desired input distribution $Q(a)$. With nonuniform-spaced mapping, the number of elements mapped to each $a \in \mathcal{A}$ is 1 and thus this requirement is typically observed.

A channel is non-degenerate if there exist no values $a_1, a_2 \in \mathcal{A}$ such that $\Pr[y|a_1] = \Pr[y|a_2]$ for all $y$ belonging to the channel output alphabet.

The proof of $\Delta < 1$ when both the mapping and the channel are non-degenerate ($\Delta$ having been defined in (24)) follows in direct lines as the one provided for $D_{\mathbf{x}} < 1$ in [1][Appendix I.A].

### APPENDIX V
### PROOF OF PART 1 OF THEOREM 5

In this section, we prove the necessity condition of Theorem 5. Our proof is a generalization of the proof provided by Richardson *et al.* [29]. An outline of the proof was provided in Section VI-C.

## A. The Erasurized Channel

We begin by defining the erasurized channel for a given cyclic-symmetric channel and examining its properties. Our development in this subsection is general, and will be put into the context of the proof in the following subsection.

*Definition 9:* Let $\Pr[\mathbf{y} \mid x]$ denote the transition probabilities of a cyclic-symmetric channel (see Definition 5). Then the corresponding *erasurized* channel is defined by the following:

The input alphabet is $\{0, ..., q-1\}$. The output alphabet is $\hat{\mathcal{Y}} \triangleq \mathcal{Y} \bigcup \{0, ..., q-1\}$ where $\mathcal{Y}$ is the output alphabet of the original (cyclic-symmetric) channel. The transition probabilities $\hat{\Pr}[y \mid x]$ are defined as follows:

For all probability vectors $\mathbf{y} \in \mathcal{Y}$,

$$\hat{\Pr}[\mathbf{y} \mid x = i] = \begin{cases} \Pr[\mathbf{y} \mid x = i] & y_i < \max(y_0, ..., y_{q-1}) \\ y_{\text{scnd}} n(\mathbf{y}) Q(\mathbf{y}^*) & y_i = \max(y_0, ..., y_{q-1}) \end{cases} \tag{42}$$

where

- $Q(\mathbf{y}^*)$ is defined as in Definition 5.
- $y_{\text{scnd}}$ is obtained by ordering the elements of the sequence $(y_0, ..., y_{q-1})$ in descending order and selecting the second largest. This means that if the maximum of the sequence elements is obtained more than once, then $y_{\text{scnd}}$ would be equal to this maximum.

For output alphabet elements $j \in \{0, ..., q-1\}$ we define

$$\hat{\Pr}[j \mid x = i] = \begin{cases} 0 & j \neq i \\ 1 - \hat{\epsilon} & j = i \end{cases} \tag{43}$$

where $\hat{\epsilon}$ is defined

$$\hat{\epsilon} = \sum_{\mathbf{y} \in \mathcal{Y}} \hat{\Pr}[\mathbf{y} \mid x = 0]$$

The following lemma discusses the erasurized channel:

*Lemma 24:* The erasurized channel satisfies the following properties

1) The transition probability function is valid.
2) The original cyclic-symmetric channel can be represented as a degraded version of the erasurized channel. That is, it can be represented as a concatenation of the erasurized channel with another channel, whose input would be the erasurized channel's output.

**Proof:**

1) It is easy to verify that $\hat{\epsilon} \leq 1$, and hence $\hat{\Pr}[y \mid x = i] \geq 0$ for all $i$ by definition. The rest of the proof follows from the observation that for all vectors $\mathbf{y} \in \mathcal{Y}$ (recall that $\mathcal{Y} \subset \hat{\mathcal{Y}}$) $\hat{\Pr}[\mathbf{y} \mid x = i] = \hat{\Pr}[\mathbf{y}^{+i} \mid x = 0]$.

$$\begin{aligned} \sum_{\hat{y} \in \hat{\mathcal{Y}}} \hat{\Pr}[\hat{y} \mid x = i] &= \sum_{\mathbf{y} \in \mathcal{Y}} \hat{\Pr}[\mathbf{y} \mid x = i] + \hat{\Pr}[i \mid x = i] = \sum_{\mathbf{y} \in \mathcal{Y}} \hat{\Pr}[\mathbf{y}^{+i} \mid x = 0] + 1 - \hat{\epsilon} \\ &= \sum_{\mathbf{y} \in \mathcal{Y}} \hat{\Pr}[\mathbf{y} \mid x = 0] + 1 - \hat{\epsilon} = 1 \end{aligned}$$

2) We define a transition probability function $q(\mathbf{y} \mid \hat{y})$ where $\hat{y} \in \hat{\mathcal{Y}}$ and $\mathbf{y} \in \mathcal{Y}$.

$$q(\mathbf{y} \mid \hat{y}) = \begin{cases} 1 & \hat{y} = \mathbf{y} \\ 1/(1 - \hat{\epsilon}) \cdot (\Pr[\mathbf{y} \mid x = j] - \hat{\Pr}[\mathbf{y} \mid x = j]) & \hat{y} = j \in \{0, ..., q-1\} \\ 0 & \text{otherwise} \end{cases}$$

It is easy to verify that the concatenation of the erasurized channel with $q(\cdot \mid \cdot)$ produces the transition probabilities $\Pr[\mathbf{y} \mid x]$ of the original cyclic-symmetric channel.

$\square$

The erasurized channel is no longer cyclic-symmetric. Hence, if we apply a belief-propagation decoder on the outputs of an erasurized channel, Lemma 3 does not apply, and the initial messages are not identical to the channel outputs. However, the following lemma summarizes some important properties of the initial message distribution, under the all-zero codeword assumption.

*Lemma 25:* Let $Q(\mathbf{z})$ denote the message distribution at the initial iteration of belief propagation decoding over an erasurized channel (under the assumption that the zero symbol was transmitted). Then $Q(\mathbf{z})$ can be written as

$$Q(\mathbf{z}) = \hat{\epsilon} P_E(\mathbf{z}) + (1 - \hat{\epsilon}) \Delta_{[1,0,...,0]} \tag{44}$$

where $P_E(\mathbf{z})$ is a probability function that satisfies

$$P_E[\exists i > 0 : z_i \geq z_0] = 1 \tag{45}$$

and $\Delta_{[1,0,...,0]}$ is a distribution that takes the vector $[1, 0, ..., 0]$ (i.e., the vector $\mathbf{y}$ where $y_0 = 1$ and $y_i = 0 \quad \forall i \neq 0$) with probability 1 ($\Delta_{[1,0,...,0]}$ must not be confused with $\Delta$ defined by (24)).

**Proof:** For any probability vector $\mathbf{z}$, we define $P_E(\mathbf{z}) = \Pr(\mathbf{z} \mid \text{the channel output was } \mathbf{y} \in \mathcal{Y})$, and $P_2(\mathbf{z}) = \Pr(\mathbf{z} \mid \text{the channel output was } j \in \{0, ..., q-1\})$. We now have

$$Q(\mathbf{z}) = \hat{\epsilon} P_E(\mathbf{z}) + (1 - \hat{\epsilon}) P_2(\mathbf{z}) \tag{46}$$

We first examine $P_E(\mathbf{z})$. Let $\mathbf{y} \in \mathcal{Y}$ denote the channel output. By definition we have

$$z_i = \alpha \cdot \hat{\Pr}[\mathbf{y} \mid x = i] \tag{47}$$

where $\alpha$ is a normalization constant, dependent of $\mathbf{y}$ but not on $i$, selected so that the sum of the vector elements $(z_0, ..., z_{q-1})$ is 1. We now examine all possibilities for $\mathbf{y}$.

First assume that the maximum of $\{y_0, ..., y_{q-1}\}$ is obtained at $y_0$ and at $y_0$ only. Let $i_{\text{scnd}} \neq 0$ be an index where the second-largest element of $\{y_0, ..., y_{q-1}\}$ is obtained. Then by (47) and (42),

$$z_0 = \alpha \cdot \hat{\Pr}[\mathbf{y} \mid x = 0] = \alpha \cdot y_{i_{\text{scnd}}} n(\mathbf{y}) Q(\mathbf{y}^*) = \alpha \cdot \hat{\Pr}[\mathbf{y} \mid x = i_{\text{scnd}}] = z_{i_{\text{scnd}}}$$

Now assume that the maximum is obtained at $y_0$ and also at $y_{i_{\max}}$ where $i_{\max} \neq 0$. Then it is easy to observe that $z_0 = z_{i_{\max}}$. Finally, assume that the maximum of $\{y_0, ..., y_{q-1}\}$ is *not* obtained at $y_0$. Let $i_{\max}$ be an index such that $y_{i_{\max}}$ obtains the maximum. Then

$$z_0 = \alpha \cdot \hat{\Pr}[\mathbf{y} \mid x = 0] = \alpha \cdot y_0 \cdot n(\mathbf{y}) Q(\mathbf{y}^*) \leq \alpha \cdot y_{\text{scnd}} n(\mathbf{y}) Q(\mathbf{y}^*) = \alpha \cdot \hat{\Pr}[\mathbf{y} \mid x = i_{\max}] = z_{i_{\max}}$$

In all cases, there exists an index $i \neq 0$ such that $z_i \geq z_0$, as required (45).

We now examine $P_2(\mathbf{z})$. Assuming the symbol 0 was transmitted, then by (43), the probability of obtaining any output symbol of the set $j \in \{0, ..., q-1\}$ other than $j = 0$ is zero. Also, the only input symbol capable of producing the output $j = 0$ with probability greater than zero is the input $i = 0$. Hence the decoder produces the initial message $[1, 0, ..., 0]$ with probability 1, and $P_2(\mathbf{z}) = \Delta_{[1,0,...,0]}$ as required. $\square$

Consider transmission over the original, cyclic-symmetric channel. Let $P_e$ be the uncoded MAP probability of error. Let $\hat{P}_e$ be the corresponding probability over the erasurized channel.

In the erasure decomposition lemma of [29], similarly defined $P_e$ and $\hat{P}_e$ are both equal to $1/2 \cdot \epsilon$, where $\epsilon$ is the erasure channel's erasure probability. In the following lemma we examine $\hat{\epsilon}$ of the erasurized channel.

*Lemma 26:* The following inequalities hold:

$$\frac{1}{2}\hat{\epsilon} \leq \hat{P}_e \leq P_e \leq \hat{\epsilon}$$

*Proof:*

1) The erasurized channel is symmetric (although not cyclic-symmetric): for all $\mathbf{y} \in \mathcal{Y}$ we have $\hat{\Pr}[\mathbf{y} \mid x = i] = \hat{\Pr}[\mathbf{y}^{+i} \mid x = 0]$, and for all $j \in \{0, ..., q-1\}$ we have $\hat{\Pr}[j \mid x = i] = \hat{\Pr}[j - i \mid x = 0]$. Hence, the decoding error is independent of the transmitted symbol, and we may assume that the symbol was 0.

Consider the erasurized channel output $\hat{Y}$. The MAP decoder decides on the symbol with the maximum APP value. If more than one such symbol exists, a random decision among the maximizing symbols is made. Let $\mathbf{Z}$ denote the vector of APP values corresponding to $\hat{Y}$. By Lemma 25, we have that with probability $\hat{\epsilon}$, $\mathbf{Z}$ is distributed as $P_E(\mathbf{z})$. Recalling (45), we have that for messages distributed as $P_E(\mathbf{z})$, an error is made with probability at least $1/2$. Therefore, $\hat{P}_e \geq (1/2)\hat{\epsilon}$.

2) By Lemma 24, the cyclic-symmetric channel is a degraded version of the erasurized channel. Hence $P_e \geq \hat{P}_e$.

3) We now prove $P_e \leq \hat{\epsilon}$. Let us assume once more that the symbol 0 was transmitted. Recall that we are now examining the decoder's performance over the cyclic-symmetric channel (and not the erasurized channel). Therefore, by Lemma 3, the vector of APP values (according to which the MAP decision is made) is identical to the channel output. Let $P_e(\mathbf{y})$ be defined as in Definition 6. We will now show that the following inequality holds,

$$\hat{\Pr}[\mathbf{y} \mid x = 0] \geq P_e(\mathbf{y}) \cdot \Pr[\mathbf{y} \mid x = 0] \tag{48}$$

- If $\mathbf{y}$ is such that the maximum of $\{y_0, ..., y_{q-1}\}$ is obtained only at $y_0$ we have from (42) that $\hat{\Pr}[\mathbf{y} \mid x = 0] < \Pr[\mathbf{y} \mid x = 0]$. However, in this case the decoder correctly decides 0. Hence $P_e(\mathbf{y}) = 0$ and (48) is satisfied.

- In any other case, we have $\hat{\Pr}[\mathbf{y} \mid x = 0] = \Pr[\mathbf{y} \mid x = 0]$. Using $P_e(\mathbf{y}) \leq 1$ we obtain (48) trivially.

We now have

$$P_e = \sum_{\mathbf{y} \in \mathcal{Y}} P_e(\mathbf{y}) \cdot \Pr[\mathbf{y} \mid x = 0] \leq \sum_{\mathbf{y} \in \mathcal{Y}} \hat{\Pr}[\mathbf{y} \mid x = 0] = \hat{\epsilon}$$

$\square$

## B. The Remainder of the Proof

To complete the proof, we would like to show that the probability of error at iteration $t$ cannot be too small. Let $\mathbf{R}_{t+n}$, denote the rightbound messages at iteration $t+n$, where $n = 0, 1, \dots$. By Lemma 4 (in a manner similar to [29]), $\mathbf{R}_t$ may equivalently be obtained as the initial message of a cyclic-symmetric channel. We now replace this channel with the corresponding erasurized channel, and obtain a lower bound on the probability of error at subsequent iterations. We let $\hat{\mathbf{R}}_{t+n}$, $n = 0, 1, \dots$, denote the respective messages following the replacement.

In the remainder of the proof, we switch to log-likelihood representation of messages. We let $\hat{\mathbf{R}}'_{t+n}$ denote the LLR-vector representation of $\hat{\mathbf{R}}_{t+n}$, $n = 1, \dots$. Adopting the notation of [29], we let $Q_n(\mathbf{w})$ denote the distribution of $\hat{\mathbf{R}}'_{t+n}$. $P_0$ denotes the distribution of the initial message $\mathbf{R}'^{(\mathbf{0})}$ of the true cyclic-symmetric channel.

Using LLR messages, Lemma 25 becomes

$$Q_0(\mathbf{w}) = \hat{\epsilon} P_E(\mathbf{w}) + (1 - \hat{\epsilon}) \Delta_{[\infty, \dots, \infty]}$$

$P_E(\mathbf{w})$ now satisfies

$$P_E[\exists i > 0 : w_i \leq 0] = 1 \tag{49}$$

After $n$ iterations of density evolution, the density becomes (in a manner similar to the equivalent binary case [29])

$$Q_n = \hat{\epsilon}(\lambda'(0)\rho'(1))^n \overline{P}_E \otimes \overline{P}_0^{\otimes(n-1)} \otimes P_0 + (1 - \hat{\epsilon}(\lambda'(0)\rho'(1))^n) \Delta_{[\infty, \dots, \infty]} + O(\hat{\epsilon}^2)$$

where $P_0$ is defined in Theorem 5. $\overline{P}_0$ and $\overline{P}_E$ correspond to the random-permutations of $P_0$ and $P_E$ (resulting from the effect of randomly selected labels), respectively and $\otimes$ denotes convolution. Let $\overline{Q}_n$ denote the distribution of $(\hat{\mathbf{R}}'_i)^{\times g}$, where $g$ is the random label on the edge along which $\hat{\mathbf{R}}'_i$ is sent. Then

$$\overline{Q}_n = \hat{\epsilon}(\lambda'(0)\rho'(1))^n \overline{P}_E \otimes \overline{P}_0^{\otimes n} + (1 - \hat{\epsilon}(\lambda'(0)\rho'(1))^n) \Delta_{[\infty, \dots, \infty]} + O(\hat{\epsilon}^2)$$

where we have used Lemma 23 (Appendix IV-E) to obtain that a random-permutation of $\overline{P}_E \otimes \overline{P}_0^{\otimes(n-1)} \otimes P_0$ is distributed as $\overline{P}_E \otimes \overline{P}_0^{\otimes n}$. Using Lemma 19 (Appendix IV-E), the probability of error (assuming the zero symbol was selected) is the same for $\overline{Q}_n$ and $Q_n$. Letting $P_e(Q_n)$ denote this probability of error, we have

$$P_e(Q_n) = \hat{\epsilon}(\lambda'(0)\rho'(1))^n P_e(\overline{P}_E \otimes \overline{P}_0^{\otimes n}) + O(\hat{\epsilon}^2)$$

Defining the probability function $T = \overline{P}_E \otimes \overline{P}_0^{\otimes n}$, we have

$$
\begin{aligned}
P_e(Q_n) &\geq \hat{\epsilon}(\lambda'(0)\rho'(1))^n \frac{1}{2} T[\exists i > 0 : W_i \leq 0] + O(\hat{\epsilon}^2) \\
&\geq \frac{1}{2}\hat{\epsilon}(\lambda'(0)\rho'(1))^n T[W_1 \leq 0] + O(\hat{\epsilon}^2) \\
&\geq \frac{1}{2}\hat{\epsilon}(\lambda'(0)\rho'(1))^n \left(\overline{P}_E[W_1 \leq 0] \cdot \overline{P}_0^{\otimes n}[W_1 \leq 0]\right) + O(\hat{\epsilon}^2)
\end{aligned}
\tag{50}
$$

Recalling (49), $P_E$ satisfies that with probability 1 there exists at least one index $i \neq 0$ such that $W_i \leq 0$. A random-permutation would transfer $W_i$ to index 1 with probability $1/(q-1)$. Hence

$$\overline{P}_E[W_1 \leq 0] \geq \frac{1}{q-1} \tag{51}$$

Let $\overline{P}_0^{(1)}$ denote the marginal distribution of the $\bar{R}_1^{\prime\,(0)}$ element of $\bar{\mathbf{R}}^{\prime\,(0)}$. By Lemma 9, $\overline{P}_0^{(1)}$ is symmetrically distributed in the binary sense. Following the development of [29] (similarly relying on results from [32][page 14]), we obtain

$$\lim_{n\to\infty} \frac{1}{n} \log \overline{P}_0^{\otimes n}[W_1 \leq 0] = \log E \exp(-\frac{1}{2}\bar{R}_1^{\prime\,(0)}) \tag{52}$$

For the above limit to be valid, we first need (see [32]) that $E \exp(s \cdot \bar{R}_1^{\prime\,(0)}) < \infty$ in some neighborhood of zero, as appears in the conditions of the theorem. We also need to show that $E\bar{R}_1^{\prime\,(0)} > 0$ (also see [32]). This will be proven shortly. We first examine $E \exp(-\frac{1}{2}\bar{R}_1^{\prime\,(0)})$.

*Lemma 27:*

$$E \exp(-\frac{1}{2}\bar{R}_1^{\prime\,(0)}) = \Delta \tag{53}$$

*Proof:* Recalling that $\bar{\mathbf{R}}^{\prime\,(0)}$ is a random-permutation of the initial message, we first observe

$$E \exp(-\frac{1}{2}\bar{R}_1^{\prime\,(0)}) = E_g\left\{ E\left[ \exp(-\frac{1}{2}R_1^{\prime\,(0)\times g}) \mid g \right] \right\} = \frac{1}{q-1} \sum_{k=1}^{q-1} E \exp(-\frac{1}{2}R_k^{\prime\,(0)}) \tag{54}$$

We now examine $E \exp(-\frac{1}{2}R_k^{\prime\,(0)})$. Recalling (14), where $Y$ denotes the random channel output and $V$ denotes the random coset symbol,

$$
\begin{aligned}
E \exp(-\frac{1}{2}R_k^{\prime\,(0)}) &= E\sqrt{\frac{\Pr[Y \mid \delta(k+V)]}{\Pr[Y \mid \delta(0+V)]}} \\
&= \sum_{v,y} \sqrt{\frac{\Pr[y \mid \delta(k+v)]}{\Pr[y \mid \delta(0+v)]}} \Pr[y \mid \delta(0+v)] \cdot \Pr[V=v] \\
&= \frac{1}{q} \sum_{v=0}^{q-1} \sum_y \sqrt{\Pr[y \mid \delta(k+v)] \Pr[y \mid \delta(v)]}
\end{aligned}
\tag{55}
$$

Combining (54), (55) and the definition (24) we obtain (53). □

We are now ready to show $E\bar{R}_1^{\prime\,(0)} > 0$. Recall from the discussion in Section VI-C that $\Delta < 1$. Using (53) and the Jensen inequality, we obtain

$$-\frac{1}{2}E\bar{R}_1^{\prime\,(0)} = \log \exp(-\frac{1}{2}E\bar{R}_1^{\prime\,(0)}) \leq \log E \exp(-\frac{1}{2}\bar{R}_1^{\prime\,(0)}) = \log \Delta < 0$$

We now proceed with the proof. By (53), (52) becomes

$$\lim_{n\to\infty} \frac{1}{n} \log \overline{P}_0^{\otimes n}[W_1 \leq 0] = \log \Delta \tag{56}$$

The remainder of the proof follows in direct lines as in [29] and is provided primarily for completeness. Combining (50) with (51) and (56) we obtain that for arbitrary $\eta > 0$ and large enough $n$,

$$P_e(Q_n) \geq \frac{1}{2(q-1)}\hat{\epsilon}(\lambda'(0)\rho'(1) \cdot (\Delta - \eta))^n + O(\hat{\epsilon}^2)$$

If $\lambda'(0)\rho'(1) > 1/\Delta$, by appropriately selecting $\eta$ we obtain that for $n$ large enough

$$P_e(Q_n) \geq 2\hat{\epsilon} + O(\hat{\epsilon}^2) \tag{57}$$

$O(\cdot)$ denotes a function, dependent on $\lambda$, $\rho$ and $n$ such that $|O(x)| < cx$ for some constant $c$. Hence there exists a constant $\hat{\epsilon}(\lambda, \rho, n)$ such that if $\hat{\epsilon} < \hat{\epsilon}(\lambda, \rho, n)$, then

$$P_e(Q_n) > \hat{\epsilon} \tag{58}$$

We now return to examine $P_e^t$ and $P_e^{t+n}$, the probabilities of error over the true channel, prior to the replacement of messages with those of an erasurized channel. Since the true channel is degraded in relation to the erasurized channel, we must have for $\hat{\epsilon} < \hat{\epsilon}(\lambda, \rho, n)$, $P_e^{t+n} \geq P_e(Q_n)$.

By Lemma 26, $\hat{\epsilon} \leq 2P_e^t$. Hence there exists $\xi(\rho, \lambda, P_0)$ such that if $P_e^t \leq \xi$, then $\hat{\epsilon} < \hat{\epsilon}(\rho, \lambda, n)$ and hence (58) is satisfied. However, Lemma 26 also asserts $P_e^t \leq \hat{\epsilon}$. Hence $P_e(Q_n) > P_e^t$ and consequently $P_e^{t+n} > P_e^t$. This contradicts Theorem 2. Thus we obtain our desired result of $P_e^t > \xi(\rho, \lambda, P_0)$ for all $t$. $\square$

## Appendix VI
## Proof of Part 2 of Theorem 5

In this section, we prove the sufficiency condition of Theorem 5. Our proof is a generalization of the proof provided by Khandekar [20] from binary to coset GF($q$) LDPC. An outline of the proof was provided in Section VI-C.

Note that throughout the proof we denote by $O(\cdot)$ functions for whom there exists a constant $c > 0$, not dependent on the iteration number $t$, such that $|O(x)| < c \cdot x$.

We are interested in $P_e(\mathbf{R}_t)$ (defined as in (22)) where $\mathbf{R}_t$ is the rightbound message as defined in Section VI-A. We begin, however, by analyzing a differently defined $D(\mathbf{R}_t)$.

Let $\mathbf{X}$ be a probability-vector random variable. The operator $D(\mathbf{X})$ is defined as follows:

$$D(\mathbf{X}) \triangleq E\sqrt{\frac{\tilde{X}_1}{\tilde{X}_0}} = \frac{1}{q-1}\sum_{i=1}^{q-1} E\sqrt{\frac{X_i}{X_0}} \tag{59}$$

Where $\tilde{\mathbf{X}}$ is a random-permutation of $\mathbf{X}$. By definition of the random-permutation, the above definition is equivalent to

$$D(\mathbf{X}) = E\sqrt{\frac{\tilde{X}_k}{\tilde{X}_0}} \tag{60}$$

for all $k = 1, ..., q-1$. Letting $\mathbf{W} = \text{LLR}(\mathbf{X})$ we obtain that

$$D(\mathbf{X}) = Ee^{-\frac{1}{2}\tilde{W}_1}$$

Note that when $q = 2$, this equation coincides with the Bhattacharya parameter that is used in [20], equation (4.4). From Lemma 27 (Appendix V-B) we obtain that,

$$D(\mathbf{R^{(0)}}) = \Delta \tag{61}$$

where $\mathbf{R^{(0)}}$ is the initial message as defined in Section VI-A. We now develop a convenient expression for $D(\mathbf{X})$.

*Lemma 28:* Let $\mathbf{X}$ denote a probability-vector symmetric random variable. Then $D(\mathbf{X}) = Ef(\mathbf{X})$ where $f(\mathbf{x})$ is given by

$$f(\mathbf{x}) \triangleq \frac{1}{q-1}\sum_{i,j\in\text{GF}(q), i\neq j}\sqrt{x_i x_j} \tag{62}$$

**Proof:** From (59) we have

$$
\begin{aligned}
D(\mathbf{X}) &= \frac{1}{q-1}\sum_{i=1}^{q-1} E\sqrt{\frac{X_i}{X_0}} \\
&= E\left[ E\left( \frac{1}{q-1}\sum_{i=1}^{q-1} \sqrt{\frac{X_i}{X_0}} \mid \mathbf{X}\in\mathbf{x}^* \right) \right]
\end{aligned}
\tag{63}
$$

The outer expectation is over all sets $\mathbf{x}^*$. The inner expectation is conditioned on a particular set $\mathbf{x}^*$. We first focus on the inner expectation.

$$
\begin{aligned}
E\left( \frac{1}{q-1}\sum_{i=1}^{q-1} \sqrt{\frac{X_i}{X_0}} \mid \mathbf{X}\in\mathbf{x}^* \right) &= \frac{1}{q-1}\sum_{\mathbf{x}\in\mathbf{x}^*}\sum_{i=1}^{q-1} \sqrt{\frac{x_i}{x_0}}\, \Pr[\mathbf{X}=\mathbf{x} \mid \mathbf{X}\in\mathbf{x}^*] \\
&= \frac{1}{q-1}\sum_{k=0}^{q-1} \frac{1}{n(\mathbf{x})}\sum_{i=1}^{q-1} \sqrt{\frac{x_i^{+k}}{x_0^{+k}}}\, \Pr[\mathbf{X}=\mathbf{x}^{+k} \mid \mathbf{X}\in\mathbf{x}^*]
\end{aligned}
\tag{64}
$$

The last equality was obtained in the same way as (31). In the following, we use the fact that $n(\mathbf{x}^{+k}) = n(\mathbf{x})$ (Lemma 13, Appendix I).

$$
\begin{aligned}
E\left( \frac{1}{q-1}\sum_{i=1}^{q-1} \sqrt{\frac{X_i}{X_0}} \mid \mathbf{X}\in\mathbf{x}^* \right) &= \frac{1}{q-1}\sum_{k=0}^{q-1} \frac{1}{n(\mathbf{x})}\sum_{i=1}^{q-1} \sqrt{\frac{x_i^{+k}}{x_0^{+k}}} \cdot x_0^{+k}\cdot n(\mathbf{x}^{+k}) \\
&= \frac{1}{q-1}\sum_{k=0}^{q-1}\sum_{i=1}^{q-1} \sqrt{\frac{x_i^{+k}}{x_0^{+k}}} \cdot x_0^{+k} \\
&= \frac{1}{q-1}\sum_{k=0}^{q-1}\sum_{i=1}^{q-1} \sqrt{x_{k+i}x_k} = f(\mathbf{x})
\end{aligned}
$$

$f(\mathbf{x})$ is invariant under any permutation of the elements. It is therefore constant for all vectors of the set $\mathbf{x}^*$. Thus we can rewrite the above as

$$
E\left( \frac{1}{q-1}\sum_{i=1}^{q-1} \sqrt{\frac{X_i}{X_0}} \mid \mathbf{X}\in\mathbf{x}^* \right) = E\left( f(\mathbf{X}) \mid \mathbf{X}\in\mathbf{x}^* \right)
$$

Plugging the above into (63) completes the proof. $\square$

We now examine the function $f(\cdot)$.

*Lemma 29:* For any probability vector $\mathbf{x}$, $0 \le f(\mathbf{x}) \le 1$.

*Proof:* $f(\mathbf{x}) \ge 0$ is obtained trivially from (62) by observing that all elements of the sum are nonnegative. To prove $f(\mathbf{x}) \le 1$ we have,

$$
\begin{aligned}
f(\mathbf{x}) &= \frac{1}{q-1}\sum_i \sqrt{x_i}\sum_{j\neq i}\sqrt{x_j} \\
&= \frac{1}{q-1}\sum_i \sqrt{x_i}\big(\sum_j \sqrt{x_j} - \sqrt{x_i}\big) \\
&= \frac{1}{q-1}\left[ \big(\sum_i \sqrt{x_i}\big)^2 - \sum_i x_i \right] \\
&= \frac{1}{q-1}\left[ \big(q\cdot\sum_i \frac{1}{q}\sqrt{x_i}\big)^2 - 1 \right]
\end{aligned}
$$

Applying Jensen's inequality we obtain

$$
\begin{aligned}
f(\mathbf{x}) &\leq \frac{1}{q-1}\left[\left(q\cdot\sqrt{\sum_i \frac{1}{q}x_i}\right)^2 - 1\right]\\
&= \frac{1}{q-1}\left[(q/\sqrt{q})^2 - 1\right] = 1
\end{aligned}
$$

$\square$

Given a probability vector $\mathbf{x}$, we define $\varepsilon(\mathbf{x}) \triangleq 1 - \max(x_0, ..., x_{q-1})$. The following lemma relates the functions $\varepsilon(\cdot)$ and $f(\cdot)$.

*Lemma 30:*

$$
\frac{1}{(q-1)\sqrt{q(q-1)}}\sqrt{\varepsilon(\mathbf{x})} \leq f(\mathbf{x}) \leq q\cdot\sqrt{\varepsilon(\mathbf{x})}
$$

*Proof:* Let $i_{\max}$ be an index that achieves the maximum in $(x_0, ..., x_{q-1})$.

Consider (62). For a particular element $x_i x_j$, assume without loss of generality $i \neq i_{\max}$. By definition of $\mathbf{x}$, we have $x_i \leq \sum_{k \neq i_{\max}} x_k = 1 - x_{i_{\max}} = \varepsilon(\mathbf{x})$. By definition we also have $x_j \leq 1$. Therefore $\sqrt{x_i x_j} \leq \sqrt{\varepsilon(\mathbf{x})}$. We now have,

$$
f(\mathbf{x}) \leq \frac{1}{q-1}\sum_{i,j\in\mathrm{GF}(q)i\neq j}\sqrt{\varepsilon(\mathbf{x})} = q\cdot\sqrt{\varepsilon(\mathbf{x})}
$$

By definition of $\mathbf{x}$, $x_{i_{\max}} \geq 1/q$. Also, there must exist $i \neq i_{\max}$ such that $x_i \geq (1-x_{i_{\max}})/(q-1) = \varepsilon(\mathbf{x})/(q-1)$. We now have

$$
f(\mathbf{x}) \geq \frac{1}{q-1}\sqrt{x_i\cdot x_{i_{\max}}} \geq \frac{1}{q-1}\sqrt{\frac{\varepsilon(\mathbf{x})}{q(q-1)}}
$$

Combining both inequalities proves the lemma. $\square$

We now state our main lemma of the proof:

*Lemma 31:* Let $\mathbf{x}^{(1)},...,\mathbf{x}^{(K)}$ be a set of probability vectors. Then

$$
1 - f\left(\bigodot_{k=1}^{K}\mathbf{x}^{(k)}\right) \geq \prod_{k=1}^{K}\left(1 - f(\mathbf{x}^{(k)})\right) + O\left(\sum_{m,n=1,...,K\ m\neq n}f(\mathbf{x}^{(m)})f(\mathbf{x}^{(n)})\right)
$$

where $\bigodot$ denotes $\mathrm{GF}(q)$ convolution, defined in (11) and used in (13).

**Proof:** We begin by examining the case of $K = 2$.

We denote $\mathbf{x}^{(1)}$ and $\mathbf{x}^{(2)}$ by $\mathbf{a}$ and $\mathbf{b}$. To simplify our analysis, we assume that $a_0 = \max(a_0, ..., a_{q-1})$. We may assume this, because otherwise we can apply a shift by $-i_{\max}$ to move the maximum to zero. This operation does not affect $f(\mathbf{a})$. It is easy to verify that $\mathbf{a}^{-i_{\max}} \odot \mathbf{b} = (\mathbf{a}\odot\mathbf{b})^{-i_{\max}}$ and hence the operation does not affect $f(\mathbf{a}\odot\mathbf{b})$ either. Similarly, we assume $b_0 = \max(b_0, ..., b_{q-1})$.

By the definition of $f(\cdot)$, we have

$$
f(\mathbf{a}\odot\mathbf{b}) = \frac{1}{q-1}\sum_{i\neq j}\sqrt{(\mathbf{a}\odot\mathbf{b})_i\cdot(\mathbf{a}\odot\mathbf{b})_j} \tag{65}
$$

We now examine elements of the sum. We first examine the case that $i = 0$ and $j \neq 0$.

$$\sqrt{(\mathbf{a} \odot \mathbf{b})_0 \cdot (\mathbf{a} \odot \mathbf{b})_j} = \sqrt{(a_0 b_0 + \sum_{k \neq 0} a_k b_{-k}) \cdot (a_j b_0 + a_0 b_j + \sum_{k \neq 0,j} a_k b_{j-k})}$$

$$= \sqrt{[a_0 b_0 + O(\varepsilon(\mathbf{a})\varepsilon(\mathbf{b}))] \cdot [a_j b_0 + a_0 b_j + O(\varepsilon(\mathbf{a})\varepsilon(\mathbf{b}))]}$$

$$= \sqrt{a_0 b_0 \cdot a_j b_0 + a_0 b_0 \cdot a_0 b_j + O(\varepsilon(\mathbf{a})\varepsilon(\mathbf{b}))}$$

$$\leq \sqrt{a_0 a_j + b_0 b_j + O(\varepsilon(\mathbf{a})\varepsilon(\mathbf{b}))}$$

$$\leq \sqrt{a_0 a_j} + \sqrt{b_0 b_j} + O(\sqrt{\varepsilon(\mathbf{a})}\sqrt{\varepsilon(\mathbf{b})})$$

The result for the case of $i \neq 0$ and $j = 0$ is similarly obtained. We now assume $i, j \neq 0$ (the element $i = j = 0$ does not participate in the sum).

$$\sqrt{(\mathbf{a} \odot \mathbf{b})_i \cdot (\mathbf{a} \odot \mathbf{b})_j} = \sqrt{(a_i b_0 + a_0 b_i + \sum_{k \neq 0,i} a_k b_{i-k}) \cdot (a_j b_0 + a_0 b_j + \sum_{k \neq 0,j} a_k b_{j-k})}$$

$$\leq \sqrt{[a_i + b_i + O(\varepsilon(\mathbf{a})\varepsilon(\mathbf{b}))] \cdot [a_j + b_j + O(\varepsilon(\mathbf{a})\varepsilon(\mathbf{b}))]}$$

$$\leq \sqrt{a_i a_j + b_i b_j + O(\varepsilon(\mathbf{a})\varepsilon(\mathbf{b}))}$$

$$\leq \sqrt{a_i a_j} + \sqrt{b_i b_j} + O(\sqrt{\varepsilon(\mathbf{a})}\sqrt{\varepsilon(\mathbf{b})})$$

Inserting the above into (65) we obtain

$$f(\mathbf{a} \odot \mathbf{b}) \leq \frac{1}{q-1} \sum_{i \neq j} \left( \sqrt{a_i a_j} + \sqrt{b_i b_j} + O(\sqrt{\varepsilon(\mathbf{a})}\sqrt{\varepsilon(\mathbf{b})}) \right) = f(\mathbf{a}) + f(\mathbf{b}) + O(\sqrt{\varepsilon(\mathbf{a})}\sqrt{\varepsilon(\mathbf{b})})$$

$$= f(\mathbf{a}) + f(\mathbf{b}) + O(f(\mathbf{a})f(\mathbf{b}))$$

The last equality having been obtained from Lemma 30. Finally, from the above we easily obtain the desired result of

$$1 - f(\mathbf{a} \odot \mathbf{b}) \geq (1 - f(\mathbf{a})) \cdot (1 - f(\mathbf{b})) + O(f(\mathbf{a})f(\mathbf{b}))$$

For the case of $K > 2$ we begin by observing that

$$1 - f\left( \bigodot_{k=1}^{K} \mathbf{x}^{(k)} \right) = 1 - f\left( (\bigodot_{k=1}^{K-1} \mathbf{x}^{(k)}) \bigodot \mathbf{x}^{(K)} \right)$$

The remainder of the proof is obtained by induction, using Lemma 29. □

We now use the above lemma to obtain the following results

*Lemma 32:* $D(\mathbf{R}_{t+1})$ satisfies,

$$D(\mathbf{R}_{t+1}) \leq \Delta \cdot \lambda \left( 1 - \rho(1 - D(\mathbf{R}_t)) + O(D(\mathbf{R}_t)^2) \right) \tag{66}$$

*Proof:* Consider $\mathbf{R}_t$. Since $\bar{\mathbf{R}}_t$ is obtained from it by applying a random permutation $\times g^{-1}$, we obtain, using Lemma 28 and the fact that $f(\mathbf{x})$ is invariant under a permutation on $\mathbf{x}$, that $D(\bar{\mathbf{R}}_t) = Ef(\bar{\mathbf{R}}_t) = Ef(\mathbf{R}_t) = D(\mathbf{R}_t)$. Thus we may instead examine $\bar{\mathbf{R}}_t$. Similarly, we examine $\bar{\mathbf{L}}_t$ instead of $\mathbf{L}_t$.

Assume the right-degree at a check-node is $d$. By (13) we have,

$$1 - D(\bar{\mathbf{L}}_{t+1}) = 1 - D(\bigodot_{k=1}^{d-1} \bar{\mathbf{R}}^{(k)})$$

where $\{\bar{\mathbf{R}}^{(k)}\}_{k=1}^{d-1}$ are i.i.d. and distributed as $\bar{\mathbf{R}}_t$. In the following, we make use of Lemma 31.

$$
\begin{aligned}
1 - D(\bar{\mathbf{L}}_{t+1}) &= 1 - Ef(\bigodot_{k=1}^{d-1} \bar{\mathbf{R}}^{(k)}) \\
&\geq E\left[\prod_{k=1}^{d-1}(1 - f(\bar{\mathbf{R}}^{(k)})) + O(\sum_{m \neq n} f(\bar{\mathbf{R}}^{(m)})f(\bar{\mathbf{R}}^{(n)}))\right] \\
&= \left[\prod_{k=1}^{d-1}(1 - Ef(\bar{\mathbf{R}}^{(k)})) + O(\sum_{m \neq n} Ef(\bar{\mathbf{R}}^{(m)})Ef(\bar{\mathbf{R}}^{(n)}))\right] \\
&= \prod_{k=1}^{d-1}(1 - D(\bar{\mathbf{R}}^{(k)})) + O(\sum_{m \neq n} D(\bar{\mathbf{R}}^{(m)})D(\bar{\mathbf{R}}^{(n)})) \\
&= (1 - D(\bar{\mathbf{R}}_t))^{d-1} + O(D(\bar{\mathbf{R}}_t)^2)
\end{aligned}
$$

Averaging over all possible values of $d$, we obtain,

$$
\begin{aligned}
1 - D(\bar{\mathbf{L}}_{t+1}) &\geq \sum_d \rho_d \cdot \left[(1 - D(\bar{\mathbf{R}}_t))^{d-1} + O(D(\bar{\mathbf{R}}_t)^2)\right] \\
&= \sum_d \rho_d \cdot (1 - D(\bar{\mathbf{R}}_t))^{d-1} + O(D(\bar{\mathbf{R}}_t)^2) \\
&= \rho(1 - D(\bar{\mathbf{R}}_t)) + O(D(\bar{\mathbf{R}}_t)^2)
\end{aligned}
\tag{67}
$$

We now turn to examine $D(\mathbf{R}_{t+1})$. Assume the variable-node degree at which $\mathbf{R}_t$ is produced is $deg$. Applying (59) and (8) we have

$$
\begin{aligned}
D(\mathbf{R}_{t+1}) &= \frac{1}{q-1}\sum_{i=1}^{q-1} E\sqrt{\frac{R_{t+1,i}}{R_{t+1,0}}} \\
&= \frac{1}{q-1}\sum_{i=1}^{q-1} E\sqrt{\frac{R_i^{(0)}}{R_0^{(0)}}\prod_{n=1}^{deg-1}\frac{L_i^{(n)}}{L_0^{(n)}}}
\end{aligned}
$$

where $\{\mathbf{L}^{(n)}\}_{n=1}^{d_i-1}$ are i.i.d. and distributed as $\mathbf{L}_{t+1}$. By Theorem 4, $\{\mathbf{L}^{(n)}\}_{n=1}^{d_i-1}$ are permutation-invariant, and thus, by Lemma 21 (Appendix IV-E), are distributed identically with their random-permutations $\{\tilde{\mathbf{L}}^{(n)}\}_{n=1}^{d_i-1}$. Thus we obtain

$$
D(\mathbf{R}_{t+1}) = \frac{1}{q-1}\sum_{i=1}^{q-1} E\sqrt{\frac{R_i^{(0)}}{R_0^{(0)}}}\prod_{n=1}^{deg-1} E\sqrt{\frac{\tilde{L}_i^{(n)}}{\tilde{L}_0^{(n)}}}
$$

Applying (60) and reordering the elements, we obtain

$$
\begin{aligned}
D(\mathbf{R}_{t+1}) &= E\left(\frac{1}{q-1}\sum_{i=1}^{q-1}\sqrt{\frac{R_i^{(0)}}{R_0^{(0)}}}\right)\prod_{n=1}^{deg-1} D(\mathbf{L}^{(n)}) \\
&= D(\mathbf{R^{(0)}}) \cdot D(\mathbf{L}_{t+1})^{deg-1} \\
&= \Delta \cdot D(\mathbf{L}_{t+1})^{deg-1}
\end{aligned}
$$

The second equality was obtained from (59). The last equality is obtained from (61). Averaging over all values of $deg$, we obtain,

$$
D(\mathbf{R}_{t+1}) = \Delta \cdot \lambda(D(\mathbf{L}_{t+1}))
\tag{68}
$$

The function $\lambda(x)$ is by definition a polynomial with non-negative coefficients. It is thus nondecreasing in the range $0 \leq x \leq 1$. Using (67) and (68) we obtain (66). $\qquad \blacksquare$

The following lemma examines convergence to zero of $D(\mathbf{R}_t)$.

*Lemma 33:* If $\lambda'(0)\rho'(1) < 1/\Delta$ then there exists $\alpha > 0$ such that if $D(\mathbf{R}_{t_0}) < \alpha$ at some iteration $t_0$, then $\lim_{t \to \infty} D(\mathbf{R}_t) = 0$.

*Proof:* Using the Taylor expansion of the function $\rho(1-x)$ around $x = 0$

$$
\begin{aligned}
\rho(1 - D(\mathbf{R}_t)) &= \rho(1) - \rho'(1) \cdot D(\mathbf{R}_t) + O(D(\mathbf{R}_t)^2) \\
&= 1 - \rho'(1) \cdot D(\mathbf{R}_t) + O(D(\mathbf{R}_t)^2)
\end{aligned}
$$

where the equality $\rho(1) = 1$ is obtained by the definition of the function $\rho(x)$. Plugging the above into (66) we obtain,

$$
D(\mathbf{R}_{t+1}) \leq \Delta \cdot \lambda \left( \rho'(1) \cdot D(\mathbf{R}_t) + O(D(\mathbf{R}_t)^2) \right)
$$

Using the Taylor expansion of $\lambda(x)$ around $x = 0$, we obtain

$$
\begin{aligned}
D(\mathbf{R}_{t+1}) &\leq \Delta \cdot \left[ \lambda(0) + \lambda'(0) \cdot \left( \rho'(1) \cdot D(\mathbf{R}_t) + O(D(\mathbf{R}_t)^2) \right) + O\left( \left( \rho'(1) \cdot D(\mathbf{R}_t) + O(D(\mathbf{R}_t)^2) \right)^2 \right) \right] \\
&= \Delta \cdot \lambda'(0)\rho'(1) \cdot D(\mathbf{R}_t) + O(D(\mathbf{R}_t)^2)
\end{aligned}
$$

Since $\Delta \cdot \lambda'(0)\rho'(1) < 1$, there exists $\alpha$ such that if $D(\mathbf{R}_{t_0}) < \alpha$ then

$$
D(\mathbf{R}_{t_0+1}) < K \cdot D(\mathbf{R}_{t_0}) < D(\mathbf{R}_{t_0}) < \alpha
$$

where $K$ is a positive constant smaller than 1. By induction, this holds for all $t > t_0$. We have $D(\mathbf{R}_t) \geq 0$ by definition, and therefore the sequence $\{D(\mathbf{R}_t)\}_{t=t_0}^{\infty}$ converges to zero. $\qquad \blacksquare$

Finally, the following lemma links the operator $D(\cdot)$ with our desired $P_e(\cdot)$, defined as in (22).

*Lemma 34:* Let $\mathbf{X}$ be a symmetric probability-vector random-variable. Then

$$
1/q^2 \cdot D(\mathbf{X})^2 \leq P_e(\mathbf{X}) \leq (q-1) \cdot D(\mathbf{X})
$$

*Proof:* We begin by showing that $P_e(\mathbf{X}) = E\varepsilon(\mathbf{X})$.

$$
\begin{aligned}
P_e(\mathbf{X}) &= \sum_{\mathbf{x}} P_e(\mathbf{x}) \Pr[\mathbf{X} = \mathbf{x}] \\
&= \sum_{\mathbf{x}^*} \left( \frac{1}{n(\mathbf{x})} \sum_{i=0}^{q-1} P_e(\mathbf{x}^{+i}) \Pr[\mathbf{X} = \mathbf{x}^{+i} \mid \mathbf{X} \in \mathbf{x}^*] \right) \Pr[\mathbf{X} \in \mathbf{x}^*]
\end{aligned}
$$

The last result was obtained in the same way as (63) and (64). The outer sum is over all sets $\mathbf{x}^*$. Let $i_1, ..., i_m$ denote the indices that achieve $\max(x_0, ..., x_{q-1})$. Then $P_e(\mathbf{x}^{+i}) = (m-1)/m$ if $i = i_1, ..., i_m$ and 1 otherwise. Using this and the symmetry of $\mathbf{X}$, we obtain

$$
P_e(\mathbf{X}) = \sum_{\mathbf{x}^*} \left( \frac{1}{n(\mathbf{x})} \sum_{i=i_1,...,i_m} \frac{m-1}{m} \cdot x_i \cdot n(\mathbf{x}^{+i}) + \sum_{i \neq i_1,...,i_m} 1 \cdot x_i \cdot n(\mathbf{x}^{+i}) \right) \Pr[\mathbf{X} \in \mathbf{x}^*]
$$

By Lemma 13 (Appendix I), $n(\mathbf{x}^{+i}) = n(\mathbf{x})$. We thus continue our development,

$$
\begin{aligned}
P_e(\mathbf{X}) &= \sum_{\mathbf{x}^*}\left(\sum_{i=0}^{q-1} 1 \cdot x_i - \frac{1}{m}\sum_{i=i_1,\dots,i_m} x_{\max}\right)\Pr[\mathbf{X}\in\mathbf{x}^*] \\
&= \sum_{\mathbf{x}^*}(1-x_{\max})\Pr[\mathbf{X}\in\mathbf{x}^*] \\
&= \sum_{\mathbf{x}^*}\varepsilon(\mathbf{x})\Pr[\mathbf{X}\in\mathbf{x}^*]
\end{aligned}
$$

The result $P_e(\mathbf{X}) = E\varepsilon(\mathbf{X})$ is obtained from the fact that $\varepsilon(\cdot)$ is constant over all vectors in $\mathbf{x}^*$.

We now have, using Lemmas 28 and 30 and the Jensen inequality

$$
D(\mathbf{X}) = Ef(\mathbf{X}) \leq qE\sqrt{\varepsilon(\mathbf{X})} \leq q\sqrt{E\varepsilon(\mathbf{X})} = q\sqrt{P_e(\mathbf{X})}
$$

This proves $1/q^2 \cdot D(\mathbf{X})^2 \leq P_e(\mathbf{X})$. For the second inequality, we observe

$$
P_e(\mathbf{X}) \leq \Pr[\exists i \neq 0 \; : \; X_i \geq X_0] \leq \sum_{i=1}^{q-1}\Pr[X_i \geq X_0] = \sum_{i=1}^{q-1}\Pr[\sqrt{\frac{X_i}{X_0}} \geq 1] \leq \sum_{i=1}^{q-1}(E\sqrt{\frac{X_i}{X_0}})/1
$$

The last inequality is obtained by Markov's inequality. Combining the above with (59) we obtain our desired result of $P_e(\mathbf{X}) \leq (q-1)\cdot D(\mathbf{X})$. $\qquad\square$

Finally, consider the value $\alpha$ of Lemma 33. Setting $\xi = \alpha^2/q^2$ we have from Lemma 34 that if $P_e(\mathbf{R}_{t_0}) < \xi$ then $D(\mathbf{R}_{t_0}) < \alpha$ and thus $D(\mathbf{R}_t)$ converges to zero. Applying Lemma 34 again, this implies that $P_e(\mathbf{R}_t)$ converges to zero, and thus completes the proof of Part 2 of the theorem. $\square$

# Appendix VII

## Proof of Theorem 6

We begin by observing that since $\mathbf{W}$ is Gaussian, $\mathbf{W}$ is symmetric if and only if for all $i = 1, \dots, q-1$ and arbitrary LLR vector $\mathbf{w}$,

$$
\begin{aligned}
2w_i &= 2\log\frac{e^{w_i}f(\mathbf{w}^{+i})}{f(\mathbf{w}^{+i})} = 2\log\frac{f(\mathbf{w})}{f(\mathbf{w}^{+i})} \\
&= 2\log\frac{1}{\sqrt{\det(2\pi\Sigma)}}\exp\left(-\frac{1}{2}(\mathbf{w}-\mathbf{m})^T\Sigma^{-1}(\mathbf{w}-\mathbf{m})\right) \\
&\quad -2\log\frac{1}{\sqrt{\det(2\pi\Sigma)}}\exp\left(-\frac{1}{2}(\mathbf{w}^{+i}-\mathbf{m})^T\Sigma^{-1}(\mathbf{w}^{+i}-\mathbf{m})\right) \\
&= (\mathbf{w}^{+i}-\mathbf{m})^T\Sigma^{-1}(\mathbf{w}^{+i}-\mathbf{m}) - (\mathbf{w}-\mathbf{m})^T\Sigma^{-1}(\mathbf{w}-\mathbf{m})
\end{aligned} \tag{69}
$$

We first assume that $\mathbf{W}$ is symmetric and permutation-invariant and prove (25). Since $\mathbf{W}$ is permutation-invariant, by Lemma 8 we have $m_i = EW_i = EW_j = m_j$ for all $i, j = 1, \dots, q-1$. We therefore denote $m \overset{\triangle}{=} m_1 = \dots = m_{q-1}$.

We begin by proving that $m \neq 0$. We prove this by contradiction, and hence we first assume $m = 0$. Consider the marginal distribution of $W_i$ for $i = 1, \dots, q-1$, which must also be Gaussian. Since $m_i = 0$, the pdf of $W_i$ satisfies $f_i(w) = f_i(-w)$. By Lemma 9, $W_i$ is symmetric in the binary sense. Hence $f_i(w) = e^{-w}f_i(-w)$. Combining both equations yields $f_i(w) = 0$ for all $w \neq 0$. Hence $W_i$ is deterministic, with zero variance, for all $i$. This leads to $\Sigma = 0$, which contradicts the theorem's condition that $\Sigma$ is nonsingular.

We now show that conditions (69) $i = 1, ..., q-1$ uniquely define $\Sigma$. Since $\Sigma$ is symmetric, so is $\Sigma^{-1}$. Assume $A$ and $B$ are two symmetric matrices such that (69) is satisfied, substituting $\Sigma^{-1}$ with $A$ and with $B$, respectively. We now show that $A = B$. Let $D \triangleq A - B$. Subtracting the equation for $B$ from that of $A$ we obtain, for $i = 1, ..., q-1$,

$$0 = (\mathbf{w}^{+i} - \mathbf{m})^T D(\mathbf{w}^{+i} - \mathbf{m}) - (\mathbf{w} - \mathbf{m})^T D(\mathbf{w} - \mathbf{m}) \tag{70}$$

For convenience, we let $L_i$ denote the matrix corresponding to the linear transformation $L_i \mathbf{w} = \mathbf{w}^{+i}$. Differentiating (70) twice with respect to $\mathbf{w}$, we obtain that $L_i^T D L_i = D$. (70) may now be rewritten as

$$
\begin{aligned}
(\mathbf{w}^{+i} - \mathbf{m})^T D(\mathbf{w}^{+i} - \mathbf{m}) &= (\mathbf{w} - \mathbf{m})^T L_i^T D L_i (\mathbf{w} - \mathbf{m}) \\
&= (\mathbf{w}^{+i} - \mathbf{m}^{+i})^T D(\mathbf{w}^{+i} - \mathbf{m}^{+i})
\end{aligned}
$$

Let $\mathbf{x} \triangleq \mathbf{w}^{+i}$. Observe that $\mathbf{x}$, like $\mathbf{w}$, is arbitrary. Simple algebraic mainpulations lead us to

$$
\begin{aligned}
2\mathbf{x}^T D(\mathbf{m}^{+i} - \mathbf{m}) &= (\mathbf{m}^{+i})^T D\mathbf{m}^{+i} - \mathbf{m}^T D\mathbf{m} \\
&= (\mathbf{m}^{+i})^T D\mathbf{m}^{+i} - \mathbf{m}^T L_i^T D L_i \mathbf{m} = 0
\end{aligned}
$$

Letting $\mathbf{x} = D(\mathbf{m}^{+i} - \mathbf{m})$ we obtain that $\|D(\mathbf{m}^{+i} - \mathbf{m})\|^2 = 0$ where $\| \cdot \|$ denotes Euclidean norm. Thus $D(\mathbf{m}^{+i} - \mathbf{m}) = 0$. Consider the vectors $\{\mathbf{m}^{+i} - \mathbf{m}\}_{i=1}^{q-1}$. We wish to show that these vectors are linearly independent. From (5), we have $(\mathbf{m}^{+i} - \mathbf{m})_k = m_{i+k} - m_i - m_k$. Recall from Section II that $i + k$ is evaluated over GF($q$) and that $m_0 = 0$. From our previous discussion, $m_i = m$ for all $i = 1, ..., q-1$. Therefore, for all $i \neq 0, k \neq 0$.

$$m_k^{+i} - m_k = \begin{cases} -m & k \neq -i \\ -2m & k = -i \end{cases}$$

We now put the vectors $\{\mathbf{m}^{+i} - \mathbf{m}\}_{i=1}^{q-1}$ in a matrix $M$ such that $M_{k,i} \triangleq (\mathbf{m}^{-i} - \mathbf{m})_k$. The matrix $M$ is now given by,

$$M = \begin{bmatrix} -2m & -m & ... & -m \\ -m & -2m & & \\ & & ... & \\ -m & & & -2m \end{bmatrix}$$

Let the matrix $V$ be defined by,

$$V = \frac{1}{m} \cdot \begin{bmatrix} \frac{1}{q} - 1 & \frac{1}{q} & ... & \frac{1}{q} \\ \frac{1}{q} & \frac{1}{q} - 1 & & \\ & & ... & \\ \frac{1}{q} & & & \frac{1}{q} - 1 \end{bmatrix}$$

That is, $V_{i,j} = (1/q - \delta[i - j])/m$. It is easy to verify that $V$ is the inverse of $M$. Hence $M$ is nonsingular, and its columns, the vectors $\{\mathbf{m}^{+i} - \mathbf{m}\}_{i=1}^{q-1}$, are thus linearly independent. We now have $q - 1$ linearly-independent vectors that satisfy $D(\mathbf{m}^{+i} - \mathbf{m}) = \mathbf{0}$. Hence $D = 0$, and we obtain that $A = B$ as desired.

Consider the matrix $M$. If we could show that $\Sigma = -M$, we would obtain (25) for $\sigma = \sqrt{2m}$ ($m > 0$ would be implied by $\Sigma_{1,1} = 2m$). For this purpose, we show that the choice $\Sigma^{-1} = (-M)^{-1} = -V$ satisfies (69).

$$
\begin{aligned}
(\mathbf{w}^{+i} - \mathbf{m})^T(-V)(\mathbf{w}^{+i} - \mathbf{m}) &- (\mathbf{w} - \mathbf{m})^T(-V)(\mathbf{w} - \mathbf{m}) = \\
&= (\mathbf{w}^{+i} - \mathbf{m} + \mathbf{w} - \mathbf{m})^T(-V)(\mathbf{w}^{+i} - \mathbf{m} - \mathbf{w} + \mathbf{m}) \\
&= (\mathbf{w}^{+i} + \mathbf{w} - 2\mathbf{m})^T(-V)(\mathbf{w}^{+i} - \mathbf{w}) \\
&= \sum_{k,j=1}^{q-1} (w_k^{+i} + w_k - 2m)(w_j^{+i} - w_j)\frac{1}{m}[\delta[k-j] - 1/q] \\
&= \frac{1}{m}\sum_{k=1}^{q-1}(w_k^{+i} + w_k - 2m)(w_k^{+i} - w_k) - \frac{1}{qm}\sum_{k=1}^{q-1}(w_k^{+i} + w_k - 2m)\cdot\sum_{j=1}^{q-1}(w_j^{+i} - w_j)
\end{aligned}
\tag{71}
$$

We now treat each of the above sums separately

$$
\begin{aligned}
\sum_{k=1}^{q-1}(w_k^{+i} + w_k - 2m)(w_k^{+i} - w_k) &= \\
&= \sum_{k=1}^{q-1}\left[(w_k^{+i})^2 - (w_k)^2 - 2m\cdot(w_k^{+i} - w_k)\right] \\
&= \sum_{k=1}^{q-1}w_{k+i}^2 + \sum_{k=1}^{q-1}w_i^2 - 2w_i\sum_{k=1}^{q-1}w_{k+i} - \sum_{k=1}^{q-1}w_k^2 - 2m\sum_{k=1}^{q-1}w_{k+i} + 2m\sum_{k=1}^{q-1}w_i + 2m\sum_{k=1}^{q-1}w_k
\end{aligned}
\tag{72}
$$

The set of indices $\{k + i : k = 1, ..., q - 1\} = \{0, ..., q - 1\}\setminus\{i\}$. Recalling $w_0 = 0$, we have:

$$
\begin{aligned}
\sum_{k=1}^{q-1}w_{k+i}^2 &= \sum_{k=1}^{q-1}w_k^2 - w_i^2 \\
\sum_{k=1}^{q-1}w_{k+i} &= \sum_{k=1}^{q-1}w_k - w_i
\end{aligned}
\tag{73}
$$

(72) now becomes

$$
\begin{aligned}
\sum_{k=1}^{q-1}(w_k^{+i} + w_k - 2m)(w_k^{+i} - w_k) &= \\
&= \left(\sum_{k=1}^{q-1}w_k^2 - w_i^2\right) + (q-1)w_i^2 - 2w_i\left(\sum_{k=1}^{q-1}w_k - w_i\right) - \sum_{k=1}^{q-1}w_k^2 - 2m\left(\sum_{k=1}^{q-1}w_k - w_i\right) + \\
&\quad + 2m(q-1)w_i + 2m\sum_{k=1}^{q-1}w_k \\
&= q\cdot w_i^2 - 2w_i\sum_{k=1}^{q-1}w_k + 2mq\cdot w_i
\end{aligned}
\tag{74}
$$

We now turn to the second sum of (71). In a development similar to that of the first sum, we obtain

$$
\sum_{k=1}^{q-1}(w_k^{+i} + w_k - 2m) = 2\sum_{k=1}^{q-1}w_k - q\cdot w_i - 2(q-1)m
\tag{75}
$$

Finally, the last sum of (71) becomes

$$
\sum_{j=1}^{q-1}(w_j^{+i} - w_j) = -q\cdot w_i
\tag{76}
$$

Combining (71), (74), (75) and (76) we obtain

$$
\begin{aligned}
(\mathbf{w}^{+i} - \mathbf{m})^T(-V)(\mathbf{w}^{+i} - \mathbf{m}) &- (\mathbf{w} - \mathbf{m})^T(-V)(\mathbf{w} - \mathbf{m}) = \\
&= \frac{1}{m}\left( q \cdot w_i^2 - 2w_i \sum_{k=1}^{q-1} w_k + 2mq \cdot w_i \right) - \frac{1}{qm}\left( 2\sum_{k=1}^{q-1} w_k - q \cdot w_i - 2(q-1)m \right)(-q \cdot w_i) \\
&= 2w_i
\end{aligned}
\tag{77}
$$

Thus $\Sigma^{-1} = -V$ satisfies (69) as desired. This completes the proof of (25).

We now assume (25) and prove that $\mathbf{W}$ is symmetric and permutation-invariant. From (25) it is clear that any reordering of the elements of $\mathbf{W}$ has no effect on its distribution, and thus $\mathbf{W}$ is permutation-invariant. To prove symmetry, we observe that the development ending with (77) relies on (25) alone, and thus remains valid. $\quad\square$

## APPENDIX VIII
## PROOFS FOR SECTION VII

### A. Proof of Lemma 11

By Lemma 17 (Appendix III-A),

$$
\begin{aligned}
I(C; \mathbf{W}) &= \sum_{k=0}^{q-1} \sum_{\mathbf{w}} \Pr[C = k]\Pr[\mathbf{W} = \mathbf{w} \mid C = k]\log_q \frac{\Pr[\mathbf{W} = \mathbf{w} \mid C = k]}{\Pr[\mathbf{W} = \mathbf{w}]} \\
&= \frac{1}{q}\sum_{k=0}^{q-1} \sum_{\mathbf{w}} \Pr[\mathbf{W} = \mathbf{w}^{+k} \mid C = 0]\log_q \frac{\Pr[\mathbf{W} = \mathbf{w}^{+k} \mid C = 0]}{\frac{1}{q}\sum_{j=0}^{q-1} \Pr[\mathbf{W} = \mathbf{w}^{+j} \mid C = 0]}
\end{aligned}
$$

The second summation in the above equations is over all LLR vectors $\mathbf{w}$ with nonzero probability.

By the lemma's condition, the tree assumption is satisfied. Thus by Theorem 1, the conditional distribution of $\mathbf{W}$ given $C = 0$ is symmetric (recalling Lemma 16, Appendix III-A). Using (19), we have

$$
\begin{aligned}
I(C; \mathbf{W}) &= \frac{1}{q}\sum_{k=0}^{q-1} \sum_{\mathbf{w}} \Pr[\mathbf{W} = \mathbf{w}^{+k} \mid C = 0]\log_q \frac{e^{-w_k}\Pr[\mathbf{W} = \mathbf{w} \mid C = 0]}{\frac{1}{q}\sum_{j=0}^{q-1} e^{-w_j}\Pr[\mathbf{W} = \mathbf{w} \mid C = 0]} \\
&= \frac{1}{q}\sum_{k=0}^{q-1} \sum_{\mathbf{w}} \Pr[\mathbf{W} = \mathbf{w}^{+k} \mid C = 0]\left( 1 - \log_q \sum_{j=0}^{q-1} e^{-(w_j - w_k)} \right) \\
&= 1 - \frac{1}{q}\sum_{k=0}^{q-1} \sum_{\mathbf{w}} \Pr[\mathbf{W} = \mathbf{w}^{+k} \mid C = 0]\log_q \sum_{j=0}^{q-1} e^{-(w_j - w_k)}
\end{aligned}
$$

By (5), $w_j - w_k = w_{j-k}^{+k}$. Since the third summation is over all $j$, we obtain by changing variables $j' = j - k$ (evaluated over GF($q$)),

$$
I(C; \mathbf{W}) = 1 - \frac{1}{q}\sum_{k=0}^{q-1} \sum_{\mathbf{w}} \Pr[\mathbf{W} = \mathbf{w}^{+k} \mid C = 0]\log_q \sum_{j'=0}^{q-1} e^{-w_{j'}^{+k}}
$$

Changing variables in the second summation, $\hat{\mathbf{w}} \triangleq \mathbf{w}^{+k}$, we obtain

$$
I(C; \mathbf{W}) = 1 - \frac{1}{q}\sum_{k=0}^{q-1} \sum_{\hat{\mathbf{w}}} \Pr[\mathbf{W} = \hat{\mathbf{w}} \mid C = 0]\log_q \sum_{j=0}^{q-1} e^{-\hat{w}_j}
$$

Since the sum over $\hat{\mathbf{w}}$ is independent of $k$, we obtain,

$$
I(C; \mathbf{W}) = 1 - \sum_{\mathbf{w}} \Pr[\mathbf{W} = \mathbf{w} \mid C = 0]\log_q \sum_{j=0}^{q-1} e^{-w_j}
$$

(26) now follows from the fact that $w_0 = 0$ by definition (see Section II). $\quad\square$

## B. The Permutation-Invariance Assumption with EXIT Method 1

In this section, we discuss a fine-point of the assumption of permutation-invariance used in the development of EXIT charts by Method 1 (Section VII-C). Strictly speaking, the initial message $\mathbf{R'}^{(0)}$ and rightbound messages $\mathbf{R'}_t$ are not permutation-invariant. However, we now show that we may shift our attention to $\mathbf{\bar{R}'}^{(0)}$ and $\bar{\mathbf{R}}'_t$, defined as in Theorem 4, which are symmetric and permutation-invariant.

We first show that $I(C; \mathbf{R'}^{(0)})$ and $I(C; \mathbf{R'}_t)$, evaluated using (26), are equal to $I(C; \mathbf{\bar{R}'}^{(0)})$ and $I(C; \bar{\mathbf{R}}'_t)$ (respectively). It is straightforward to observe that the right-hand-side of (26) is invariant to any fixed permutation of the elements of the random vector $\mathbf{W}$. Thus, a random-permutation will also have no effect on its value. By the discussion in Appendix IV-F, $\mathbf{\bar{R}'}^{(0)}$ and $\bar{\mathbf{R}}'_t$ are random-permutations of $\mathbf{R'}^{(0)}$ and $\mathbf{R'}_t$, respectively. Thus, we have obtained our desired result.

We proceed to show that the derivation of the approximation of $I_{E,VND}$ in Section VII-C is justified if we replace $\mathbf{R'}^{(0)}$ and $\mathbf{R'}_t$ with $\mathbf{\bar{R}'}^{(0)}$ and $\bar{\mathbf{R}}'_t$. By the discussion in Appendix IV-F, $\bar{\mathbf{R}}'_t$ may be obtained by replacing the instantiation $\mathbf{r'}^{(0)}$ of $\mathbf{R'}^{(0)}$ in (15) with an instantiation of $\mathbf{\bar{R}'}^{(0)}$. Thus, $\bar{\mathbf{R}}'_t$ is obtained from $\mathbf{L}'_t$ and $\mathbf{\bar{R}'}^{(0)}$ using the same expressions through which $\mathbf{R'}_t$ is obtained from $\mathbf{L}'_t$ and and $\mathbf{R'}^{(0)}$. Therefore, the discussion of the derivation of the approximation for $I_{E,VND}$ (see Appendix VII-C) remains justified.

By the discussion in Appendix IV-F, the distribution of $\mathbf{L}_t$ is obtained from $\bar{\mathbf{R}}_t$ using (10), and the distribution of $\mathbf{R}_t$ is not required for its computation. Finally, the approximation for $I_{E,CND}$ in Section VII-C has been verified empirically, and therefore does not require any further justification.

## C. Gaussian Messages as Initial Messages of an AWGN Channel

Let $\mathbf{W}$ be a Gaussian LLR-vector random variable defined as in Theorem 6. Let $\Pr[\mathbf{w} \mid x]$ be the transition probabilities of the cyclic-symmetric channel defined by $\mathbf{W}$ (see Lemma 6 and Remark 1, Section V-C). We will now show that this channel is in effect a $q-1$-dimensional AWGN channel.

We begin by examining $\Pr[\mathbf{w} \mid x = i]$.

$$\Pr[\mathbf{w} \mid x = i] = \Pr[\mathbf{W} = \mathbf{w}^{+i}] = \Pr[\mathbf{W}^{-i} = \mathbf{w}]$$

Thus the channel output, conditioned on transmission of $i$, is distributed as $\mathbf{W}^{-i}$. The operation $-i$, as defined by (5), is linear. Thus $\mathbf{W}^{-i}$ is Gaussian with a mean of $\mathbf{m}^{-i}$ ($\mathbf{m}$ being defined by (25)) and a covariance matrix which we will denote by $\Sigma^{(-i)}$. Let $k, l = 1, ..., q-1$.

$$\Sigma_{k,l}^{(-i)} = \text{cov}(W_k^{-i}, W_l^{-i}) = \text{cov}(W_{k-i} - W_{-i}, W_{l-i} - W_{-i}) = \Sigma_{k-i,l-i} - \Sigma_{-i,l-i} - \Sigma_{k-i,-i} + \Sigma_{-i,-i} \quad (78)$$

where $\Sigma$ is given by (25) and we define, for convenience, $\Sigma_{0,j} = \Sigma_{j,0} \overset{\Delta}{=} 0$ for all $j = 0, ..., q-1$ (also, recall from Section II, that $k-i$ and $l-i$ are evaluated over GF($q$)). Evaluating (78) for all $k, l = 1, ..., q-1$, it is easily observed that $\Sigma^{(-i)} = \Sigma$.

The above implies that the cyclic-symmetric channel defined by $\mathbf{W}$ is distributed as a $q-1$ dimensional AWGN channel whose noise is distributed as $\mathcal{N}(\mathbf{0}, \Sigma)$ and whose input alphabet is given by $\delta(i) = \mathbf{m}^{-i}$. Both the noise and the input alphabet are functions of $\sigma$. By definition, this channel is cyclic-symmetric and thus the LLR-vector initial messages of LDPC decoding satisfy $\mathbf{r'^{(0)}} = \mathbf{w}$ where $\mathbf{w}$ is the channel output.

In the sequel, we would like to consider channels whose input alphabet is independent of $\sigma$. For this purpose, we consider a channel whose output $\mathbf{y}$ is obtained from $\mathbf{w}$ by $\mathbf{y} = (2/\sigma^2) \cdot \mathbf{w}$. The result is equivalent to an AWGN channel whose input alphabet is given by $\delta(i) = (2/\sigma^2) \cdot \mathbf{m}^{-i} = \mathbf{1}^{-i}$ where $\mathbf{1} \overset{\triangle}{=} [1, ..., 1]^T$ and whose noise is distributed as $\mathcal{N}(\mathbf{0}, \Sigma_z)$ where $\Sigma_z = (2/\sigma^2)^2 \Sigma$. Letting $\sigma_z \overset{\triangle}{=} 2/\sigma$, we obtain that $\Sigma_z$ is defined as the matrix $\Sigma$ of (25) with $\sigma$ substituted by $\sigma_z$.

The multiplication by $2/\sigma^2$ does not fool the initial messages of LDPC decoding, and thus $\mathbf{r'^{(0)}} = \mathbf{w} = (\sigma^2/2) \cdot \mathbf{y} = (2/\sigma_z^2) \cdot \mathbf{y}$. We summarize these results in the following lemma.

*Lemma 35:* Consider transmission over a $q - 1$-dimensional AWGN channel, and assume zero-mean noise with a covariance matrix $\Sigma_z$ defined as the matrix $\Sigma$ of (25) with $\sigma$ substituted by $\sigma_z$. Assume the following mapping from the code alphabet $\delta(i) = \mathbf{1}^{-i}, i = 0, ..., q - 1$, where $-i$ is defined using LLR representation and $\mathbf{1}$ is defined above.

1) Let $\mathbf{y}$ denote the $q - 1$ dimensional channel output and $\mathbf{r'^{(0)}}$ denote the LLR-vector initial message. Then $\mathbf{r'^{(0)}} = 2/\sigma_z^2 \cdot \mathbf{y}$.

2) Let the random variable $\mathbf{R'^{(0)}}$ denote the initial message, conditioned on the all-zero codeword assumption. Then $\mathbf{R'^{(0)}}$ is Gaussian distributed, and satisfies (25) with $\sigma = 2/\sigma_z$.

### D. Properties and Computation of $J(\cdot)$

We examine $J(\sigma)$ in lines analogous to the development of ten Brink [36] for binary codes. In Appendix VIII-C, we showed that a Gaussian $\mathbf{W}$ distributed as in Theorem 6 and characterized by $\sigma$, may equivalently be obtained as the initial message, under the all-zero codeword assumption, of a $q - 1$ dimensional AWGN channel characterized by a parameter $\sigma_z = 2/\sigma$. The capacity of this channel is $J(\sigma) = I(C; \mathbf{W})$. The parameter $\sigma_z$ infers an ordering on the AWGN channels such that channels with a greater $\sigma_z$ are degraded with respect to channels with a lower $\sigma_z$. Thus $J(\sigma)$ is monotonically increasing and $J^{-1}(\cdot)$ is well-defined. As $\sigma \to \infty$, $\sigma_z$ approaches zero. Thus

$$\lim_{\sigma \to \infty} J(\sigma) = 1$$

Similarly,

$$\lim_{\sigma \to 0} J(\sigma) = 0$$

To compute $J(\cdot)$ and $J^{-1}(\cdot)$, we need to evaluate (26) for a Gaussian random variable as defined in Theorem 6. Following [35], we evaluate (26) for values of $\sigma$ along a fine grid in the range $\sigma \in (0, ..., 6.5)$ (6.5 being selected because $J(6.5) \sim 1$), and then applied a polynomial best-fit to obtain an approximation of $J(\cdot)$ and $J^{-1}(\cdot)$ (note that this operation is performed once: the resulting polynomial approximations of $J(\cdot)$ and $J^{-1}(\cdot)$ are the same for all codes).

In [35] the equivalent $J(\cdot)$ was evaluated by numerically computing the one-dimensional integral by which the expectation is defined. In our case, the distribution of $\mathbf{W}$ is multidimensional, and is more difficult to evaluate.

We therefore evaluate the right-hand side of (26) empirically, by generating random samples of $\mathbf{W}$ according to Theorem 6.

### E. Computation of $J_R(\sigma;\ \sigma_z, \delta)$

The computation of $J_R(\sigma;\ \sigma_z, \delta)$ is performed in lines analogous to the computation of $J(\sigma)$ as described in Appendix VIII-D. We compute $J_R(\sigma;\ \sigma_z, \delta)$ for fixed values of $\sigma_z$ and $\delta$ and for values of $\sigma$ along a fine grid in the range $\sigma \in (0, ..., 6.5)$. We then apply a polynomial best-fit to obtain an approximation of $J_R(\sigma;\ \sigma_z, \delta)$ for all $\sigma$ and an approximation of $J_R^{-1}(I;\ \sigma_z, \delta)$.

To compute $J_R(\sigma;\ \sigma_z, \delta)$ at a point of the above discussed grid, we evaluate the right-hand side of (26) (replacing $\mathbf{W}$ with a rightbound LLR-vector message $\mathbf{R}'$) empirically. Samples of $\mathbf{R}'$ are obtained by adding samples of initial messages to those of intermediate values. The samples of the initial messages are produced using Lemma 12 (with the coset symbol $v \in \{0, ..., q-1\}$ randomly selected with uniform probability). The samples of the intermediate values, for a given $\sigma$, are produced using Theorem 6.

Note that unlike $J(\sigma)$, which satisfies $J(0) = 0$, $J_R(0;\ \sigma_z, \delta)$ is greater than zero. This results from the fact that the distribution of the rightbound message $\mathbf{R}'$ corresponding to $\sigma = 0$ is equal to the initial message $\mathbf{R}'^{(\mathbf{0})}$, and $I(C;\ \mathbf{R}'^{(\mathbf{0})}) > 0$. Letting $I^{(0)} = I(C;\ \mathbf{R}'^{(\mathbf{0})})$, we have that $J_R^{-1}(I;\ \sigma_z, \delta)$ is not defined in the range $I \in [0, I^{(0)})$.

### F. Computation of $I_{E,CND}(I_A;\ j, \sigma_z, \delta)$

Our development begins in the lines of Appendices VIII-D and VIII-E. We compute $I_{E,CND}(I_A;\ j, \sigma_z, \delta)$ for fixed values of $\sigma_z$ and $\delta$ and for values of $I_A$ along a fine grid. We then apply a polynomial best-fit to obtain an approximation of $I_E(\sigma;\ j, \sigma_z, \delta)$ for all $\sigma$ in this range.

To compute $I_{E,CND}(I_A;\ j, \sigma_z, \delta)$ at a point of the above discussed grid, we again evaluate the right-hand side of (26) empirically. We begin by applying $J_R^{-1}(I_A;\ \sigma_z, \delta)$ to obtain the value of $\sigma$ which (together with $\sigma_z$ and $\delta$) characterizes the LLR-vector rightbound message distribution. We then produce samples of rightbound messages as described in Appendix VIII-E. We also produce samples of labels $g \in \mathrm{GF}(q) \backslash \{0\}$ that are required to compute the leftbound samples $\bar{\mathbf{l}}'$ of $\bar{\mathbf{L}}'$. The label samples are generated by uniform random selection. We use the samples $\bar{\mathbf{l}}'$ of $\bar{\mathbf{L}}'$ to empirically evaluate the right-hand side of (26) (replacing $\mathbf{W}$ with $\bar{\mathbf{L}}'$) and obtain $I_{E,CND}(I_A;\ j, \sigma_z, \delta)$. Note that computing (26) with $\bar{\mathbf{L}}'$ instead of $\mathbf{L}'$ had no effect on the final result.

Finally, $I_{E,CND}(I_A;\ j, \sigma_z, \delta)$ as defined in Section VII-E, like $J_R^{-1}(I;\ \sigma_z, \delta)$ (discussed in Appendix VIII-E), is not defined for $I \in [0, I^{(0)})$. This interval is not used in the EXIT chart analysis of Section VII-E.

### ACKNOWLEDGMENT

The authors would like to thank the anonymous reviewers and the associate editor for their comments and help.